\documentclass[prb,aps,preprint]{revtex4}
\usepackage[T1]{fontenc}
\usepackage{lmodern}
\usepackage{graphicx}
\usepackage{wrapfig}
\usepackage{amsmath}
\usepackage{color}
\usepackage{amssymb}

\newcommand{\Fs}{{FeSCs}}

\def\k{{\bf k}}
\def\p{{\bf p}}
\newcommand{\be}{\begin{equation}}
\newcommand{\ee}{\end{equation}}
\newcommand{\bea}{\begin{eqnarray}}
\newcommand{\eea}{\end{eqnarray}}
\newcommand{\beq}{\begin{equation}}
\newcommand{\eeq}{\end{equation}}

\def\k{{\bf k}}

\newcommand{\KFA}{KFe$_2$As$_2$}

\begin{document}

\title{Pairing mechanism in Fe-based  superconductors}

\author{Andrey Chubukov}

\affiliation {Department of Physics, University of Wisconsin,
Madison, Wisconsin 53706, USA}
\date{\today}

\pacs{74.20.Rp}
\date{August 17, 2011}

\begin{abstract}
I review recent works on the symmetry and the structure of the superconducting  gap in Fe-based superconductors and
 on the underlying pairing mechanism in these systems.  The experimental data on superconductivity show very rich behavior,
  with potentially different symmetry of a superconducting state for different compositions of the same material.
 The variety of different pairing states raised the issue whether
 the physics of Fe-based superconductors is model-dependent or is universal, governed by a single underlying pairing mechanism.
 I argue that the physics is universal and that all pairing states obtained so far can be understood within the same universal pairing
    scenario and are well described by the effective low-energy model with a small number of input parameters.
 \end{abstract}

\maketitle

\section{Introduction}
\label{sec:1}
The discovery, in 2008, of superconductivity in $Fe$-based pnictides [\onlinecite{bib:Hosono}] (binary compounds of the elements from the 5th group: N, P, As, Sb, Bi)
 was, arguably, among the most significant breakthroughs in condensed matter physics during the past decade. A lot of efforts by the condensed-matter community have
been devoted in the few years  after the discovery
to  understand  normal state properties
 of these materials, the pairing mechanism, and the symmetry and the structure
 of the pairing gap.

The family of $Fe$-based superconductors (FeSCs)  is already quite large and
keeps growing. It includes various Fe-pnictides such as $1111$ systems RFeAsO
($R=$rare earth element)
[\onlinecite{bib:Hosono,bib:X Chen,bib:G Chen,bib:ZA Ren}], $122$ systems XFe$_2$As$_2$(X=alkaline earth
metals) [\onlinecite{bib:Rotter,bib:Sasmal,bib:Ni}],  111
systems like LiFeAs [\onlinecite{bib:Wang}], and
 also Fe-chalcogenides (Fe-based compounds with elements from the 16th group:
 S, Se, Te) such as FeTe$_{1-x}$Se$_x$
~~[\onlinecite{bib:Mizuguchi}] and AFe$_s$Se$_2$
($A = K, Rb, Cs$)~~[\onlinecite{exp:AFESE,exp:AFESE_ARPES}].

Parent compounds of FeSCs are  metals, in distinction to cuprate superconductors for which parent compounds are Mott
 insulators. Still, in similarity with the cuprates, in most cases these parent compounds are antiferromagnetically ordered~~[\onlinecite{Cruz}].
  Because  electrons which carry magnetic moments still travel relatively freely from site to site,
   the magnetic order is often termed as a spin-density-wave (SDW), by analogy with e.g., antiferromagnetic  $Cr$, rather than "Heisenberg antiferromagnetism" -- the latter term is reserved for systems in which electrons are "nailed down" to particular lattice sites by very strong Coulomb repulsion.

  Superconductivity (SC) in FeSCs  emerges upon either hole or electron doping  (see Fig. 1), but can also be induced  by pressure or by
isovalent replacement of one pnictide element by another, e.g., As by
 P (Ref.~[\onlinecite{nakai}]). In some systems, like LiFeAs~~[\onlinecite{bib:Wang}] and
LaFePO~~[\onlinecite{bib:Kamihara}], SC emerges already at zero doping, instead of of a magnetic order.

\begin{figure}[tbp]
\includegraphics[angle=0,width=\linewidth]{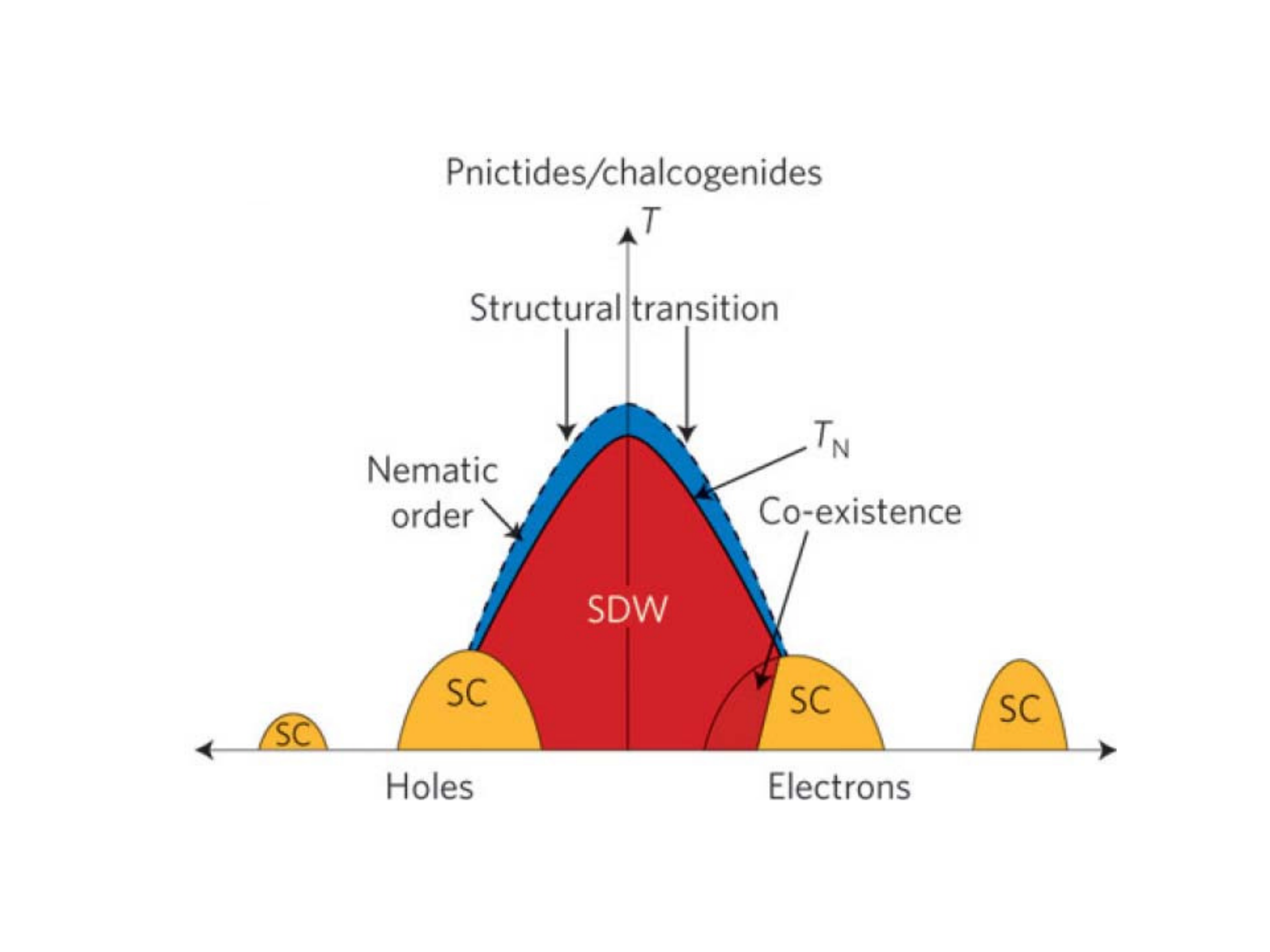}
\caption{Schematic phase diagram of Fe-based pnictides upon hole or electron doping.  In the shaded region, superconductivity and  antiferromagnetism co-exist. Not all details/phases are shown.
Superconductivity can be initiated not only by doping but also by pressure and/or isovalent replacement of one pnictide element by another~~[\onlinecite{nakai}].
 Nematic phase at $T > T_N$ is subject of debates.
 Superconductors at large doping are  KFe$_2$As$_2$ for hole doping~
~[\onlinecite{KFeAs_ARPES_QO,KFeAs_exp_nodal}] and A$_x$Fe$_{2-y}$Se$_2$ (A = K, Rb, Cs) for electron doping~~[\onlinecite{exp:AFESE,exp:AFESE_ARPES}].
 Whether superconductivity in pnictides exists at all
 intermediate dopings is not clear yet.  From Ref.~[\onlinecite{review_we}].}
\label{fig1}
\end{figure}

The magnetism, the electronic structure, the normal state properties
 of  FeSCs, and the interplay between FeSCs and cuprate superconductors have been reviewed in several recent publications~~[\onlinecite{review,review_2,review_3,review_4,Graser,peter,Kuroki_2,rev_physica,mazin_schmalian,review_we,korshunov}]. This review is an attempt to summarize our current understanding of the pairing mechanism and the symmetry and the structure of the
 pairing gap at various hole and electron dopings.

The phenomenon of SC has a long history.  SC has been discovered by Kamerlingh Onnes
 exactly a century ago~~[\onlinecite{Onnes}]).  It has been explained in general terms nearly fifty years later, in 1957, by Bardeen, Cooper, and
Schrieffer (BCS),  who demonstrated that an arbitrary weak attractive interaction between low-energy fermions is sufficient to pair them into a
  bound state. At weak coupling,  paired fermions immediately form Bose-Einstein condensate and behave as one single macroscopic quantum object
   and move coherently under the applied electric field, i.e superconduct.  In $d$-dimensional electronic systems low-energy fermonic states are located, in momentum space, near particular $d-1$ dimensional surfaces, called Fermi surfaces (FS) on which fermionic energy is zero relative to the chemical potential. At weak/moderate coupling, the pairing problem is confined to a near vicinity of a FS. The interaction between fermions is generally non-singular with respect to variations of the distance to the FS and can be approximated by its value right on the FS.

 What causes the attraction  between fermions is a more subtle question, and the nature and the origin of the pairing glue have been
  the subject of great debates in condensed-matter community over the last 50 years.  BCS  attributed the attraction between fermions
 to the underlying interaction between electrons and phonons~~[\onlinecite{BCS}] (the two electrons effectively interact with each other  by emitting and absorbing the same phonon which then serves as a glue which binds electrons into pairs).
 Electron-phonon mechanism has been successfully applied to explain SC in a large variety of materials, from $Hg$ and $Al$ to recently discovered and extensively studied $MgB_2$ with the transition temperature $T_c =39K$~~[\onlinecite{mgb2}].
 Non-phononic  mechanisms of the pairing have also been discussed, most notably in connection with
  superfluidity in $^3He$~~[\onlinecite{3he}], but didn't become the mainstream before the  discovery of SC in $LaBaCuO$
in 1986 ~[\onlinecite{bednortz}].
 That discovery, and subsequent discoveries of superconductivity at higher $T_c$ in other cuprates
 signaled the beginning of the new
era of ``high-temperature superconductivity'' to which FeSCs added a new avenue with  quite high traffic over the last three years.

Superconductivity is quite robust phenomenon. It has been known from early 60th~~[\onlinecite{stat_phys}] that in isotropic systems the equation for superconducting  $T_c$ factorizes  if one expands the interaction between the two fermions  in partial components  corresponding to interactions in the subspaces with
   a given angular momentum of the two interacting fermions $l =0, 1,2,3$, etc [in spatially isotropic systems $l=0$ component is called $s-$wave, $l=1$ component is called $p-$wave, $l=2$ component is called $d-$wave, and so on]. If just one component with some $l$ is attractive, the system undergoes a SC transition at some temperature $T=T_c$.  For phonon-mediate superconductors, $s-$wave superconductivity is the most likely outcome. In the cuprates, however, the pairing symmetry has been firmly established as $d-$wave.
    The vast majority of researches believe that such pairing is not caused by phonons and emerges instead due to screened Coulomb interaction between electrons. The screened Coulomb interaction $U(r)$ is constant and repulsive at short distances but has a complex dependence on $r$  at large distances and may develop an attractive component at some $l$.  One solid reason for the attraction, at least at large $l$, has been identified by Kohn and Luttinger back in 1965 (Ref. ~[\onlinecite{KL}]).

In lattice systems, angular momentum is no longer a good quantum number, and the  equation for $T_c$ only factorizes between different
 irreducible representations of the lattice space group. In tetragonal systems, which include both cuprates and FeSCs , there are four one-dimensional irreducible representations $A_{1g}$, $B_{1g}$, $B_{2g}$, and $A_{2g}$ and one two-dimensional representation $E_{2g}$.
Each representation has infinite set of eigenfunctions. The eigenfunctions from $A_{1g}$ are invariant under symmetry transformations in a tetragonal lattice: $x \to -x, ~y \to -y,~ x \to y$, the eigenfunctions from $B_{1g}$ change sign under $x \to y$, and so on.
If a superconducting gap has $A_{1g}$ symmetry, it is often called $s-$wave because the first eigenfunction from $A_{1g}$ group is just a constant in momentum space (a $\delta-$function in real space).  If the gap has $B_{1g}$ or $B_{2g}$ symmetry, it is called $d-$wave ($d_{x^2-y^2}$ or $d_{xy}$, respectably),
because in momentum space the leading eigenfunctions in $B_{1g}$ and $B_{2g}$ are
 $\cos k_x - \cos k_y$ and $\sin k_x \sin k_y$, respectively, and these two reduce to $l=2$ eigenfunctions $\cos 2 \theta$ and $\sin 2 \theta$ in the isotropic limit.

\begin{figure}[tbp]
\includegraphics[angle=0,width=\linewidth]{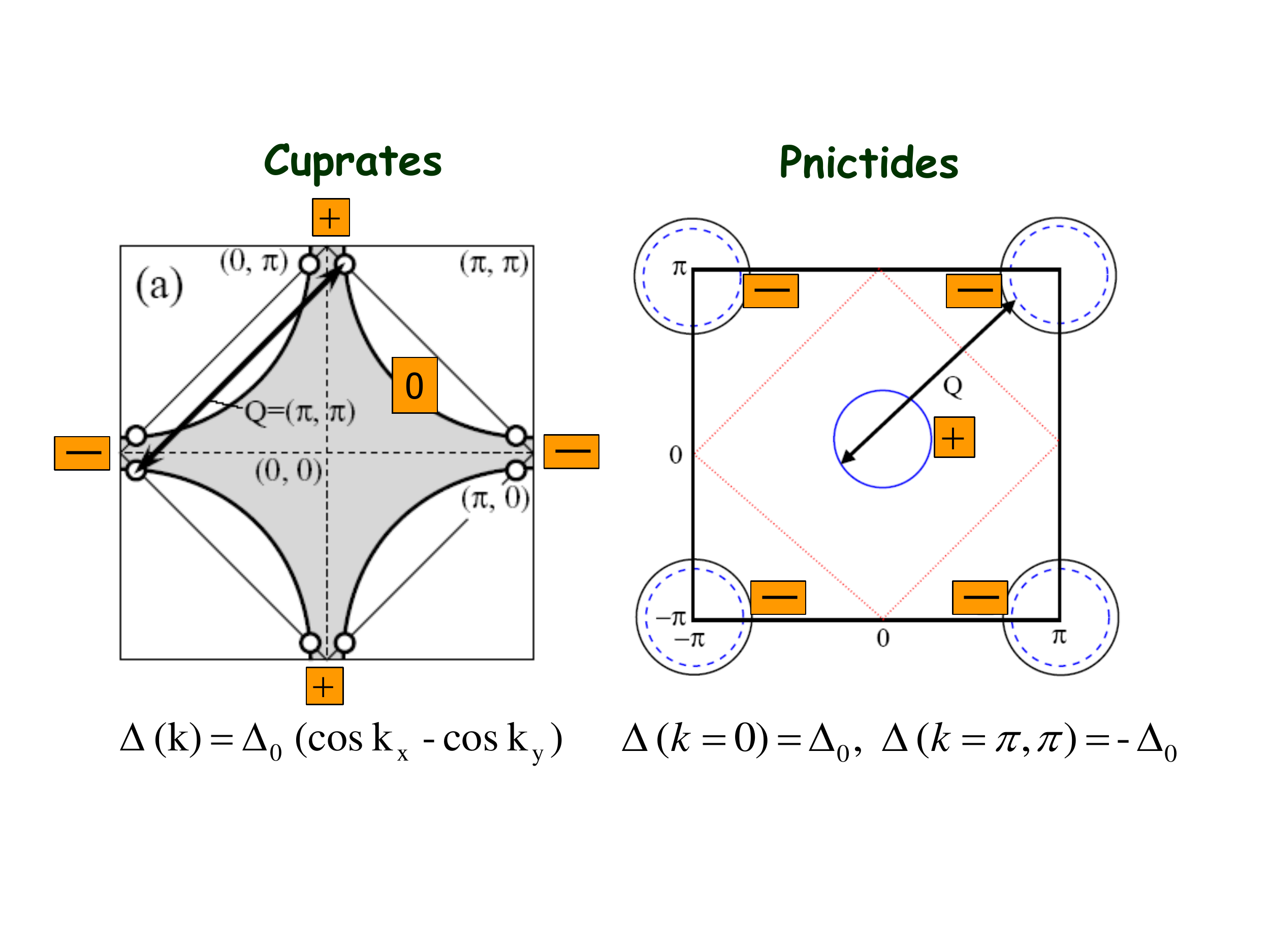}
\caption{A comparison of the pairing state from spin-fluctuation exchange in cuprate SCs and in FeSCs. In the cuprates (left panel) the FS is large, and antiferromagnetic ${\bf Q} = (\pi,\pi)$ connects points on the same FS. Because spin-mediated interaction is positive (repulsive), the
 gap must change sign between FS points separated by ${\bf Q}$. As the consequences, the gap changes sign twice along the FS. This implies a $d-$wave gap symmetry. In FeSCs (left panel) scattering by ${\bf Q}$ moves fermions from one FS to the other. In this situation, the gap must change sign between different FS, but to first approximation remains a constant on a given FS. By symmetry, such a gap is an $s-$wave gap. It is called $s^{+-}$ because it changes sign between different FSs}
\label{fig:comparison}
\end{figure}

In the cuprates, the superconducting gap has been proved experimentally to have $B_{1g}$ symmetry~~[\onlinecite{d-wave}].
 This gap symmetry appears quite naturally in the cuprates,  in the doping range where they are metals,
  if one assumes that the glue that binds fermions together is a spin-fluctuation exchange rather than a phonon (see Fig.\ref{fig:comparison}).
   The notion of a spin fluctuation is actually nothing but the convenient
  way to describe multiple Coulomb interactions between fermions. It is believed, although not proved rigorously, that in systems located
 reasonably close to a magnetic instability, the fully screened Coulomb interaction between fermions can be approximated by an effective interaction in which fermions exchange quanta of their collective fluctuations in the spin channel.  That $B_{1g}$ gap is selected is not
 a surprise because such gap $\Delta (k) \propto cos k_x - \cos k_y$ changes sign not only under $k_x \to k_y$ but also
  between ${\bf k}$  and ${\bf k}' = {\bf k} + {\bf Q}$ where ${\bf Q} = (\pi,\pi)$ is the momenta at which spin fluctuation-mediated pairing interaction $U({\bf k}, {\bf k}')$ is peaked.  This sign change is the crucial element for any electronic mechanism
  of superconductivity because one needs to extract an attractive (negative) component from repulsive (positive) screened Coulomb interaction. For $B_{1g}$ gap  such a component is $\int d{\bf k} d{\bf k'} \Delta (k) U({\bf k}, {\bf k}')\Delta (k')$, and the integral  obviously has a negative value when $U({\bf k}, {\bf k}')$ is peaked at $(\pi,\pi)$.

In FeSCs, magnetism and superconductivity are also close neighbors on the phase diagram, and it has been proposed~~[\onlinecite{Mazin,Kuroki}]
 at the very beginning of the $Fe$ era that
  the pairing  mechanism in FeSCs is also a spin-fluctuation exchange.  However,  the geometry of low-energy states in FeSCs and in the
  cuprates is different, and in most FeSCs the momentum ${\bf Q}$ connects low-energy fermionic states near the center and the corner of the Briilouin zone (see Fig.\ref{fig:comparison}). A simple experimentation with trigonometry then tell us that the SC  gap $\Delta (k)$ must be symmetric with respect to $k_x \to k_y$ and $k_x \to -k_x$, but still must change sign under ${\bf k} \to  {\bf k} + {\bf Q}$.
Such gap belongs to  $A_{1g}$ representation, but it only has contributions from a particular subset of $A_{1g}$ states with the form
$\cos k_x + \cos k_y$, $\cos {3 k_x}  +  \cos (3 k_y)$, etc which all change sign under  ${\bf k} \to  {\bf k} + {\bf Q}$.  Such gap is generally called an extended $s-$wave gap, or $s^{+-}$ gap.

 Majority of researches  do believe that in weakly/moderately doped FeSCs the gap does have $s^{+-}$ symmetry. However, numerous studies of superconductivity in FeSCs over the last three years demonstrated that   the physics of the pairing is
 more involved than it was originally thought because of multi-orbital/multi-band nature of low-energy fermionic excitations in FeSCs (see below). It turns out that both the symmetry and the structure of the pairing gap
 result from  rather non-trivial interplay between spin-fluctuation exchange, intraband Coulomb repulsion, and momentum structure of the interactions. In particular, an $s^\pm$wave gap can be with or without nodes, depending on the orbital content of low-energy excitations.
 In addition, the structure of low-energy spin fluctuations evolves with doping, and the same spin-fluctuation
mechanism that gives rise to  $s^{+-}$ gap at small/moderate doping in a particular material can give rise  to a $d-$wave gap at strong hole or electron doping.

There is more uncertainly on the theory side.  In addition to spin fluctuations,  FeSCs also possess charge fluctuations whose strength
is the subject of debates. There are proposals~~[\onlinecite{jap,ku}] that in multi-orbital FeSCs charge fluctuations are strongly enhanced because the system is reasonably close to a transition into a state with an orbital order (e.g., a spontaneous symmetry breaking between the occupation of different orbitals). A counter-argument is that orbital order does not develop on its own but is induced by a magnetic order~~[\onlinecite{rafael_we}].
 If charge fluctuations are relevant, one should consider, in addition to spin-mediated pairing interaction, also the
 pairing interaction mediated  by charge fluctuations. The last interaction can give rise to a conventional, sign-preserving  $s-$wave pairing~~[\onlinecite{jap}].  A "p-wave" gap scenario (a gap belonging to $E_{2g}$ representation) has also been put forward~~[\onlinecite{p_lee}].

 From experimental side, $s$-wave gap symmetry is consistent with ARPES data on moderately doped KFe$_2$As$_2$ and BaFe$_2$(As$_{1-x}$P$_x$)$_2$,
 which detected only a small variation of the gap along the FSs
centered at $(0,0)$ (Ref.~[\onlinecite{laser_arpes}]), and with the evolution of the tunneling data in a magnetic field~~[\onlinecite{hanaguri}.
However, for  heavily hole-doped \KFA  various experimental
probes~~[\onlinecite{KFeAs_exp_nodal}] indicate the presence of gap nodes,
which for the FS geometry in these materials~~[\onlinecite{KFeAs_ARPES_QO}] are consistent with a $d$-wave gap.
For the doping  range where the gap is very likely an $s-$wave,  the data on some  FeSCs were interpreted as evidence for the full gap~~[\onlinecite{ding,chen,osborn1,carrington1}, while the data for other FeSCs were interpreted
 as evidence that the gap has nodes~~[\onlinecite{nodes,carrington}] or deep minima~~[\onlinecite{proz,proz_1,moler}]. In addition,  recent
nuclear magnetic resonance (NMR) experiments on LiFeAs have been interpreted in favor of a $p$-wave gap~~[\onlinecite{buechner}].

In this paper, I  argue that  all these seemingly very different gap structures
 (with the exception of a $p-$wave), actually follow quite naturally from the same underlying physics idea that FeSCs can be treated as moderately interacting itinerant fermionic systems with multiple FS sheets and effective four-fermion
  intra-band and inter-band interactions in the band basis.
  I introduce the effective low-energy model with small numbers of input parameters~~[\onlinecite{maiti_last}] and use it to study the doping evolution of the pairing in hole and electron-doped FeSCs. It has been argued~~[\onlinecite{maiti_last}] that various approaches based on underlying microscopic model in the orbital basis reduce  to this model at low energies.

  The paper is organized as follows. In Sec. \ref{sec:2} I discuss general aspects of the pairing in FeSCs. I briefly review
   the band structure of FeSCs and show that it contains several bands of low-energy excitations. I then present generic symmetry considerations of the pairing in a multi-band superconductor.
  I show that a ``conventional wisdom'' that an s-wave gap
 is nodeless along the FSs, d-wave gap has 4 nodes, etc, has only  limited applicability in
  multi-band superconductors, and there are cases when the gap with four
nodes has an $s-$wave symmetry, and the gap without nodes has a $d-$wave symmetry.  In Sec. \ref{sec:3} I discuss the interplay between intra-band and inter-band interactions, first for a toy two-pocket model and then for realistic multi-pocket models, and set the conditions for an attraction in an $s-$wave or a
 $d-$wave channel. I consider 5-orbital model with local interactions, convert it into a band basis, and argue that for most of input parameters the bare interaction
 is repulsive in all channels  due to  strong intra-pocket Coulomb repulsion.
 In Sec. \ref{sec:4} I discuss the ways to overcome Coulomb repulsion.
 I review  random phase approximation (RPA) and renormalization group (RG) approaches and show that
 magnetic fluctuations
 enhance inter-pocket interaction, if this interaction is positive,
 what gives rise to an attraction in both $s^\pm$ and $d_{x^2-y^2}$  channels.
I briefly discuss $s^{++}$ pairing, which emerges if input parameters are such that  inter-pocket interaction is negative.
In Sec.\ref{sec:5} I  use  the combination of RPA and leading angular harmonic approximation (LAHA)  to analyze the pairing in $s-$ and $d-$wave channels at different dopings.  I show that  magnetically-mediated pairing
 leads to (i) an $s^\pm$ superconductivity with nodes on electron pockets for moderate electron doping, (ii)  an $s^\pm$ superconductivity without nodes for moderate hole dopings, (iii)  a nodeless  $d-$wave superconductivity for strong electron doping (except, possibly, small nodal regions near $k_z = \pi/2$), and (iv)  a nodal $d-$wave superconductivity  for strong hole doping.
 I  briefly review  the experimental situation in Sec. \ref{sec:7} and present concluding remarks in Sec. \ref{sec:7}.
I list separately the summary points and the list of future issues.

\section{Generic aspects of  pairing in {\text FeSCs}}
\label{sec:2}

\subsection{The electronic structure}
\begin{figure}[tbp]
\includegraphics[angle=0,width=0.8\linewidth]{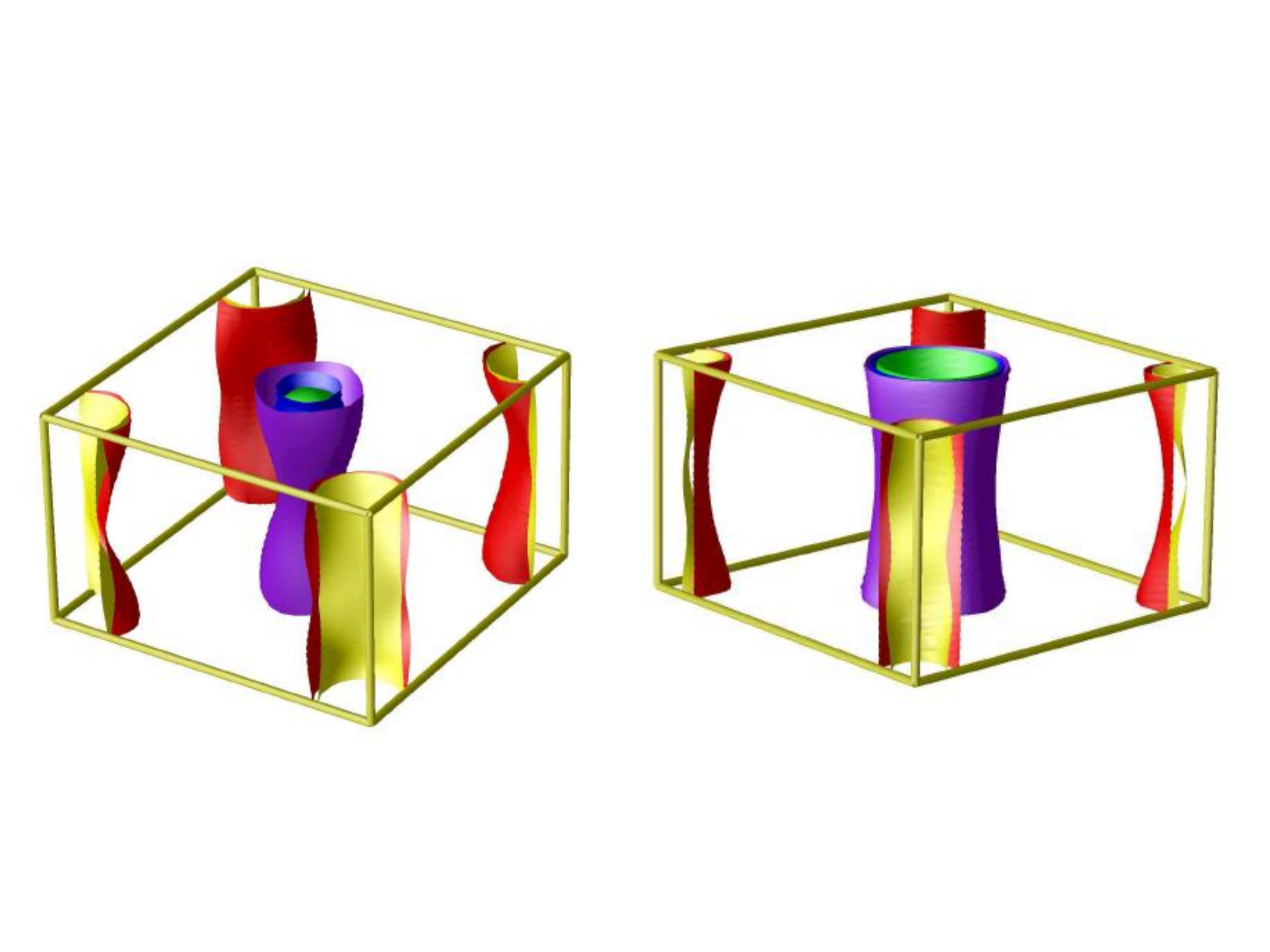}
\caption{The electronic structure of FeSCs. In  weakly and moderately electron-doped  materials (left panel)
  the FS consists of quasi-2D warped cylinders centered at $(0,0)$ and $(\pi,\pi)$ in a 2D cross-section. The ones near $(0,0)$ are hole pockets (filled states are outside cylinders), the ones near $(\pi,\pi)$ are electron pockets (filled states are inside cylinders)
  There also exists  a quasi-3D hole pocket
 near $k_z = \pi$. In hole-doped FeSCs the electronic structure is very similar, but 3D hole pocket becomes  quasi-2D warped hole cylinder.
  From Ref. ~[\onlinecite{mazin_schmalian}]. }
\label{fig:FS}
\end{figure}
The electronic structure of FeSCs at low energies is rather well established by
 ARPES~[\onlinecite{Li}] and quantum oscillation measurements~~[\onlinecite{bib:Suchitra}].
In  weakly and moderately  electron-doped  materials, like BaFe$_{1-x}$Co$_x$Fe$_2$As$_2$
  the FS contains
 several quasi-2D warped cylinders centered  at $k=(0,0)$ and $k= (\pi,\pi)$ in a 2D cross-section, and may also contain  a quasi-3D pocket
 near $k_z = \pi$ (Fig.\ref{fig:FS}).  The fermionic dispersion is electron-like
 near the FSs at $(\pi,pi)$  (filled states are inside a FS)  and hole-like near the FSs centered at $(0,0)$
  (filled states are outside a FS).  In heavily electron-doped \Fs, like
A$_x$Fe$_{1-y}$Se$_2$ (A = K, Rb, Cs), only electron pockets remain, according to recent ARPES studies.~~[\onlinecite{exp:AFESE}]
In weakly and moderately hole-doped \Fs,
like Ba$_{1-x}$K$_x$Fe$_2$As$_2$, the electronic structure is similar to that at moderate electron doping, however the spherical FS becomes
the third quasi 2D hole FS centered  at $(2\pi,0) = (0,0)$. In addition, new low-energy hole states likely appear around $(\pi,\pi)$
 and squeeze electron pockets~~[\onlinecite{Evtush}].  At strong hole doping, electron FSs disappear and only
  only hole FSs are present~~[\onlinecite{KFeAs_ARPES_QO}] These electronic structures agree well with  first-principle calculations
~~[\onlinecite{bib:first,Boeri,review_3}], which is another argument to treat
 FeSCs as itinerant fermionic systems.
\begin{figure}[tbp]
\includegraphics[angle=0,width=\linewidth]{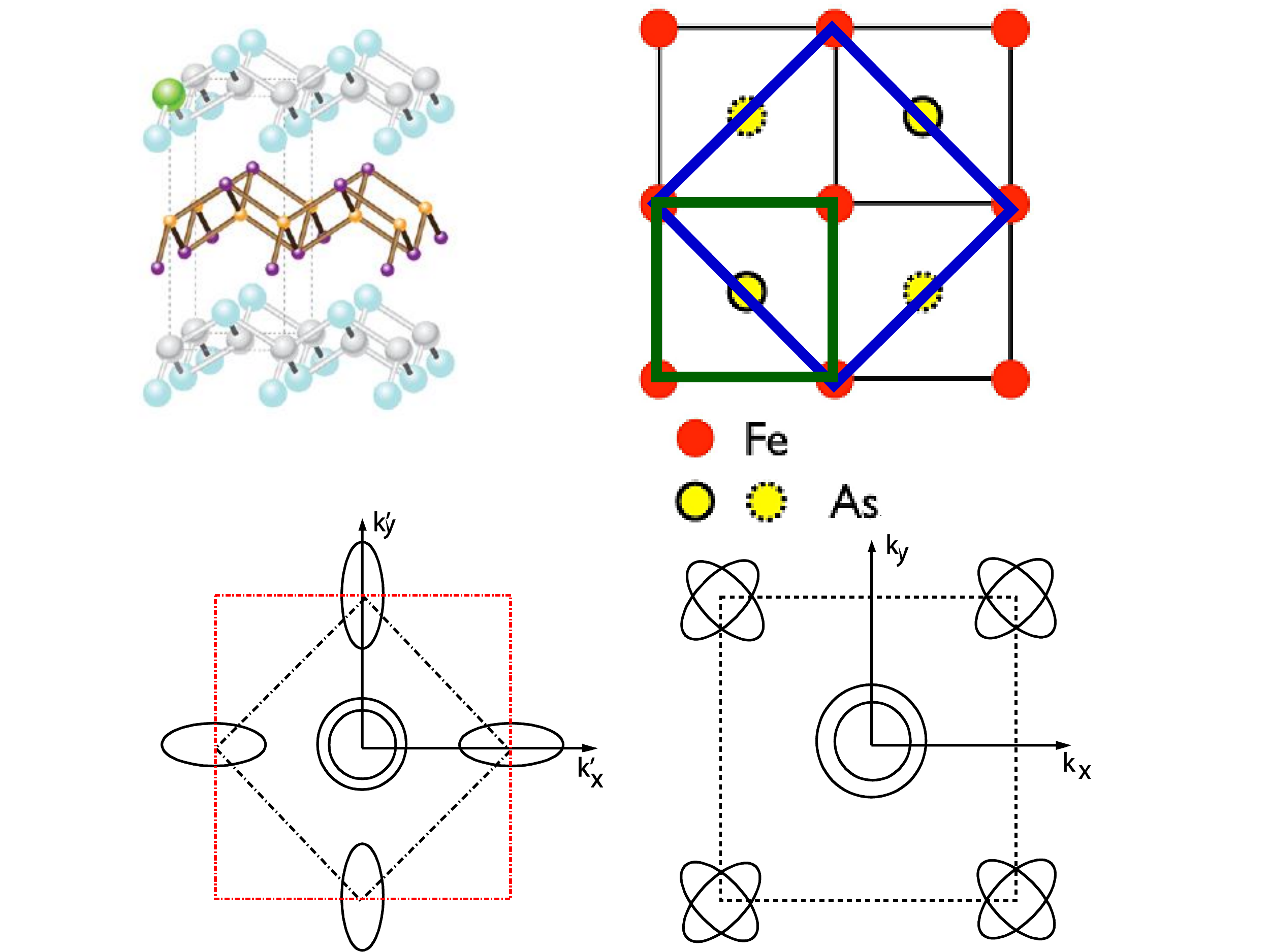}
\caption{Upper panel: 3D electronic structure of LaOFeAs (left) and its 2D cross-section (left). In only Fe states are considered, an elementary cell contains one Fe atom (green). The actual unit cell (blue) contains two Fe atoms because of two non-equivalent positions of a pnictide above and below the Fe plane.  Lower panel -- the location of hole and electron FSs in a 2D cross section  in the folded BZ (two Fe/cell, right) and
 in the unfolded BZ (one Fe/cell, left). From Refs. ~[\onlinecite{norman}], ~[\onlinecite{ising}](b) and ~[\onlinecite{vekhter_10}]b.}
\label{fig:folded}
\end{figure}
The measured FS reflects the actual crystal structure of FeSCs in which there are two non-equivalent positions of a pnictide above and below  an $Fe$ plane, and, as a result, there are two $Fe$ atoms in the unit cell (this actual Brillouin zone (BZ) is called  "folded BZ"). From theory perspective, it would be easier to work in the BZ which contains only one $Fe$ atom
in the unit cell (this theoretical BZ is called "unfolded BZ"). I illustrate the difference between folded and unfolded BZ in Fig.\ref{fig:folded}.  In general, only folded BZ is physically meaningful. However, if by some reason a potential from
 a pnictogen (or chalcogen) can be neglected, the difference between the folded and the unfolded BZ becomes  purely geometrical:
  the  momenta ${\tilde k}_x$ and ${\tilde k}_y$ in the folded BZ are linear combinations of $k_x$ and $k_y$ in the unfolded BZ: ${\tilde k}_x = k_x + k_y$, ${\tilde k}_y = k_x -k_y$.  In this situation, folded and unfolded BZ become essentially equivalent.

  Most of the existing theory works on the pairing mechanism and the structure of the SC gap analyze the pairing problem in the unfolded BZ, where which two hole pockets are centered at $(0,0)$ and one at $(\pi,pi)$, and the two electron pockets are are at $(0,\pi)$ and $(\pi,0)$.
  It became increasingly clear recently that the interaction via a pnictogen/chalcogen and also 3D effects do play some role for the pairing, particularly in strongly electron-doped systems.~~[\onlinecite{3D,mazin}] However, it is still very likely that the key aspects
  of the pairing in FeSCs can be understood  by analyzing a pure 2D electronic structure with only Fe states involved. Below I assume that this is the case and consider a 2D model in the unfolded BZ with hole FSs near $(0,0)$ and $(\pi,pi)$ and electron FSs at $(0,\pi)$ and $(\pi,0)$.

\subsection{The structure of $s-$wave and $d-$wave gaps in a multi-band SC}

I now use the multi-band electronic structure as input and consider the pairing problem at weak coupling.
I show that an $s-$wave gap generally has angle dependence on electron FSs and may even has nodes, while a d-wave gap which is normally assumed to have nodes, may in fact be nodeless on the same electron FSs.

\begin{figure}[tbp]
\includegraphics[angle=0,width=0.7\linewidth]{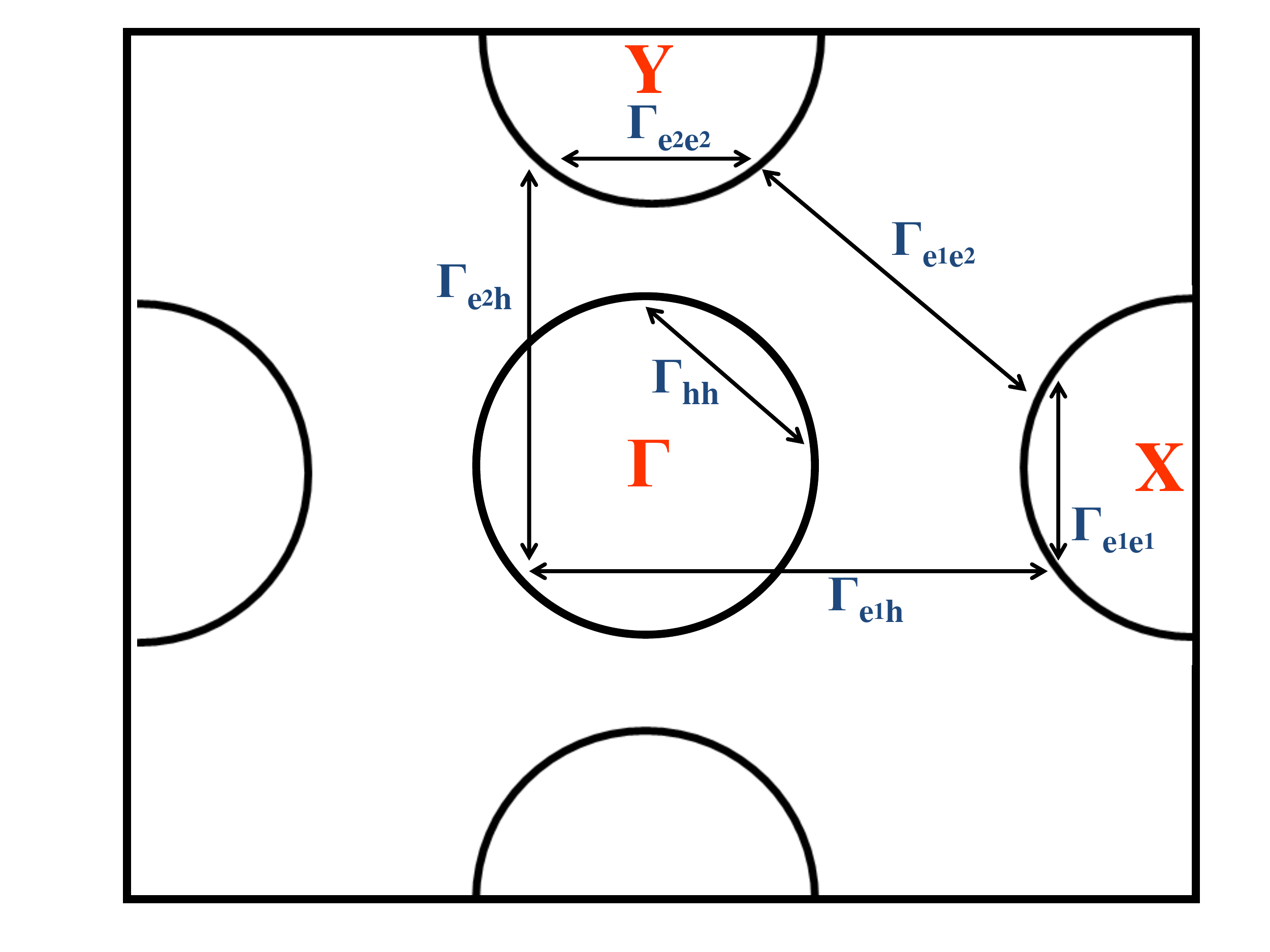}
\caption{Intra-pocket and inter-pocket interactions in 4-band 2D model for FeSCs. For simplicity, only one hole FS is shown. $\Gamma$, $X$, and $Y$ points are $(0,0), (\pi,0)$, and $(0,\pi)$, respectively.
$\Gamma_{hh}$ is the interaction within hole pocket, $\Gamma_{e_1 h}$ and $\Gamma_{e_2 h}$ are the interactions between hole and one of electron pockets, and $\Gamma_{e_1 e_1}$, $\Gamma_{e_2, e_2}$ (not shown) and $Gamma_{e_1 e_2}$ are intra-pocket and inter-pocket interactions involving the two electron pockets. Each interaction contains $s-$wave and $d-$wave component, and even for $s-$wave $\Gamma_{e_i h}$ and $\Gamma_{e_i e_j}$ depend on the angles along electron FSs.  From Ref. ~[\onlinecite{maiti_10}]. }
\label{fig:interactions}
\end{figure}

A generic low-energy BCS-type model in the band basis is  described by
 \beq {\cal H} = \sum_{i,\k}
\epsilon_{i} (\k) a^\dagger_{i \k}  a_{i \k} + \sum_{i,j, \k, \p}
\Gamma_{i,j} (\k,\p) a^\dagger_{i \k} a^\dagger_{i -\k} a_{j \p}a_{j
-\p}
 \label{rev_2}
 \eeq
The quadratic  term describes low-energy
excitations near hole and electron FSs, labeled by $i$ and $j$, and the
four-fermion term describes the scattering of a pair $(k \uparrow,
-k\downarrow)$ on the FS $i$ to a pair $(p \uparrow,
-p\downarrow)$ on the FS $j$.
These interactions are either intra-pocket interactions (hole-hole $\Gamma_{h_ih_i}$ or electron-electron $\Gamma_{e_ie_i}$ ), or inter-pocket interactions
(hole-electron $\Gamma_{e_j h_i}$, hole-hole $\Gamma_{h_i \neq h_j}$, and  electron-electron $\Gamma_{e_i \neq e_j}$).
  I illustrate this in Fig.\ref{fig:interactions}.

Assume for simplicity that
 the frequency dependence of $\Gamma$ can be neglected and low-energy fermions are Fermi-liquid quasiparticles with
   Fermi velocity $v_{k_F}$.  In this situation,  the gap $\Delta (k)$ also doesn't depend on frequency, and  the linearized gap equation becomes
 the eigenfunction/eigenvalue problem:
\beq
\lambda_i\Delta_i (k) = - \int \frac{d p_\parallel}{4\pi^2  v_{{\bf p}_F}}
\Gamma({\bf k}_F,{\bf p}_F) \Delta_i (p)
\label{gap_eq}
\eeq
where $\Delta_i$ are  eigenfunctions and $\lambda_i$ are eigenvalues.  The system is unstable towards pairing if one or more $\lambda_i$ are
  {\it positive}. The corresponding $T_{c,i}$ scale as $T_{c,i} = \Lambda_i e^{-1/\lambda_i}$. Although $\Lambda_i$ are generally different for different $i$, the exponential dependence on $1/\lambda_i$ implies that, most likely, the solution with the largest positive $\lambda_i$ emerges first and establish the pairing state, at least immediately below $T_c$.

Like we discussed in the Introduction, the vertex $\Gamma (k,p)$ can be decomposed into
 representations of the tetragonal space group (one-dimensional representations are
 $A_{1g}$, $B_{1g}$, $B_{2g}$, and $A_{2g}$).
  Basis functions from different representations do not mix,
but each contains infinite number of components.  For example, $s-$wave
pairing corresponds to  fully symmetric $A_{1g}$ representation, and the
 $s-$wave ($A_{1g}$) component of $\Gamma (k,p)$
 can be quite generally expressed
as
\beq
\Gamma^{(1g)}(k,p)= \Gamma_s(k,p)=\sum_{m,n}A^{s}_{mn}\Psi^s_m(k)\Psi^s_n(p) \label{s_1} \eeq

where $\Psi^s_m (k)$ are the basis functions of the A$_{1g}$ symmetry
group: $1$, $\cos k_x \cos k_y$, $\cos k_x + \ cos k_y$, etc, and $A^{s}_{mn}$
are coefficients. Suppose that $k$ belongs to a
hole FS and is close to $k=0$. Expanding {\it any} wave function
with A$_{1g}$ symmetry near $k=0$, one obtains along $|{\bf k}| =
k_F$,
\beq
 \Psi^s_m(k) = a_m + b_m \cos 4 \phi_k + c_m \cos 8 \phi_k +...
\label{s_2}
 \eeq
where $\phi_k$ is the angle along the hole FS (which is not necessary a circle).
Similarly, for $B_{1g}$ representation the wavefunctions are
$\cos k_x - \cos k_y$, $cos 2k_x - cos 2k_y$, etc, and expanding them near $k=0$ one obtains
\beq
 \Psi^d_m(k) = a^*_m \cos 2 \phi_k  + b^*_m \cos 6 \phi_k + c^*_m \cos 10 \phi_k +...
\label{d_2}
 \eeq
There are no fundamental reasons  to expect that $b_m,~c_m$ or $b^*_m, c^*_m$
are much smaller than $a_m$ or $a^*_m$, but sub-leading terms are often small
numerically. The known example is the numerical smallness of
$\cos (2+4n) \phi$  components with $n \geq 1$ of the $d_{x^2-y^2}-$wave gap in
the spin-fluctuation scenario for the cuprates~~[\onlinecite{mike}] (meaning that a d-wave gap can be reasonably well approximated by $\cos 2 \phi$).
 Taking this example as circumstantial
evidence, I neglect all subleading terms, i.e., assume that $s-$wave interaction between fermions on the hole FSs  can
be approximated by an angle-independent $\Gamma^s_{h_ih_j} (k,p) =
U_{h_ih_j}$ ($h_i$ label different hole FSs), while $d-$wave ($B_{1g}$) interaction can be approximated by $\Gamma^d_{h_ih_j} (k,p) = {\tilde U}_{h_ih_j} \cos 2 \phi_k \cos 2\phi_p$.

The situation changes, however, when we consider the pairing component
 involving fermions from electron FSs. Suppose
that $k$ are still near the center of the BZ, but $p$
are near one of the two electron FSs, say the one centered at
$(0,\pi)$.  Consider all possible $\Psi_n(p)$ with $A_{1g}$
symmetry A simple experimentation with trigonometry shows that
there are two different subsets of basis functions:
\bea
&&{\text subset}~ I: 1,~ \cos{p_x} \cos{p_y},~ \cos{2 p_x} + \cos{2p_y} ... \nonumber \\
&&{\text subset}~ II: \cos{p_x} + \cos{p_y},~\cos{3 p_x} + \cos{3 p_y}...
\label{s_3} \eea
For a circular FS centered at $(0,\pi)$, the functions
 from subset I can be again expanded in series of $\cos 4 l \phi_p$ with integer $l$.
The functions from subset II are different -- they all vanish
at $(0,\pi)$ and are expanded in series of $\cos (2\phi_p + 4 l
\phi_p)$ (the first term is $\cos 2 \phi_p$, the second is
$~\cos 6 \phi_p$, etc).  For elliptic  FS $\cos 4 l \phi_p$ and $\cos (2\phi_p + 4 l\phi_p)$ terms  appear in both subsets.  In both cases, the total
\bea
 \Psi^s_m(p) &=& {\bar a}_m + {\bar b}_m \cos 4 \phi_p + {\bar c}_m
\cos 8 \phi_p + ...\nonumber\\
&&+  {\bar {\bar a}}_m \cos 2 \phi_p  + {\bar {\bar b}}_m \cos 6 \phi_p + {\bar{\bar
c}}_m \cos 10 \phi_k +...
\label{s_2p}
 \eea
For the other electron FS,  $\Psi^s_m(p)$ is the same, but   momentum components  $p_x$ and $p_y$are interchanged, hence the sign of all
$\cos (2\phi +4l\phi_p)$ components changes.

Let's make the same approximation as
before and neglect all components with $l >0$.  Then
\beq
 \Psi^s_m(p) = {\bar a}_m \pm  {\bar {\bar a}}_m \cos 2 \phi_p
\label{s_2p1}
 \eeq
where the upper sign is for one electron FS and the lower for the other.
It is essential that the angle-independent term and the  $\cos 2 \phi_p$ term
 have to be treated on equal footing because each  is the {\it leading} term in the corresponding series. Combing (\ref{s_2p1})  with the fact that $ \Psi^s_m(k)$ can be approximated by a constant, we obtain a generic form of the
 $s-$wave component of the interaction between fermions near hole and electron FSs
\bea
\Gamma^s_{e_1,h_i}(k,p) &=& U_{e,h_i} \left(1 + 2\alpha_{e,h}\cos 2
\phi_{p_{e1}} +...\right) \nonumber\\
 \Gamma^s_{e_2,h_i}(k,p) &=& U_{e,h_i} \left(1 - 2\alpha_{e,h}\cos 2
\phi_{p_{e2}} + ....\right)
\label{s_4}
\eea
where dots stand for $\cos{4 \phi_k}, \cos{4\phi_p},
\cos{6\phi_p}$, etc terms.

 By the same reasoning, $s-$wave components of  inter-pocket and intra-pocket interactions between fermions from electron FSs are
\bea
\Gamma^s_{e_1,e_1}(k,p) &=& U_{e,e} \left(1+2\alpha_{ee}
\left(\cos2\phi_{k_{e1}}+ \cos2\phi_{p_{e1}}\right)\right. \nonumber \\
&&+ 4 \beta_{ee} \cos2\phi_{k_{e1}} \cos2\phi_{p_{e1}} +... \nonumber \\
\Gamma^s_{e_2,e_2}(k,p) &=& U_{e,e} \left(1 - 2\alpha_{ee}
\left(\cos2\phi_{k_{e2}}+ \cos2\phi_{p_{e2}}\right)\right. \nonumber \\
&&+ 4 \beta_{ee} \cos2\phi_{k_{e2}} \cos2\phi_{p_{e2}} +... \nonumber \\
\Gamma^s_{e_1,e_2}(k,p) &=& U_{e,e} \left(1 + 2\alpha_{ee}
\left(\cos2\phi_{k_{e1}} - \cos2\phi_{p_{e2}}\right)\right. \nonumber \\
&&- 4 \beta_{ee} \cos2\phi_{k_{e1}} \cos2\phi_{p_{e2}} +...
\label{s_5}
\eea

Once the pairing interaction has the form of Eqs. (\ref{s_4}) and
(\ref{s_5}), the  gaps along the hole FSs are angle-independent (modulo $\cos 4 \phi$ terms), but
the gaps along the two electron FSs are of the form
\beq
\Delta^{(s)}_e (k) = \Delta_e \pm \bar{\Delta}_e \cos2\phi_k.
\eeq
 When $\bar{\Delta}_e$ is small compared
to $\Delta_e$, the angle dependence is weak, but when $|\bar{\Delta}_e| > |\Delta_e|$, $s-$wave gaps
have nodes at ``accidental'' values of $\phi$, which differ between the two electron FSs.

A similar consideration holds for $d_{x^2-y^2}$ gap.  Within the same approximation of leading angular momentum harmonics, we have
\bea
 \Gamma^d_{e_1,h_i}(k,p) &=& {\tilde U}_{e,h_i} \cos 2 \phi_{h_i}
\left(1 + {\tilde \alpha}_{e,h}\cos 2
\phi_{p_{e1}}\right) +... \nonumber\\
 \Gamma^d_{e_2,h_i}(k,p) &=&  {\tilde U}_{e,h_i} \cos 2 \phi_{h_i}
 \left(-1 + {\tilde \alpha}_{e,h}\cos 2
\phi_{p_{e2}} \right) + ...
\label{s_4_d}
\eea
and
\bea
\Gamma^d_{e_1,e_1}(k,p) &=& {\tilde U}_{e,e} \left(1+2\alpha_{ee}
\left(\cos2\phi_{k_{e1}}+ \cos2\phi_{p_{e1}}\right)\right. \nonumber \\
&&+ 4 \beta_{ee} \cos2\phi_{k_{e1}} \cos2\phi_{p_{e1}} +... \nonumber \\
\Gamma^d_{e_2,e_2}(k,p) &=& {\tilde U}_{e,e} \left(1 - 2\alpha_{ee}
\left(\cos2\phi_{k_{e2}}+ \cos2\phi_{p_{e2}}\right)\right. \nonumber \\
&&+ 4 \beta_{ee} \cos2\phi_{k_{e2}} \cos2\phi_{p_{e2}} +... \nonumber \\
\Gamma^d_{e_1,e_2}(k,p) &=& {\tilde U}_{e,e} \left(-1 - 2\alpha_{ee}
\left(\cos2\phi_{k_{e1}} - \cos2\phi_{p_{e2}}\right)\right. \nonumber \\
&&+ 4 \beta_{ee} \cos2\phi_{k_{e1}} \cos2\phi_{p_{e2}} +...
\label{s_5_d}
\eea

The solution of the gap equation then yields the gap in the
 form
\bea
&&\Delta^{(d)}_h (k) = {\tilde \Delta}_h \cos 2 \phi_k \nonumber \\
&&\Delta^{(d)}_e (k) = \pm {\tilde \Delta}_e  + \bar{{\tilde \Delta}}_e \cos2\phi_k.
\eea
Along the hole FS, the gap behaves as a conventional $d-$wave gap with 4 nodes along the
diagonals.  Along electron FSs, the two gaps differ in the sign of the angle-independent terms,
and have in-phase $\cos 2 \phi$ oscillating components. When $ \bar{{\tilde \Delta}}_e <<  {\tilde \Delta}_e$ the two electron gaps
 are simply ``plus'' and ``minus'' gaps, but when  $ \bar{{\tilde \Delta}}_e >  {\tilde \Delta}_e$, each has
accidental nodes, again along different directions on the two electron FSs.

 We see therefore that the geometry of the FSs in FeSCs affects the gap structure in quite fundamental way:
  because electron FSs are centered at the $k$ points which are not along BZ diagonals, $s-$wave gaps on these FSs have $\cos 2 \phi$
 oscillations which one normally would associate with a $d-$wave symmetry, and $d-$wave gaps have  constant (plus-minus) components
 which one would normally associate with an $s-$wave symmetry.  When these ``wrong'' components are large, the gaps
 have accidental nodes. These nodes
 may be present or absent for both $s-$wave and $d-$wave gaps, i.e.,
 symmetry constraints play no role here.

An $s-$wave gap with nodes in one of the ``exotic'' options offered by
 the electronic structure of FeSCs. Another ``exotic''  option is  a $d-$wave state without nodes. In heavily electron-doped FeSCs, hole states are gapped, and
 only electron FSs remain. The $d-$wave gaps on these two FSs have no nodes
 if $\cos 2 \phi$ oscillation component is smaller than a constant term, hence the system will display a behavior typical for a fully gapped SC despite that the gap actually has a d-wave symmetry. There are even more exotic options offered by the actual three-dimensionality of the electronic structure and/or the hybridization of the electron FSs due to interaction via a pnictide/chalcogen.
One RPA calculation~~[\onlinecite{3D}] places the nodes of the $s^\pm$ gap on hole pockets, near particular $k_z$. Another study argues~~[\onlinecite{mazin}] that a
 $d-$wave gap for heavily electron doped FeSCs must have nodes near $k_z = \pi/2$.

 It is essential, however, that  either a more conventional one or a more
 exotic pairing state develops only if the corresponding eigenvalue $\lambda_i$ is
 positive. To understand  under what conditions $\lambda_i >0$ we now have to consider specific models for FeSCs.

\section{The interplay between intra-pocket and inter-pocket interactions }
\label{sec:3}

In this section we make the first step in the analysis of what causes the attraction in FeSCs and consider how the sign and magnitude of $\lambda_i$ depends on the interplay between intra-pocket and inter-pocket interactions.

\subsection{Toy two-pocket model}
\label{sec:toy}

As a warm-up, consider first an idealized two-pocket model
 consisting of two identical circular pockets: a
 hole pocket at $(0,0)$ and an electron pocket at $(\pi,\pi)$. Because the electron FS is centered at $k$ along the diagonal,  the $\cos 2 \phi$ terms in the electronic gap are no longer present,  so some part of the physics of FeSCs is lost. Still, this idealized model is a good staring point to consider the interplay between intra-pocket and inter-pocket interactions.

 Compared to Eqs. (\ref{s_4}-\ref{s_5_d}), we now have
$\Gamma^s_{hh} = U_{hh}$, $\Gamma^s_{he} = U_{he}$, $\Gamma^s_{ee} =
 U_{ee}$, and  $\Gamma^d_{hh} (k_F,p_F)= {\tilde U}_{hh} \cos{2 \phi_k}
\cos{2 \phi_p}$, $\Gamma^d_{he} (k,p) = {\tilde U}_{he} \cos 2 \theta_k$,
 $\Gamma^s_{ee} = {\tilde U}_{ee}$.

The eigenvalue problem, Eq. (\ref{gap_eq}), reduces to the set of two coupled equations in each channel. Solving them, we obtain
\beq \label{eq:lambda_1s}
\lambda^s_{1,2}  = \frac{-(u_{hh}+u_{ee}) \pm
\sqrt{(u_{hh}-u_{ee})^2+4 u^2_{he}}}{2}
 \eeq
and
\beq
 \label{eq:lambda_1d}
\lambda^d_{1,2}  = \frac{-({\tilde u}_{hh}+{\tilde u}_{ee}) \pm
\sqrt{({\tilde u}_{hh}-{\tilde u}_{ee})^2+4 u^2_{he}}}{4}
 \eeq
where $u_{ij} = U_{ij} N_F$ and  $N_F = m/(2\pi)$ is the density of states  (DoS)  per spin projection.

We see that a solution with a positive $\lambda$ in either $s-$or $d-$wave channel exists if the inter-pocket interaction is larger than intra-pocket interactions. Specifically, one needs
\beq
u^2_{he} > u_{ee} u_{hh}~{\text or}~
{\tilde u}^2_{he} > {\tilde u}_{ee} {\tilde u}_{hh}.
\label{cond}
\eeq

If this condition is not met, the system remains in the normal state down to $T=0$. If Eq. (\ref{cond}) is satisfied and $u_{he} >0$,  $s-$wave solution with $\lambda^s >0$ yields a ``plus-minus'' gap, $\Delta_e = - \Delta_h$ ($s^\pm$ state). If  $u_{he} <0$ (what requires intra-band attraction),  $s-$wave solution is a conventional one, with $\Delta_e = \Delta_h$.  Similarly, $d-$wave solutions yield $\cos 2 \phi$ gaps on the hole and electron FSs with either zero or $\pi$ phase shift.

\subsection{Multi-band models}

Consider next a more realistic case of two electron FSs centered at $(0,\pi)$ and $(\pi,0)$.  Now hole-electron and electron-electron interactions have
 $\cos 2 \phi$ terms, and the eigenvalue/eigenfunction problem, Eq. (\ref{gap_eq}),
 reduces to the set of either four (or five) coupled equations in either s-wave or d-wave channels:  two  (or three) $\Delta$'s are the gaps on the hole FSs, and two other $\Delta$'s are angle-independent and $\cos 2 \phi$ components of the gaps
on the electron FSs.  Accordingly, there are either four or five different $\lambda_s$ and $\lambda_d$.

\begin{figure}[t]
$\begin{array}{cc}
\includegraphics[width =0.5\columnwidth ]{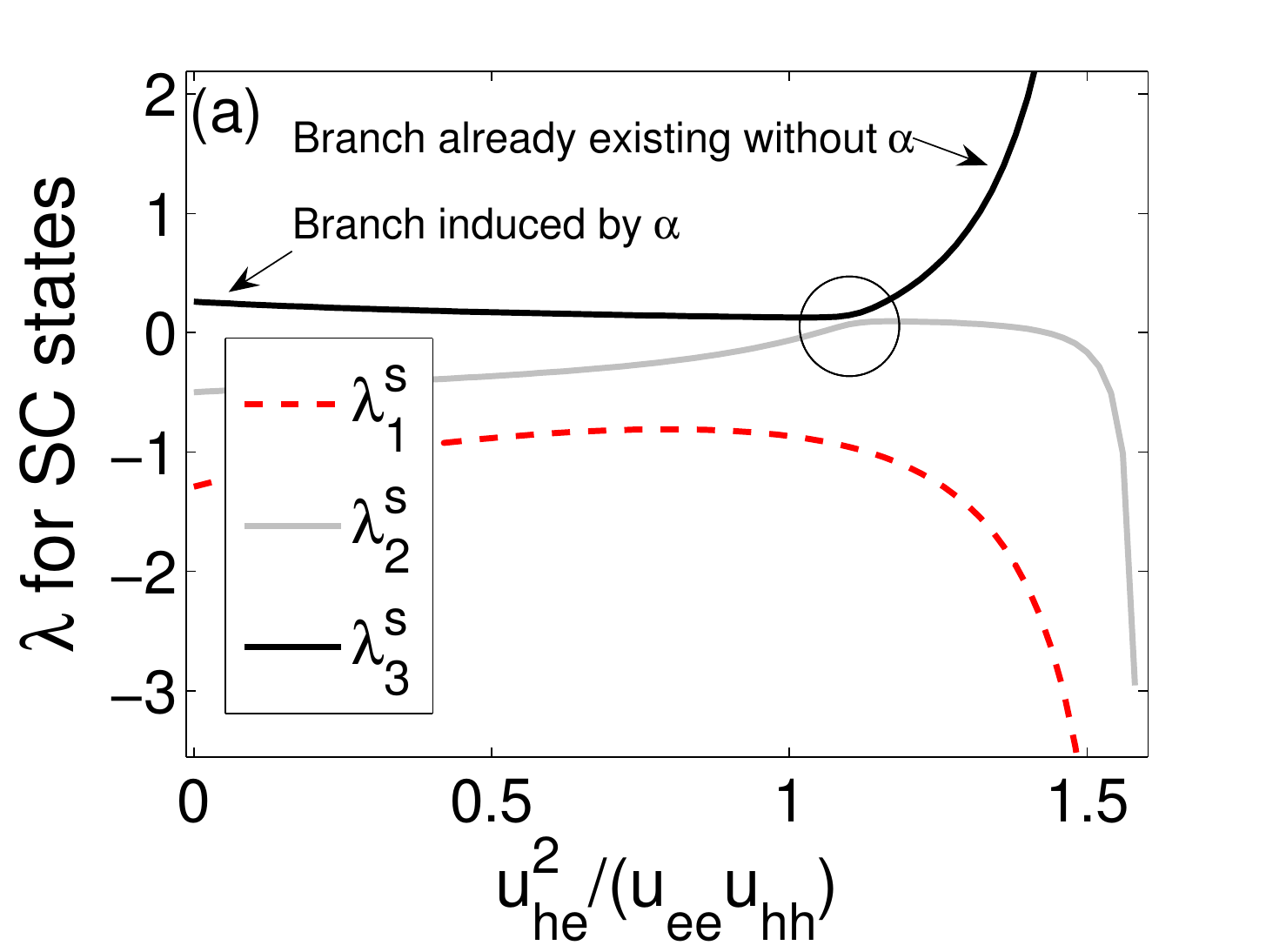}&
\includegraphics[width = 0.5\columnwidth]{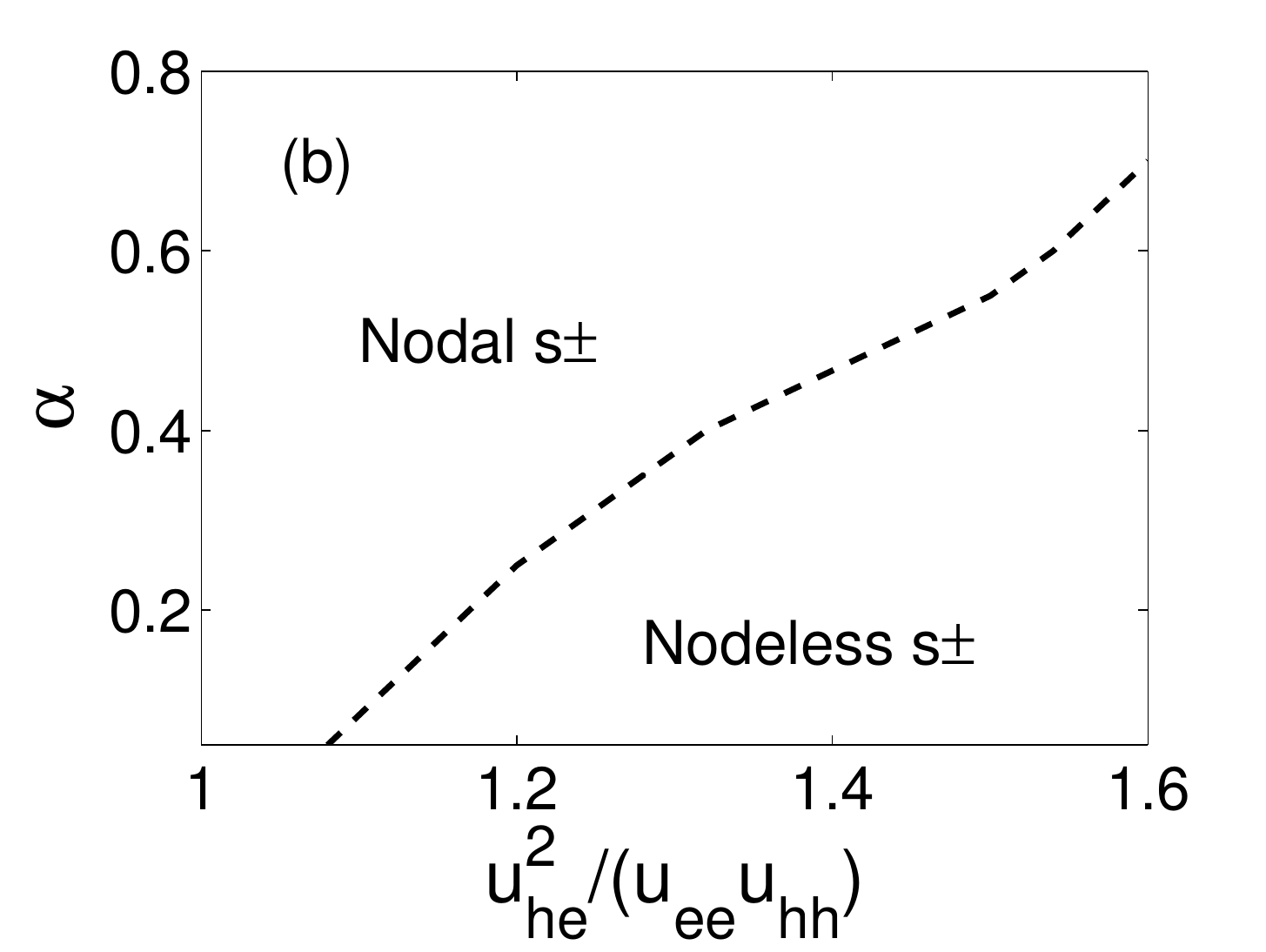}
\end{array}$
\caption{\label{fig:SC_vrtcs} (a) The three eigenvalues in the $s-$wave channel $\lambda^s_i$ as functions of $u^2_{he}/(u_{ee} u_{hh})$ for $\alpha_{ee} = \beta_{ee} =0$ and  $\alpha_{he} =0.4$. For any  $u^2_{he}/(u_{ee} u_{hh})$, one $\lambda^s_i$ is positive (attractive),  other two are negative.
 Positive $\lambda^s_i$ corresponds
to $s\pm$ pairing.
 At small  $u^2_{he}/(u_{ee} u_{hh})$ pairing is induced
by $\alpha_{he}$ and the gap has nodes on electron FSs.
At large  $u^2_{he}/(u_{ee} u_{hh})$ positive $\lambda^s_i$ exists
 already at $\alpha_{he} =0$, and the gap along electron FS has nodes only if $\alpha_{he}$ is above the threshold. The
circle marks the area where positive and negative solutions come
close to each other. The splitting between the two increases with
$\alpha_{he}$. (b) The regions of nodeless and nodal $s^\pm$ gap, depending on
$\alpha_{he}$ and $u^2_{he}/u_{ee} u_{hh}$. From Ref. ~[\onlinecite{maiti_10}].}
\end{figure}

The analysis of $4\times4$ or $5\times5$ gap equations is tedious but straightforward. I will not discuss it in length (for a detailed discussion see Refs.~~[\onlinecite{maiti_10,maiti_last}]) but rather focus on an issue of whether it is still required that the inter-pocket interaction $u_{he}$  exceeds  the threshold  set by intra-pocket hole-hole and electron-electron interactions. Interestingly enough, this may no longer be necessary. To illustrate this, consider the case of an $s-$wave pairing in a four-pocket model and assume for simplification
 that only one hole pocket is relevant to the pairing. Then the eigenvalue
 problem reduces to the set of three equations for $\Delta_h$, $\Delta_e$, and
 ${\bar \Delta}_e$  ($\Delta_e (k) = \Delta_e + {\bar \Delta}_e \cos {2\phi_k}$).
Solving the set, we find three solutions $\lambda^s_i$ ($i=1,2,3$).  In the absence of $\cos 2\phi$ terms in $\Gamma_{ij} (k,p)$, $\lambda^s_3 =0$, and
$\lambda^s_{1,2}$ are given by (\ref{eq:lambda_1s}) with $u_{ee} \to 2 u_{ee}$ and $u^2_{he} \to 2 u^2_{he}$.  Obviously, $u_{he}$ has to exceed a threshold, otherwise $\lambda^s_{1,2} <0$.
 Once the angle dependent terms in (\ref{s_4}-\ref{s_5}) become non-zero, $\lambda^s_3$ also becomes non-zero, and its sign depends on the interplay between
 $\alpha_{he}$, $\alpha_{ee}$, and $\beta_{ee}$.
 In particular, when $u^2_{he} < u_{ee}u_{hh}$ (and, hence, $\lambda_{1,2} <0$),
 $\lambda^s_3$
is positive or negative depending on whether or not $A >0$, where
\beq
A = 4 u_{ee}u_{hh} \left(\alpha^2_{ee} -\beta_{ee}\right)+u^2_{he} \left(\alpha^2_{he}+ 2 \beta_{ee} -3 \alpha_{he} \alpha_{ee}\right)
\label{inequality}
\eeq
When the angle-dependence of the electron-electron interaction
 can be neglected, i.e., $\alpha_{ee} = \beta_{ee} =0$, $\lambda^s_3 >0$ no matter what is the ratio of  $u^2_{he}$ and  $u_{ee}u_{hh}$. In particular, for $u_{hh} u_{ee} > u^2_{he}$ and $\alpha_{he} <<1$,
 \beq
\lambda^s_3 = \alpha^2_{he} \frac{2 u^2_{he} u_{hh}}{u_{hh} u_{ee} - u^2_{he}} >0
\label{rev_1}
\eeq
 In other words, for one of
 $s-$wave solutions, $\lambda^s >0$  even if intra-pocket
 repulsions are the largest.
The full solution of the $3\times3$ set with  $\alpha_{ee} = \beta_{ee} =0$ shows that two $\lambda$'s are repulsive and one is attractive
for arbitrary  $u^2_{he}/u_{ee}u_{hh}$. When the ratio is small,
the attractive solution is  close to (\ref{rev_1}), when the ratio is large, the
 attractive solution is close to $\lambda^s_1$ in (\ref{eq:lambda_1s}).
  I illustrate this in Fig.~\ref{fig:SC_vrtcs}

There is, however, one essential difference between the cases $u^2_{he}/u_{ee}u_{hh} >1$ and   $u^2_{he}/u_{ee}u_{hh} <1$. In the first case, momentum-dependence of the interaction just modifies the ``plus-minus'' solution which already existed
 for momentum-independent interaction. In this situation, the gap along electron FS gradually acquires some $\cos 2 \phi$ variation and remains nodeless for small $\alpha_{he}$. In the second case, the solution with $\lambda >0$ is induced by the momentum dependence of the interaction, and the eigenvalue corresponding to $\lambda^s_3$ necessary has ${\bar \Delta}_e > \Delta_e$, i.e., $s-$wave gap has nodes along the electron FS~~[\onlinecite{cvv}].  In other words, the pairing occurs for all parameters but whether the gap is nodal or not at small $\alpha_{he}$
 depends on the relative strength of intra-pocket and inter-pocket interactions.  When intra-pocket interaction dominates, the gap ``adjusts'' and develops strong $\cos 2 \phi$ component which does not couple to a momentum-independent
 $u_{ee}$ term and by this effectively reduces the strength of electron-electron repulsion.

The same reasoning holds  for the case of two non-equivalent hole FSs, and for 5-pocket  models, and also for the $d-$wave channel, For all cases, the solution
 with $\lambda_i>0$ may exist even when intra-pocket interactions are the largest, but in this situation the  gaps on the hole FSs have accidental nodes. The
 existence or non-existence of the solution at strong intra-pocket repulsion
 depends on the complex interplay between the prefactors of $\cos 2 \theta$  terms in electron-hole and electron-electron pairing vertices, see Eq. (\ref{inequality}).

\begin{figure}[t]
$\begin{array}{cc}
\includegraphics[width=0.45\columnwidth]{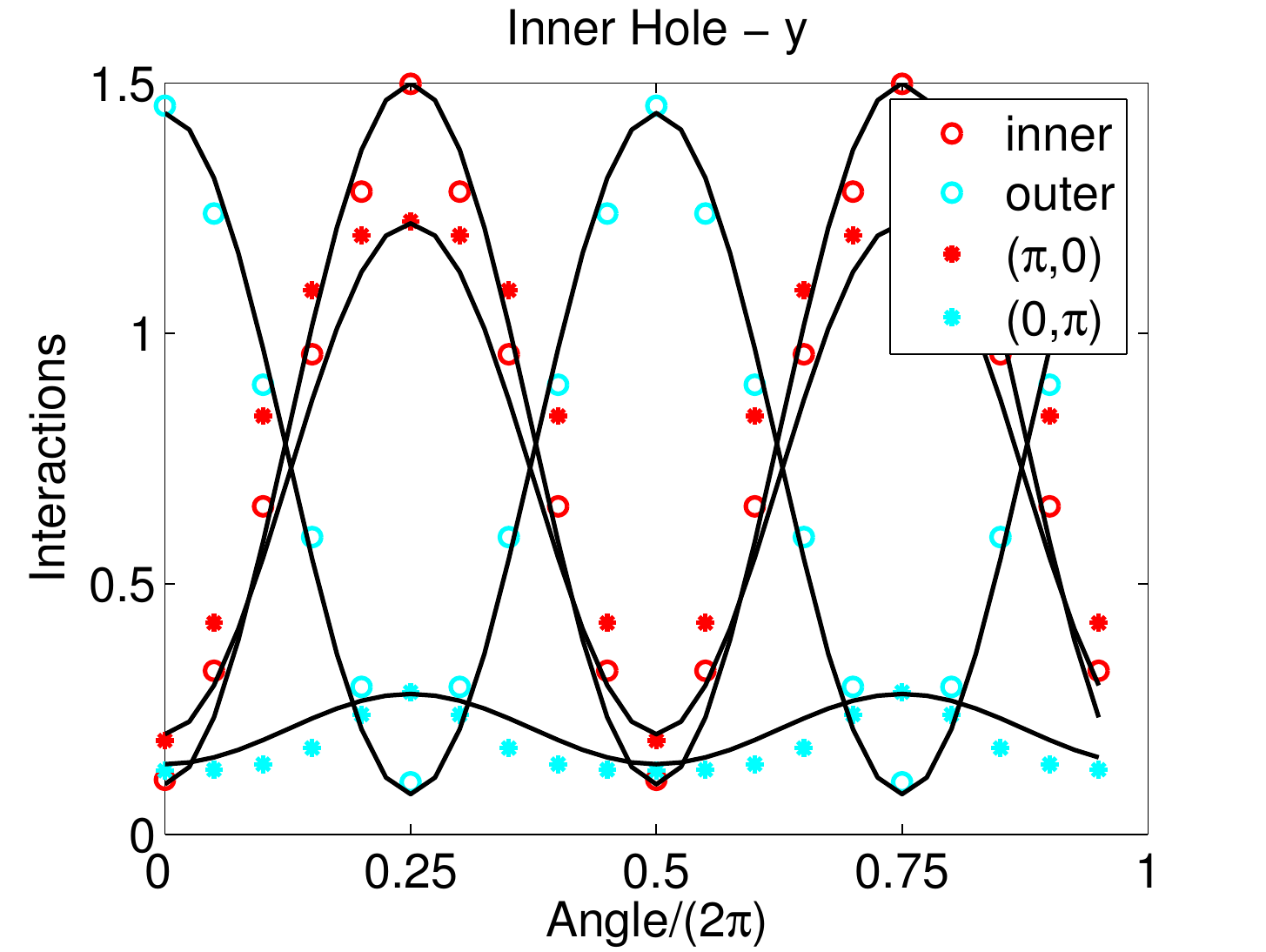}&
\includegraphics[width=0.45\columnwidth]{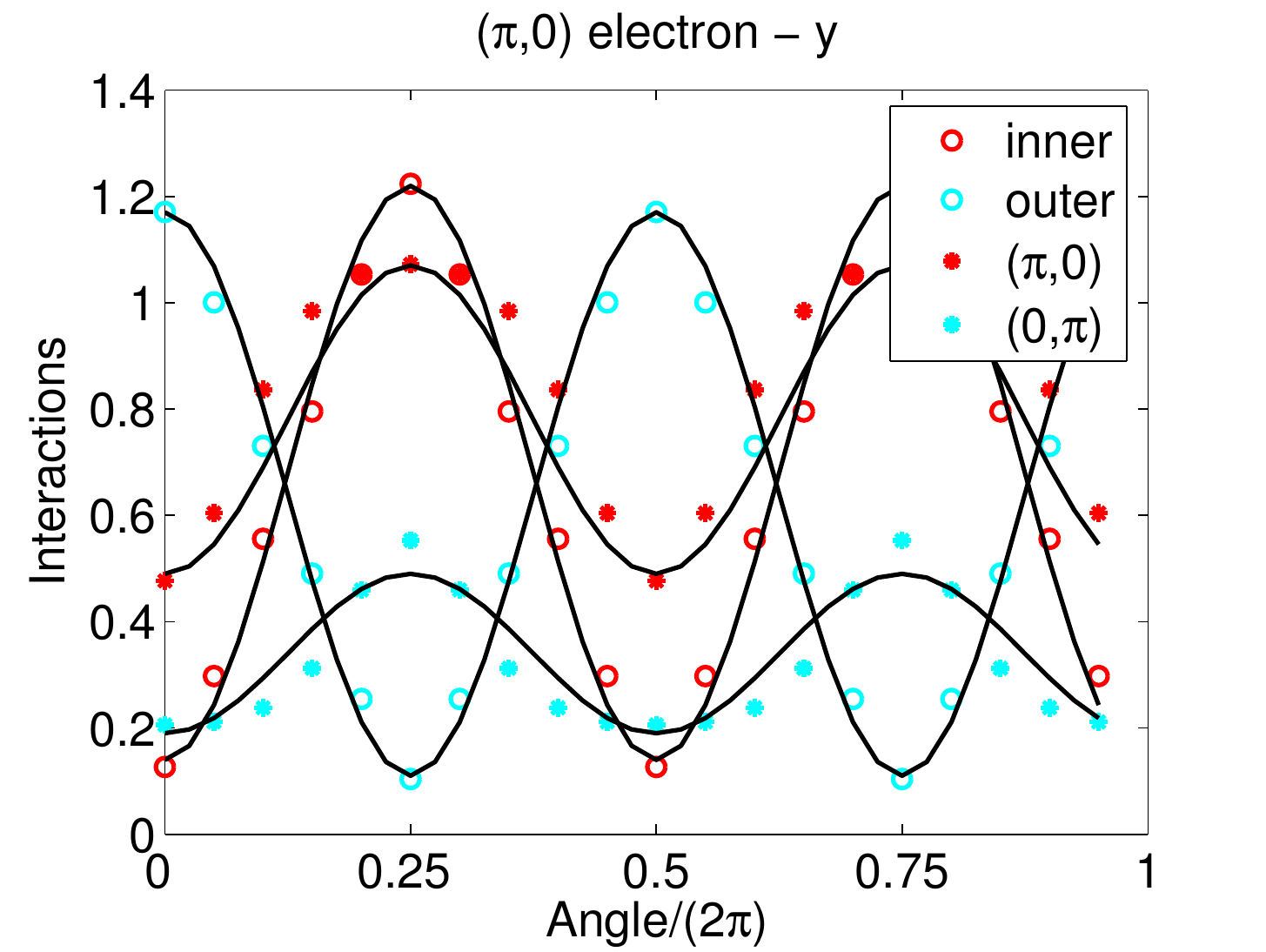}
\end{array}$
\caption{\label{fig:plots_wo_RPA} Representative fits
 of the interactions
$\Gamma_{ij} ({\bf k}_F, {\bf p}_F$) by LAHA for the 4-pocket model.
  $\Gamma_{ij}$ are obtained by
converting the Hamiltonian, Eqs. (\ref{eq:multiorb_Hubbard}), (\ref{eq:multiorb_Hubbard_1}) from the orbital to the band basis.
The symbols
represent interactions computed numerically for the 5-band orbital model using
LDA band structure, the  black lines are the fits using Eqs.
 (\ref{s_4})-(\ref{s_5_d}).  The fit is for the set $U=1.67$, $J=J'=0.21$,
$V =1.46$, and $\mu= 0.08$ (all in eV). A positive $\mu$ corresponds to electron doping.   ${\bf k}_F$ in
$\Gamma_{ij} ({\bf k}_F, {\bf p}_F)$ is selected
along $y$ direction on either an electron or a hole FS
(its location is specified on top of each figure), and ${\bf p}_F$ is varied
 along each of FSs. The angle $\phi$ is measured relative to $k_x$.}
\end{figure}

\begin{table*}[htp]
\centering
\begin{tabular}{|c|c|c|c|c|c|c|c|c|c|c|c|c|c|}
\hline
&$u_{h_1h_1}$&$u_{h_2h_2}$&$u_{h_1h_2}$&$u_{h_1e}$&$\alpha_{h_1 e}$&$u_{h_2e}$&$\alpha_{h_2 e}$&$u_{ee}$&$\alpha_{ee}$&$\beta_{ee}$\\
\hline
NSF&$0.8$&$0.76$&$0.78$&$0.46$&$-0.24$&$0.4$&$-0.30$&0.77&0.14&0.09\\
\hline
SF&$2.27$&$2.13$&$2.22$&$4.65$&$-0.34$&$2.29$&$-0.22$&3.67&0.15&0.04\\
\hline
&$\tilde{u}_{h_1 h_1}$&$\tilde{u}_{h_2 h_2}$&$\tilde{u}_{h_1 h_2}$&$\tilde{u}_{h_1 e}$&$\tilde{\alpha}_{h_1 e}$&$\tilde{u}_{h_2 e}$&$\tilde{\alpha}_{h_2 e}$&$\tilde{u}_{ee}$&$\tilde{\alpha}_{ee}$&$\tilde{\beta}_{ee}$\\
\hline
NSF&$0.7$&$0.66$&$-0.68$&$-0.25$&$-0.58$&$0.24$&$-0.42$&0.11&-0.5&0.25\\
\hline
SF&$1.50$&$1.40$&$-1.50$&$-3.73$&$-0.44$&$1.44$&$-0.32$&1.03&-0.49&-0.02\\
\hline
\end{tabular}
\caption{Table for $s$-wave and $d-$wave parameters for
 the same set as in Fig.~\ref{fig:plots_wo_RPA}.
  NSF and SF mean the bare interaction without the
spin-fluctuation component and the full interaction, respectively.} \label{tab:s-set6}
\end{table*}

\subsection{How to extract $\Gamma_{ij} (\k,\p)$ from the orbital model?}
\label{orb}

So far, in our discussion  $u_{ij}$,  $\alpha_{ij}$, etc, are
 treated as some phenomenological inputs.  To obtain the actual values of these parameters, one needs a microscopic model.
The most commonly considered model for FeSCs is an effective  5-orbital model
 for $Fe$ atoms with local intra-orbital and inter-orbital hopping
integrals and  intra-orbital
and inter-orbital density-density (Coulomb) repulsions, Hund-rule exchange, and the pair hopping term.
\begin{widetext}
\begin{equation}
H_{int} =  \sum_{is}U_{ii} n_{i,s\uparrow} n_{is\downarrow} +
\sum_{i,s,t\neq s} \frac{V_{st}}{2} n_{is} n_{it} - \sum_{i,s,t\neq s}
\frac{J_{st}}{2}\vec{S}_{is}\cdot \vec{S}_{it}  +  \frac{1}{2}  \sum_{i,s,t\neq s} J'_{st} \sum_\sigma c_{is\sigma}^\dagger
c_{is\bar{\sigma}}^\dagger c_{it\bar{\sigma}} c_{it\sigma}\label{eq:multiorb_Hubbard}
\end{equation}
where $n_{is} =  n_{i,s\uparrow} + n_{is\downarrow}$.

The  Hamiltonian $H_{int}$ can be equivalently re-expressed via spin-independent interactions, as
\begin{equation}
H_{int} =  \sum_{is}U_{ii} n_{i,s\uparrow} n_{is\downarrow} +
\sum_{i,s,t\neq s} \frac{{\bar U}_{st}}{2} n_{is} n_{it} + \sum_{i,s,t\neq s}
\frac{J_{st}}{4} c_{is\sigma}^\dagger
c_{it\sigma} c^\dagger_{it\bar{\sigma}} c_{it{\bar\sigma}}
+  \frac{1}{2}  \sum_{i,s,t\neq s} J'_{st} \sum_\sigma c_{is\sigma}^\dagger
c_{is\bar{\sigma}}^\dagger c_{it\bar{\sigma}} c_{it\sigma}\label{eq:multiorb_Hubbard_1}
\end{equation}
where ${\bar U}_{st} = V_{st} + J_{st}/4$.
\end{widetext}

The hopping integrals (36 total) are obtained
from the fit to DFT band structure.~[\onlinecite{ref:Cao}]
For the interaction parameters, the most common approximation is to assume
 that ${\bar U}_{st}$, $J_{st}$ and $J'_{st}$  are independent of the orbital
indices $s$ and $t$, as long as $s\neq t$. The model can be also extended to include non-local Fe-Fe interactions via a pnictide~~[\onlinecite{orbital_J}].

The bare parameters in  (\ref{eq:multiorb_Hubbard}) and (\ref{eq:multiorb_Hubbard_1}) are inter-related due to local spin-rotation invariance~~[\onlinecite{Graser,Kuroki}], but that invariance
 is broken if we view (\ref{eq:multiorb_Hubbard}) and (\ref{eq:multiorb_Hubbard_1}) as an effective low-energy model in which the interactions are
dressed by the renormalizations coming from fermions with energies of order bandwidth. By this reason, in most studies $U$, ${\bar U}$, $J$, and $J'$ are treated as independent parameters.

We now need to convert  (\ref{eq:multiorb_Hubbard}), (\ref{eq:multiorb_Hubbard_1}) into the
band basis and re-express it in the form of Eq. (\ref{rev_2}). This is done
by transforming into the momentum space, introducing new, hybridized operators, which diagonalize the quadratic form, and re-expressing the interaction terms in
 (\ref{eq:multiorb_Hubbard}) or (\ref{eq:multiorb_Hubbard_1}) in terms of these new operators.  The end result of this procedure is the effective Hamiltonian in the band basis which has the form of Eq. (\ref{rev_2}) with $\Gamma_{ij} (\k,\p)$
 given by
\begin{eqnarray}
{\Gamma}_{ij} (\k,\k') & = & \sum_{stpq} \alpha_{i}^{t,*}(-\k)  \alpha_{i}^{s,*}(\k)
\mathrm{Re}\left[{\Gamma}_{st}^{pq} (\k,\k') \right] \nonumber \\
&& \times \alpha_{j}^{p}(\k')  \alpha_{j}^{q}(-\k'),
\label{eq:fullGamma}
\end{eqnarray}
where
$[{\Gamma}_{st}^{pq} (\k,\k')$
 are linear combinations of $U, {\bar U}$, $J$ and ${\bar J}$,
 and  $\alpha_{i}^p$ is the matrix element connecting
 the original fermionic operator $c_p$ in the orbital basis with the
new  fermionic operator $a_i$  on FS $i$ in the band basis.
  The matrix elements $\alpha_i^p$
 contain information which orbitals mostly contribute to a particular segment
 of a  particular FS~~[\onlinecite{Graser,peter}].  Because of this,
 the interaction $\Gamma_{ij} (\k,\p)$ in the band basis
generally depends on the angles
 along different FSs and  contains components in
all representations of the tetragonal $D_{4h}$ group.

The angle dependence of $s-$wave and
$d_{x^2-y^2}$ vertices agrees by symmetry with Eqs (\ref{s_4})-(\ref{s_5_d}).
What s a'priori unknown is how well the interactions can be approximated by the
 leading angle harmonics, i.e., whether the terms labeled as $...$ in  (\ref{s_4})-(\ref{s_5_d}) can actually be neglected.
This issue was analyzed in detail in Ref.~[\onlinecite{maiti_last}], and the answer is affirmative -- the LAHA works rather well. In Fig.\ref{fig:plots_wo_RPA} I show representative fits for a particular set of parameters and in Table 1
I show $u_{eh}$ and other parameters, extracted from the fit (in the lines marked
marked ``NSF'', meaning that this is for the bare interaction, without extra spin-fluctuation component (see below).  The results somewhat
vary depending on the values of $U$, $V$, $J$, $J'$, but in general
 intra-band  interactions in the $s-$wave channel,  $u_{ee}$ and $u_{hh}$,
 exceed interband $u_{he}$.  This is not surprising because $u_{ee}$ and $u_{hh}$ are essentially Coulomb interactions at small momentum transfers, while $u_{eh}$
 is the interaction at large momentum transfer, and it should be smaller
 on general grounds.   Only when $V=J=J'=0$,
the interaction in the band basis becomes independent on the momentum~~[\onlinecite{cee}], i.e., $u_{ee} = u_{hh} = u_{he}$ (this was termed ``Coulomb avoidance'' in Ref.~[\onlinecite{mazin_schmalian}]).  According to Table \ref{tab:s-set6},
 intra-band interactions are also larger in the d-wave channel: ${\tilde u}_{h_ih_i} {\tilde u}_{ee} > {\tilde u}^2_{h_ie}$, although the reasons why this is
 the case are not transparent.

According to the analysis in Sec.\ref{sec:2}, as long as intra-orbital
 interactions exceed inter-orbital interaction, $\lambda_i$ are negative (repulsive) if we neglect $\cos 2\phi$ terms in $\Gamma_{ij}$.
 If we don't neglect these terms,  $\lambda_i$ can be
 positive even for small $u_{he}$ or ${\tilde u}_{he}$ but this requires
 that angle-dependent term satisfy the inequality in Eq. (\ref{inequality}).
 The $\alpha_{eh}, \alpha_{ee}$, and $\beta_{ee}$, and their $d-$wave analogs
 in Table \ref{tab:s-set6} are such that this condition is not me, that is
 for the bare interaction   neither $s-$wave nor $d-$wave pairing is possible.

The conditions on $\alpha$, $\beta$, etc depend on the parameters used for the
 hopings. Varying these parameters (e.g., making $J$ larger) one can, in principle, met the condition in Eq.  (\ref{inequality}) and obtain either
 $s-$wave or $d-$wave solution with a positive $\lambda_i$ and with strong $\cos 2 \phi$ oscillations of the gap along electron FSs. But the generic trend is
 that the bare interaction yields repulsive $\lambda_i$ in $s-$wave and $d-$wave channels.

\section{How to overcome intra-pocket repulsion}
\label{sec:4}

 How to overcome strong intra-pocket repulsion is the major issue for FeSCs.  A
conventional   McMillan-Tolmachev  renormalization~~[\onlinecite{mcmillan}] that
 reduces Coulomb repulsion and allows electron-phonon attraction to overcome it
 does not help at this stage because both $u_{ee} u_{hh}$ and $u^2_{he}$
 renormalize in the same way, and if repulsive  $u_{ee} u_{hh}$ part
is initially stronger, the renormalization just reduces the
strength of the total repulsive interaction, but doesn't change its
sign.

The generic idea how to overcome Coulomb repulsion goes back to Kohn-Luttinger
 approach to superconductivity~~[\onlinecite{KL}] The pairing interaction I considered so far
 is the bare interaction between a particular pair of fermions. The actual pairing interaction $\Gamma_{ij} (\k,\p)$ is given by the fully renormalized irreducible vertex in the particle-particle channel. This irreducible vertex is the sum of the original (bare) interaction and terms of second and higher order in $U, V$, etc. Part of these higher-order terms account for the screening of the original interaction by particle-hole pairs, another part can be re-cast as effective interactions mediated by collective bosonic degrees of freedom in either charge or spin channel. The issue I discuss now is
 how to obtain dressed pairing vertex.

\subsection{RPA approach}

One route is to adopt a semi-phenomenological approach and assume that the system is reasonably close to a particular density-wave ordered phase, such that the strongest component of the effective interaction is mediated by near-gapless fluctuations of this particular density-wave order.  The approach goes back to Berk and Schrieffer~~[\onlinecite{berk_schrieffer}], who considered effective pairing interaction mediated by ferromagnetic spin fluctuations. In recent years it
 has been applied to cuprates~[\onlinecite{scalapino}] and other correlated electron materials~~~[\onlinecite{chub_maslov}]. Most of parent compounds of FeSCs possess long-range SDW order, and it is natural to assume that magnetically mediated pairing interaction plays the central role.

\begin{figure}[t]
$\begin{array}{cc}
\includegraphics[width=0.4\columnwidth]{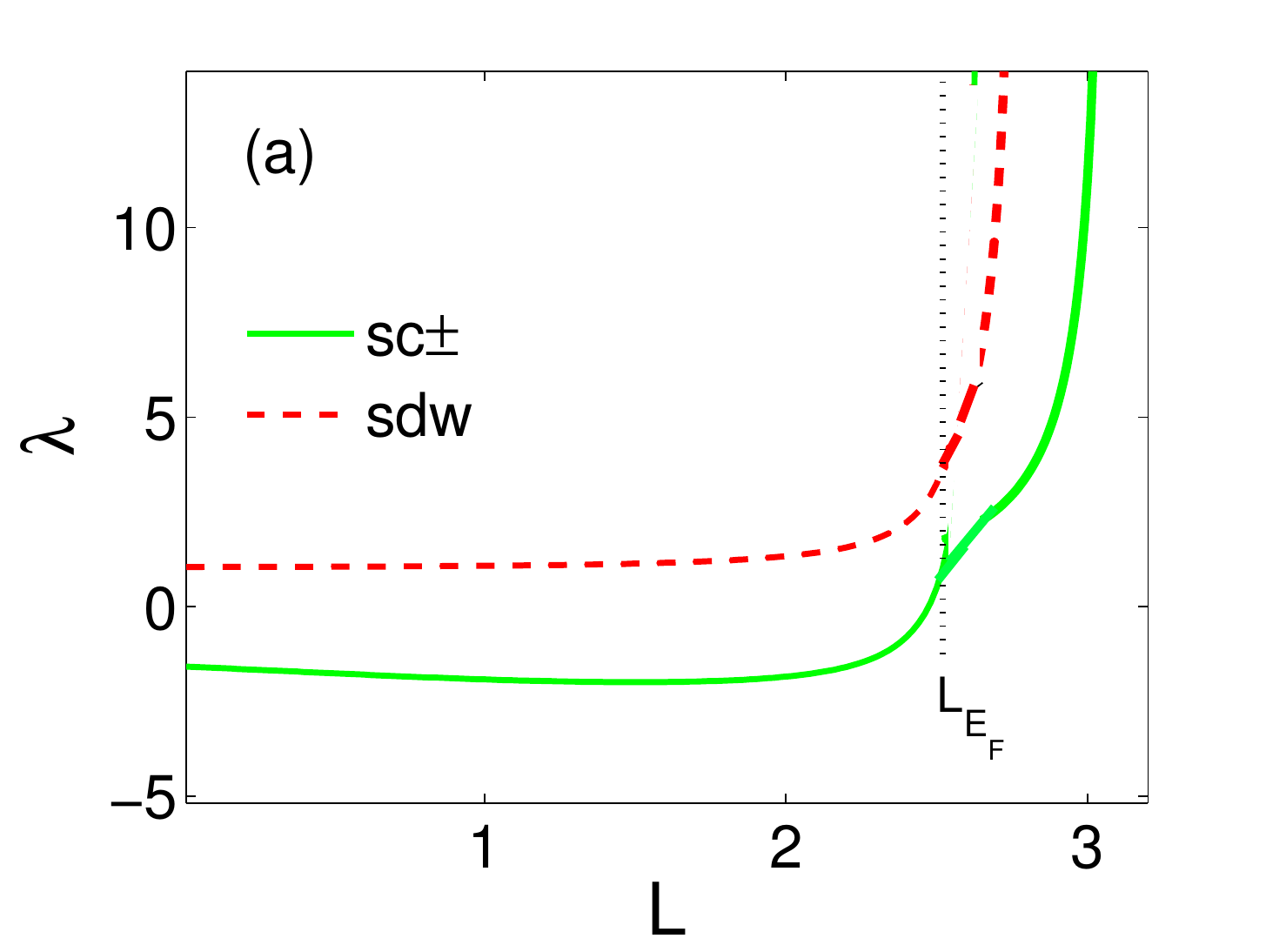}&
\includegraphics[width=0.4\columnwidth]{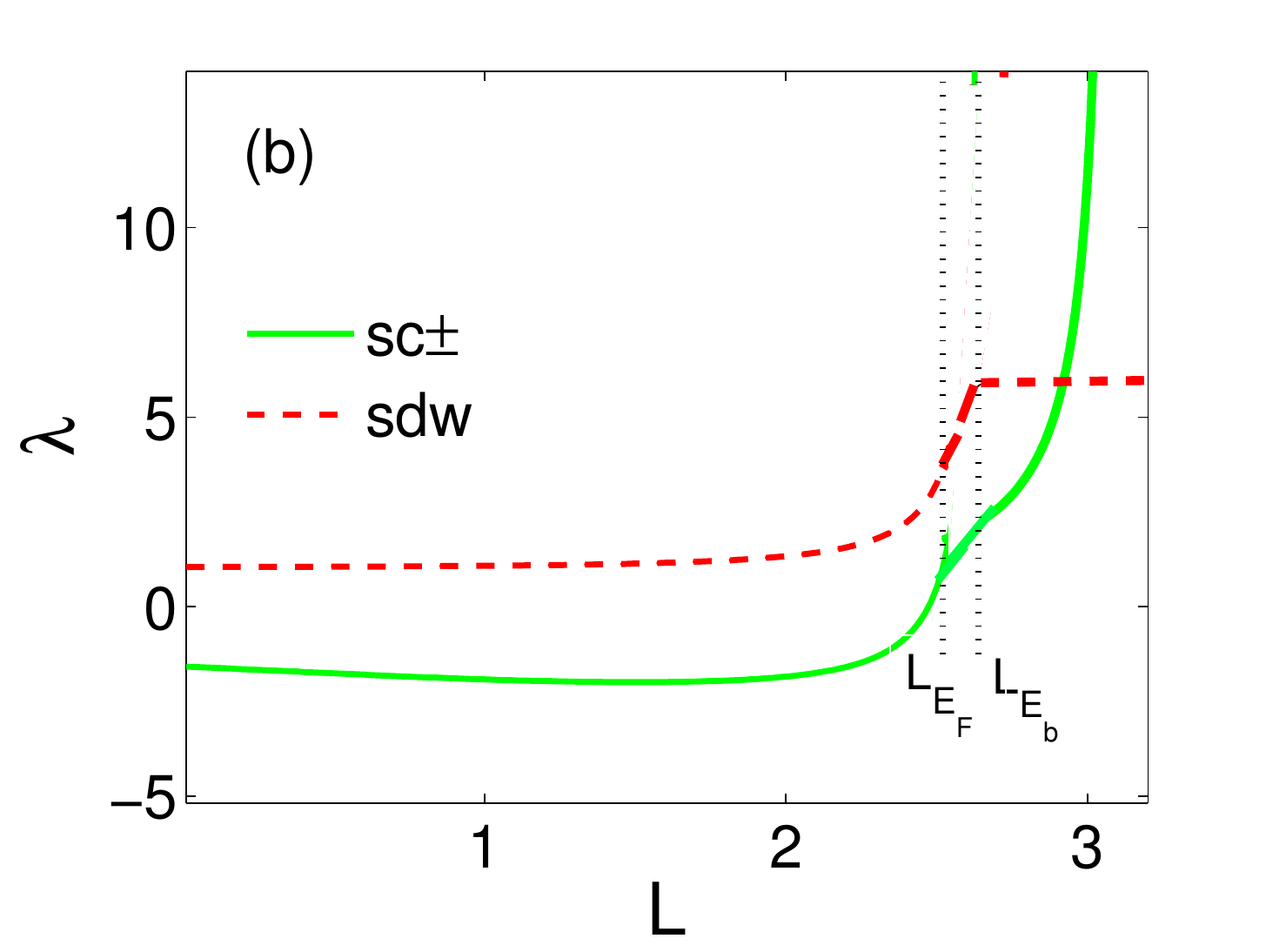}\\
\includegraphics[width=0.4\columnwidth]{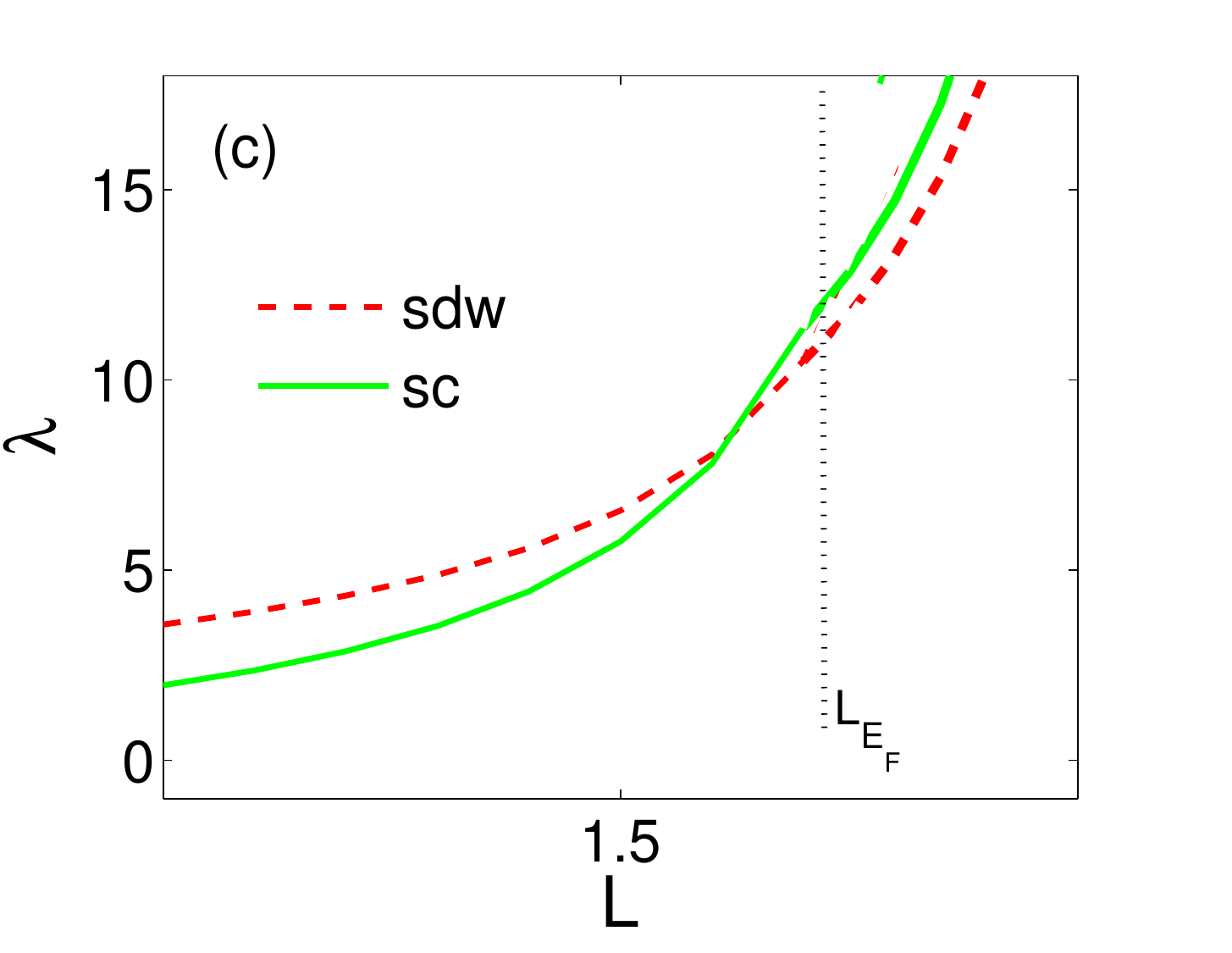}&
\includegraphics[width=0.4\columnwidth]{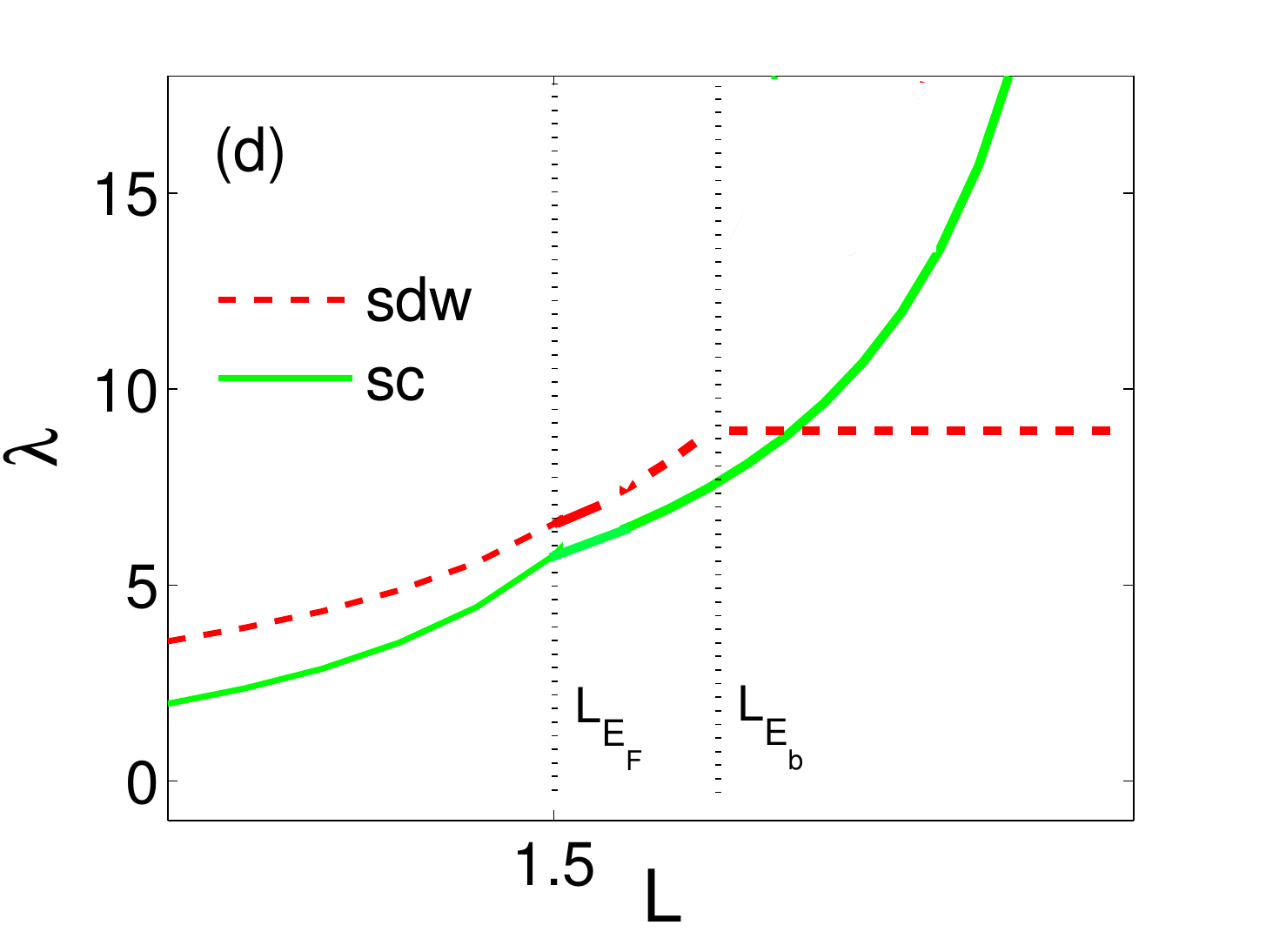}\\
\end{array}$
\caption{\label{fig:rev_1} Running vertices in SDW and SC $s\pm$
channels for the 2-pocket model, panels (a)-(b), and the 4-pocket model, panels
 (c)-(d), as functions of the RG parameter $L = \log{W/E}$.
 For the 2 pocket model $\lambda^{SDW}$ is the largest along the whole RG flow at  perfect nesting (a), but  $\lambda^{s^\pm}$  eventually wins at a finite doping when  $\lambda^{SDW}$ is cut below some $E_b$. For the 4-pocket model,
$\lambda^{s^\pm}$ can be the largest coupling already at  perfect nesting (c),
 or exceed $\lambda^{SDW}$  above some critical doping (d). Figures  are from Ref.~~[\onlinecite{maiti_10}].}
 \end{figure}
 There is no rigorously justified diagrammatic derivation of
an effective fermion-fermion interaction mediated by collective modes,
except for special cases~~[\onlinecite{chub_maslov}]. The pragmatic approach is to select a set of ladder-type or bubble-type
diagrams which  give rise to a SDW  order above some interaction threshold
  and use the same set of diagrams below the threshold. This approach is
 termed RPA by analogy with the screening problem. A generic way to apply RPA is to antisymmetrize the original interaction, decompose it into spin and charge channels and renormalize each interaction in a ladder/bubble approximation, neglecting cross-terms. Each 4-fermion interaction component
${\Gamma}_{st}^{pq}$ then contains two combinations of fermionic spin indices: $\delta_{\alpha,\beta} \delta_{\gamma,\delta}$ and ${\vec \sigma}_{\alpha\beta} {\vec \sigma}_{\gamma\delta}$, where ${\bf \sigma}$ are Pauli matrices. These two
 combinations have to be convoluted with the spin
 structure $i\sigma^y$ of the two-particle anomalous pairing vertex. Using
$\sigma^y_{\alpha \gamma} \delta_{\alpha,\beta} \delta_{\gamma,\delta} = \sigma^y_{\beta\delta}$ and  $\sigma^y_{\alpha \gamma} {\bf \sigma}_{\alpha\beta} {\bf \sigma}_{\gamma\delta}
 = -3  \sigma^y_{\beta\delta}$ and incorporating the factor $-3$ into the spin part of the effective pairing vertex, one obtains, in the orbital basis
\begin{eqnarray}
{\Gamma}_{st}^{pq} (\k,\p) &=& \left[\frac{1}{4} \Gamma^s+\frac{3}{4} \Gamma^c +\frac{3}{2} (\Gamma^s)^2
\chi_s^{RPA}  (\k-\p)  \nonumber \right.\,~~~~~~\,\\
 &-&  \left. \frac{1}{2} (\Gamma^c)^2  \chi_c^{RPA}  (\k-\p) \right]_{ps}^{tq}
\label{rev_5}
\eea
where $\Gamma^s$ and $\Gamma^c$ are spin and charge components of the bare
 antisymmetrized interaction,  $\chi_s^{RPA} $ is the dressed static spin susceptibility and $\chi_c^{RPA}$ is the dressed static charge susceptibility. The
  interaction, Eq. (\ref{rev_5}),
  is then converted into the band basis using (\ref{eq:fullGamma}), and the pairing in different channels is analyzed in the same way as it was done for the bare interaction, i.e.,
either by explicitly solving integral equation on $\lambda_i$ in a particular channel~~[\onlinecite{Kuroki_2,tom_09,Graser,peter,lara}], or by approximating $\Gamma_{ij} (\k,\p)$ by the leading angular harmonics and solving $4\times4$ or $5\times5$ gap equations (see Sec\ref{sec:5} below). The full analysis of the pairing problem often requires one to know the frequency dependence of $\chi$, but for the analysis of the pairing symmetry and the momentum dependence of the gap the frequency dependence is not overly relevant and I neglect it.

In all model calculations,  the spin susceptibility
$\chi_s^{RPA}$ is enhanced due to close proximity to SDW order,
the charge susceptibility $\chi_c^{RPA}$  remains small and can be neglected.
How $(\Gamma^s)^2\chi_s^{RPA}$ in (\ref{rev_5})  modifies $u_{ij}$ depends on the type of SDW order. For moderately doped FeSCs $\chi^{RPA}_s (q)$ is peaked at $(0,\pi)$ and $(\pi,0)$, which are the two momenta separating hole and electron FSs.  Not surprisingly, electron-hole interaction $u_{he}$ is enhanced relative to $u_{ee}$ and $u_{hh}$. By less obvious reasons, d-wave component of electron-hole
interaction ${\tilde u}_{he}$ is also enhanced, although not as strongly as $u_{he}$.  As an illustration, I show in Table~\ref{tab:s-set6}, in the lines marked as SF,  renormalized interactions for the same set of input parameters which
 were considered in the previous section. Clearly, $u_{he}$ and ${\tilde u}_{he}$ are enhanced.  If the enhancement is strong enough, either Eq. (\ref{inequality}) becomes valid, or $u^2_{he}/u_{ee} u_{hh}$ becomes larger than one (or
${\tilde u}^2_{he}/{\tilde u}_{ee} {\tilde u}_{hh}$ becomes larger than one), and
 the system becomes unstable towards s-wave or d-wave superconductivity, whichever $\lambda_i$ is the largest.
\begin{figure}[htp]
 \includegraphics[width=0.9\columnwidth]{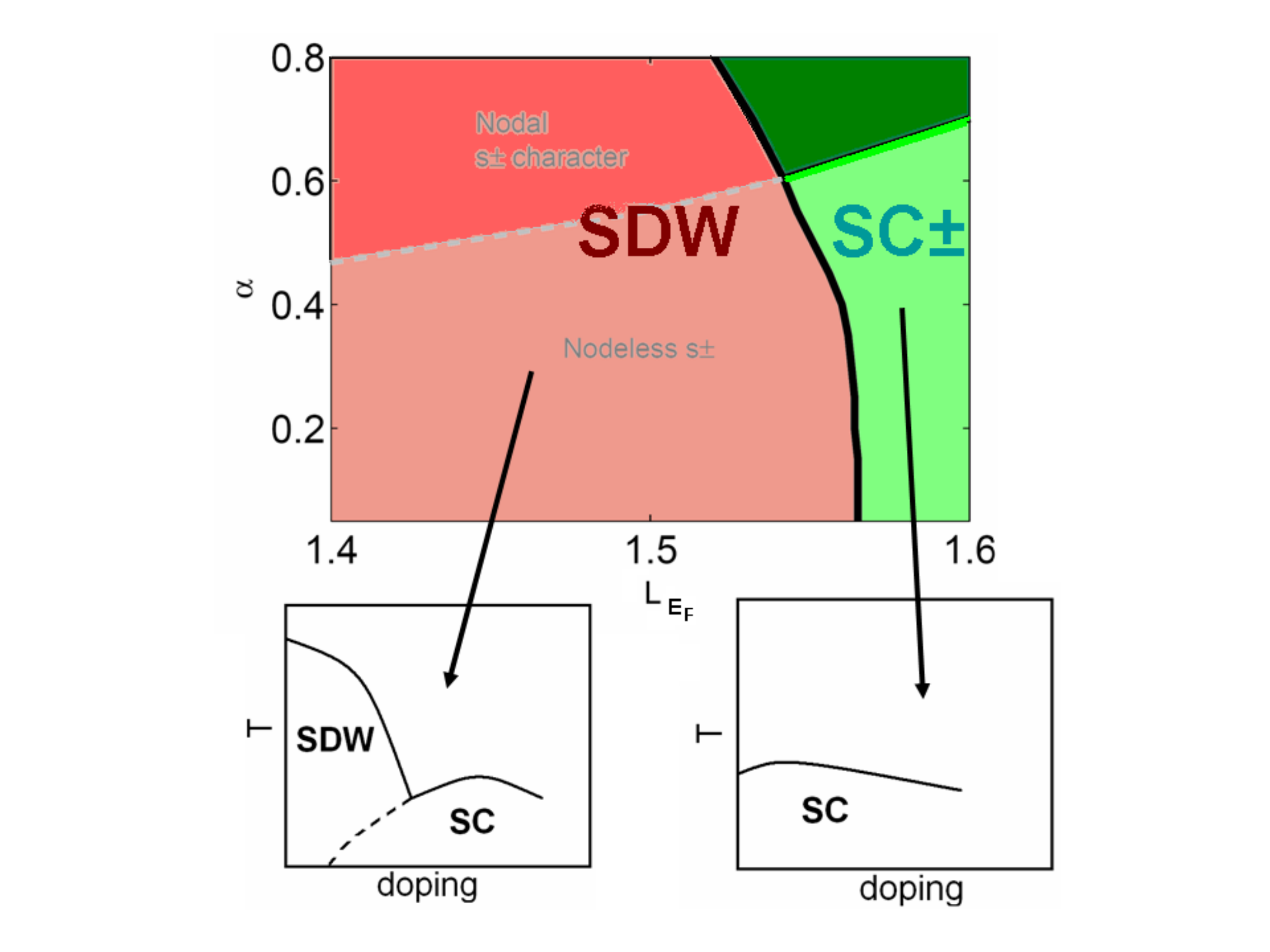}
 \caption{\label{fig:rev_1_1}
  Top panel: the phase diagram of the 4-pocket model at zero doping as the function of the strength of the angle-dependent component of the interaction, $\alpha$, and $L_{E_F} = \log W/E_F$, where $W$ is the bandwidth and $E_F$ is the Fermi energy. Lower panels -- the phase diagrams as functions of doping for different $L_{E_F}$. The superconducting gap in
   $s^{\pm}$ state  may be nodeless or have nodes along electron FSs.  In the region labeled as SDW, SC emerges only at a finite doping, when  SDW instability is cut. In the region labeled as $SC^\pm$, SC state emerges already at
 zero doping.  Figure is from Ref.~~[\onlinecite{maiti_10}].}
\end{figure}

This scenario requires bare $\Gamma^s$ to be positive. Only then spin susceptibility $\chi_s^{RPA} (q)$ is enhanced. In the orbital basis, this requirement essentially means that intra-orbital Coulomb repulsion  $U$ must be larger than
 inter-orbital ${\bar U}$.   In the band basis,
 the same requirement implies that bare $u_{he}$ must be positive~~[\onlinecite{cee}].
  What if bare $u_{he}$ is negative?  Then $|u_{he}|$ may still get an enhancement
 within RPA, but the enhancement now comes from
 charge (orbital) fluctuations~~[\onlinecite{jap}]. The latter are strong if the system is close to a state with an orbital order.
There are proposals~~[\onlinecite{ku}] that broken rotational invariance observed above $T_N$ in most of FeSCs~~[\onlinecite{nematic}] may be due to orbital order. [An alternative scenario~~[\onlinecite{ising}] is an Ising order associated with $Z_2$ degree of freedom separating $(0,\pi)$ and $(\pi,0)$ spin configurations.] If
 $u^2_{he}$ is enhanced by orbital fluctuations such that it again exceeds $u_{ee} u_{hh}$, s-wave SC again develops, but the eigenvalue now corresponds to a conventional $s^{++}$ gap, with the same sign along hole and electron FSs~~[\onlinecite{jap}]. I will compare experimental evidence for $s^{\pm}$ vs $s^{++}$ gap
 in Sec.\ref{sec:6}.  For the rest of this section I  assume that $u_{he} >0$, i.e., the attractions is due to spin fluctuations.

There are two  underlying assumptions in the RPA approach.
  The obvious one is the selection of only ladder or bubble diagrams in the absence of a small parameter.
 The less obvious assumption is the very idea that the pairing can be viewed as  mediated by collective density-wave bosonic excitations, because ``mediation''  implies that density-wave  fluctuations develop at energies well above relevant energies for the pairing. This last idea is reasonably well justified for the cuprates, in which magnetic fluctuations  develop at energies of a few hundred med, well above $T_c$. Whether this is also the case for FeSCs is less obvious as magnetism in most of these materials is itinerant (parent compounds are metals), and  the highest  $T_c$ is at dopings at which SC and SDW ordering temperatures almost coincide.  Whether in this situation spin fluctuations can be viewed as ``pre-existing'' for the SC problem is not obvious. It is then instructive to
 consider also an alternative approach which treats magnetic and SC fluctuations on equal footings.  This alternative approach is based on RG technique and is termed as ``RG''

\subsection{RG approach}

RG approach is unbiased in the sense that it doesn't assume that the pairing is mediated by a collecting bosonic degree of freedom. Rather, it departs from a
 bare Hamiltonian with the original 4-fermion interaction and studies how particle-particle and particle-hole susceptibilities
 evolve as one progressively integrates out contributions from fermions with energies larger than some running scale $E$. This should, in principle,
 address the issue whether spin fluctuation develop at a larger $E$ than pairing fluctuations.  In practice, this approach is indeed also an approximate one
because even at weak coupling one cannot explicitly sum up contributions from fermions with energies between the bandwidth $W$ and  $E$.
 This can be done rigorously only  if the renormalizations depend  on the running scale logarithmically, i.e., are the functions of $L = \log{W/E}$.  Then the flow of the couplings can be described by a set of differential equations $d u_{ij}/dL = f(u_{lm})$, where  $u_{lm}$ in $f(u_{lm})$ are renormalized interactions at the same running scale L.

Renormalizations in a superconducting channel are indeed depend on $L$, but
 In this respect FeSCs are ``gifts from nature''
 because  density-wave renormalizations involving fermions near hole and electron FSs also scale as $\log{W/E}$  at $E$ above the threshold set by non-perfect nesting~~[\onlinecite{zlatko,cee}].  Then one can obtain a set of coupled RG equations for pairing and density-wave vertices. This set is called ``parquet'' because the renormalizations in the particle-particle
 channel are often represented as  horizontal ladder diagrams, renormalizations in the particle-hole (density-wave) channel as vertical ladder diagrams, and taking both on equal footing amounts to building perturbatively a parquet pattern.

From physics perspective, the key feature of RG approach is that the pairing,
 SDW, and CDW fluctuations all ``talk to each other'' at intermediate energies. Not only pairing fluctuations are enhanced by  SDW fluctuations, but also
 SDW fluctuations are affected by pairing fluctuations and so on.

To get the idea how the RG works in FeSCs, consider momentarily a toy two-pocket model. Previously, in Sec. \ref{sec:toy}, I introduced three interactions which
 contribute to the pairing:  intra-pocket hole-hole and electron-electron interactions $u_{hh}$ and $u_{ee}$ and inter-pocket hole-electron interaction $u_{he}>0$.
 Suppose that bare electron-hole interaction is weak and the bare $\lambda^{s^\pm} <0$. Now let's add two other interactions which do not directly contribute to the pairing but
contribute to density-wave instability: density-density and exchange
 interactions between fermions from hole and electron pockets, $u_{dd}$ and
$u_{ex}$, respectively. The coupling in the SDW channel is $\lambda^{SDW} = u_{dd} + u_{he}$.  As one progressively integrates out contributions from high energies, all couplings evolve and obey
\begin{eqnarray}
&&\dot{u}_{dd} = u_{dd}^2 + u_{he}^2 \nonumber \\
&&\dot{u}_{ex} = 2 u_{ex}(u_{dd} - u_{ex} ) \nonumber \\
&&\dot{u}_{he} =  u_{he}(4 u_{dd} - 2 u_{ex}-u_{hh}-u_{ee})\nonumber \\
&&\dot{u}_{hh} = -u_{he}^2 - u_{hh}^2 \nonumber \\
&&\dot{u}_{ee} = -u_{he}^2 - u_{ee}^2 \nonumber
\label{2_a}
\end{eqnarray}
where the derivatives are with respect to ${\bar L} = (1/2) \log W/E$.
We see that $u_{he}$ gets a boost from $u_{dd}$, which is the part of the
SDW vertex,  and in turn $u_{dd}$ gets a boost from $u_{he}$. Solving the set,
 we find that $u_{he}$ grows under RG while $u_{hh}$ and $u_{ee}$ decrease (and eventually change sign), such that below some $E_0$ the running $\lambda^{s^\pm}$ changes sign and becomes attractive.

The physics behind the sign change of  $\lambda^{s^\pm}$ is essentially  the same
  as in RPA analysis -- $u_{eh}$ gets a boost from SDW fluctuations. But RG analysis also addresses the issue of the interplay between SDW and SC orders as both are determined by the running interactions which ``push'' each other.  For the two-pocket model SDW eigenvalue $\lambda^{SDW}$ remains larger than $\lambda^{s^\pm}$ along the whole RG trajectory (panels a and b in Fig. \ref{fig:rev_1}) , but in multi-pocket systems more complex behavior is possible~~[\onlinecite{Thomale_10,maiti_10}] (see panels c and d in Fig. \ref{fig:rev_1} and Fig. \ref{fig:rev_1_1}).

The ability to address in an unbiased way what kind of order develops first
 is an obvious advantage of the RG approach. At the same same time,
 parquet RG has only limited applicability range
because Eqs. (\ref{2_a}) and the analogous equations for multi-pocket models~~[\onlinecite{maiti_10}] are valid only when the running $E$ is larger than the Fermi energy $E_F$.
This wouldn't be an obstacle if the leading instability in FeSCs would
develop already at $E > E_F$ (Refs.~[\onlinecite{cee,podolsky,rice}]). But this is unlikely
given that $E_F \sim 10^2 meV$ is order of magnitude larger than $T_N$ and $T_c$.  One then has to extend the analysis to  $E<E_F$.
  It turns out that at  such $E$
renormalizations in the SDW and SC channels
 decouple from each other~~[\onlinecite{rev_physica,maiti_10}] and each $\lambda^i$
 evolves according to $d \lambda_R^i/dL = (\lambda_R^i)^2$ (where now $L = \log W/E$)  Then SDW or SC instability occurs first depending on which $\lambda$ (either $\lambda^{SDW}$ or $\lambda^{s^\pm}$)  is larger at $E_F$.
 Away from a perfect nesting, the logarithmical RG flow of $\lambda^{SDW}$ is cut below some scale $E_b$, while $\lambda^{s^\pm}$ continue to grow and
 definitely becomes the leading instability above a certain doping.
 I illustrate this in Fig. \ref{fig:rev_1_1}.

This consideration shows that what parquet RG actually provides are the values
 of the dressed interactions $u_{ij}$ at the scale of $E_F$. Below this scale,
  SDW order develops without input from  SC channel, and on the other hand
SC develops in a BCS-fashion.
 If the doping is such that SC instability is the leading instability in the problem, the symmetry
 and the structure of the SC gap is analyzed in the same way as in Sec. \ref{sec:3}), only  $u_{ij}$ and ${\tilde u}_{ij}$ are now  dressed interactions, renormalized by the RG flow between $W$ and $E_F$.  From this perspective,
RPA  and RG approaches are similar in the sense that the end result
 of both approaches is the effective BCS  Hamiltonian at $E_F$ with renormalized interactions some of which are pushed up by SDW fluctuations.

\begin{figure}[htp]
$\begin{array}{cc}
\includegraphics[width=0.4\columnwidth]{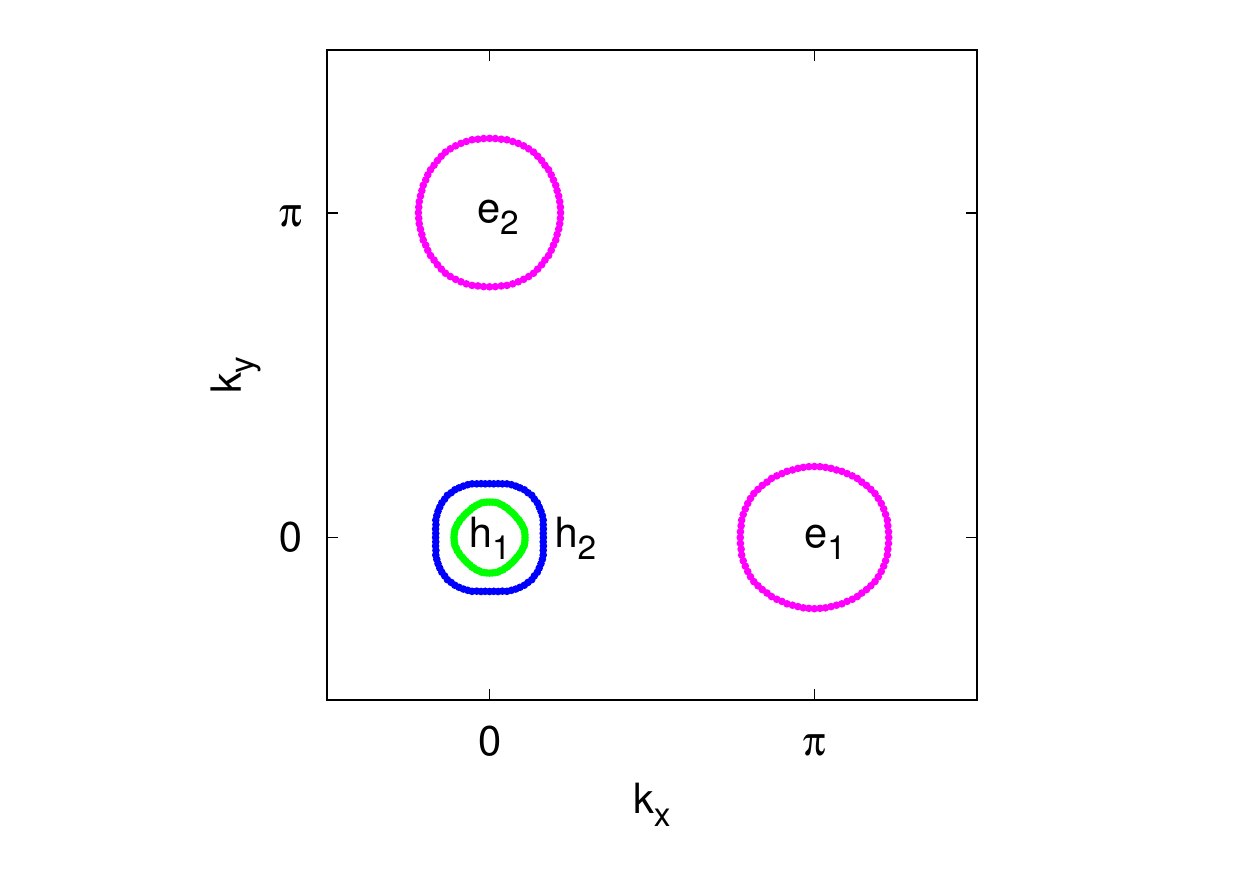}&
\includegraphics[width=0.4\columnwidth]{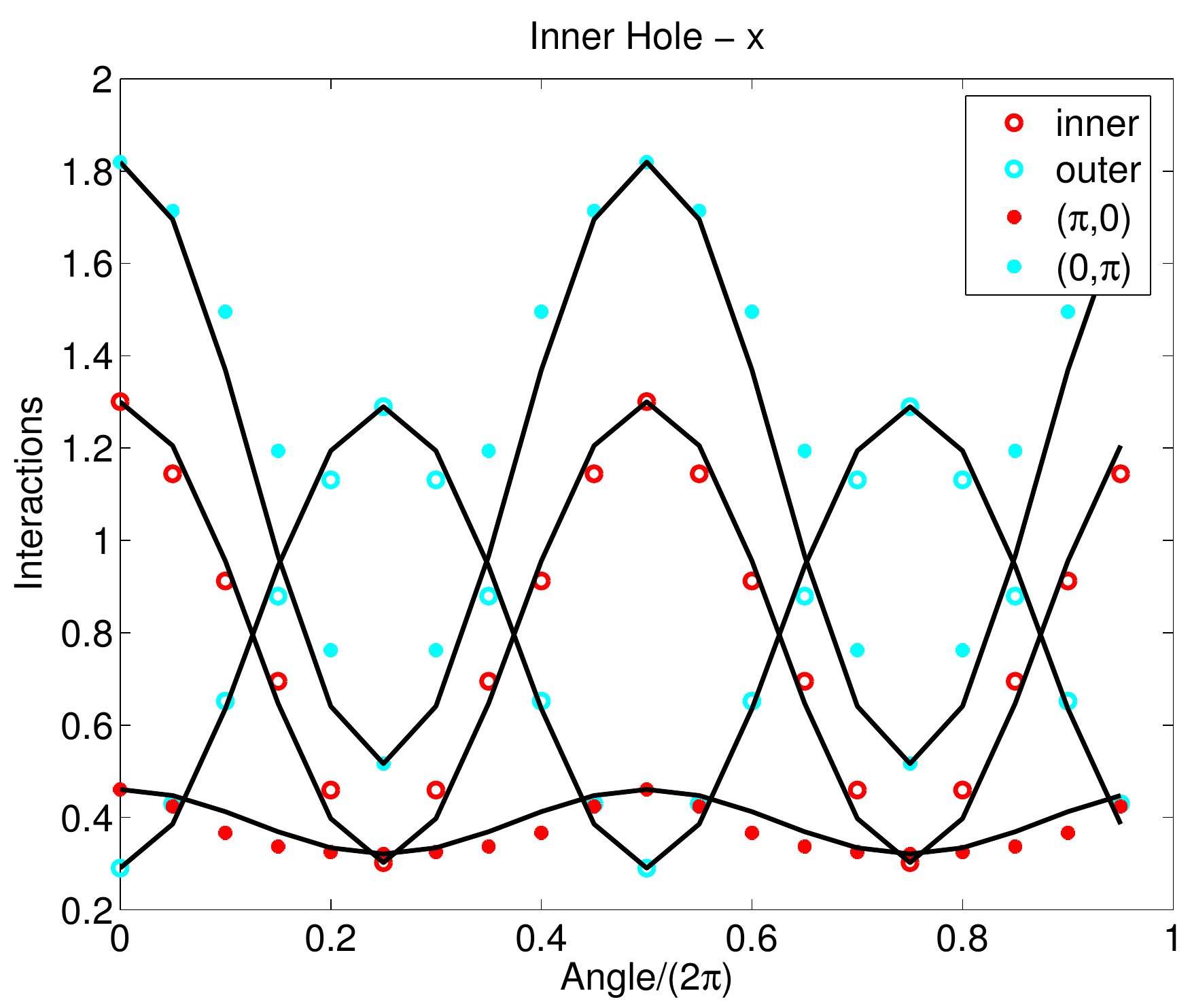}\\
\includegraphics[width=0.4\columnwidth]{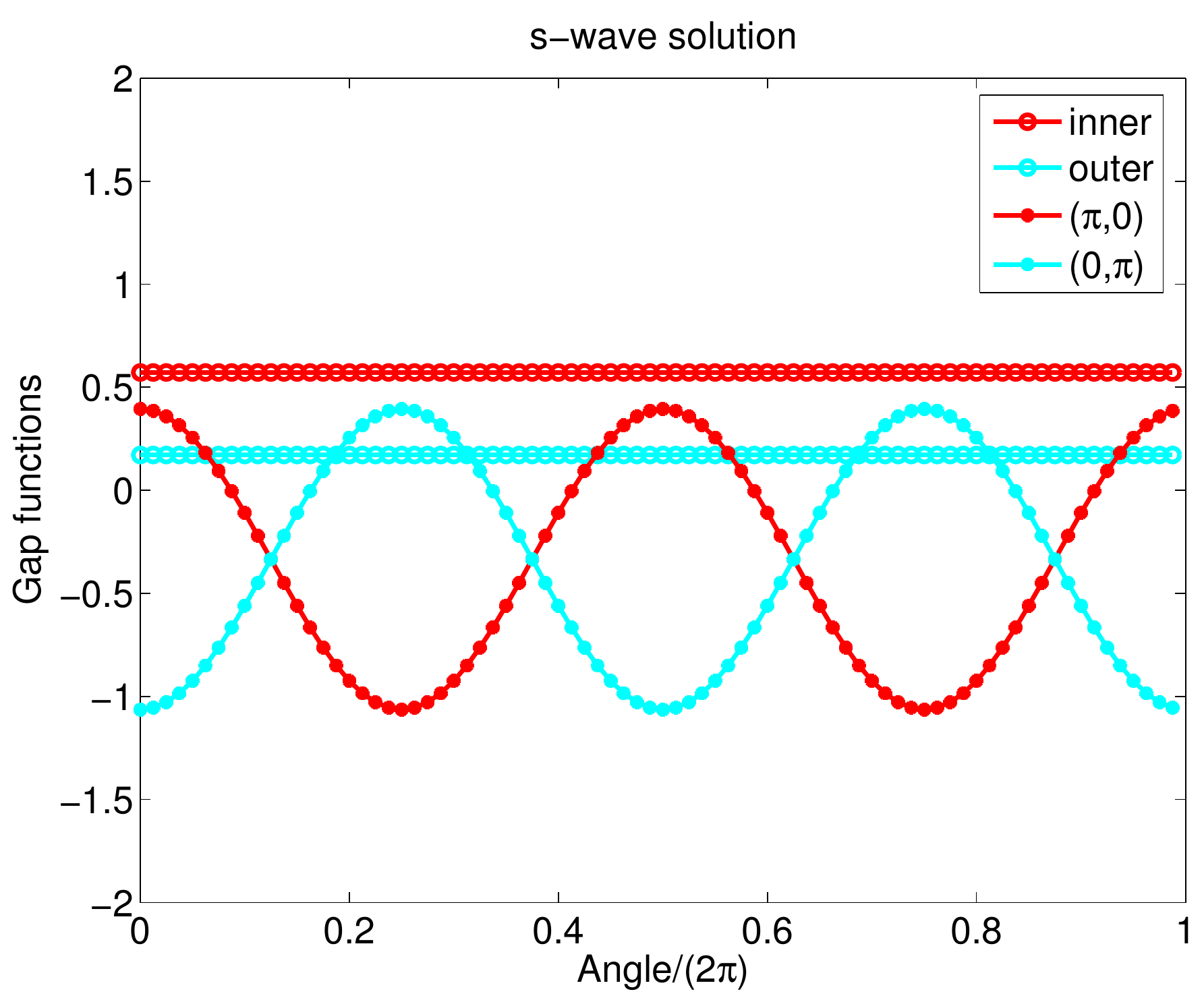}&
\includegraphics[width=0.4\columnwidth]{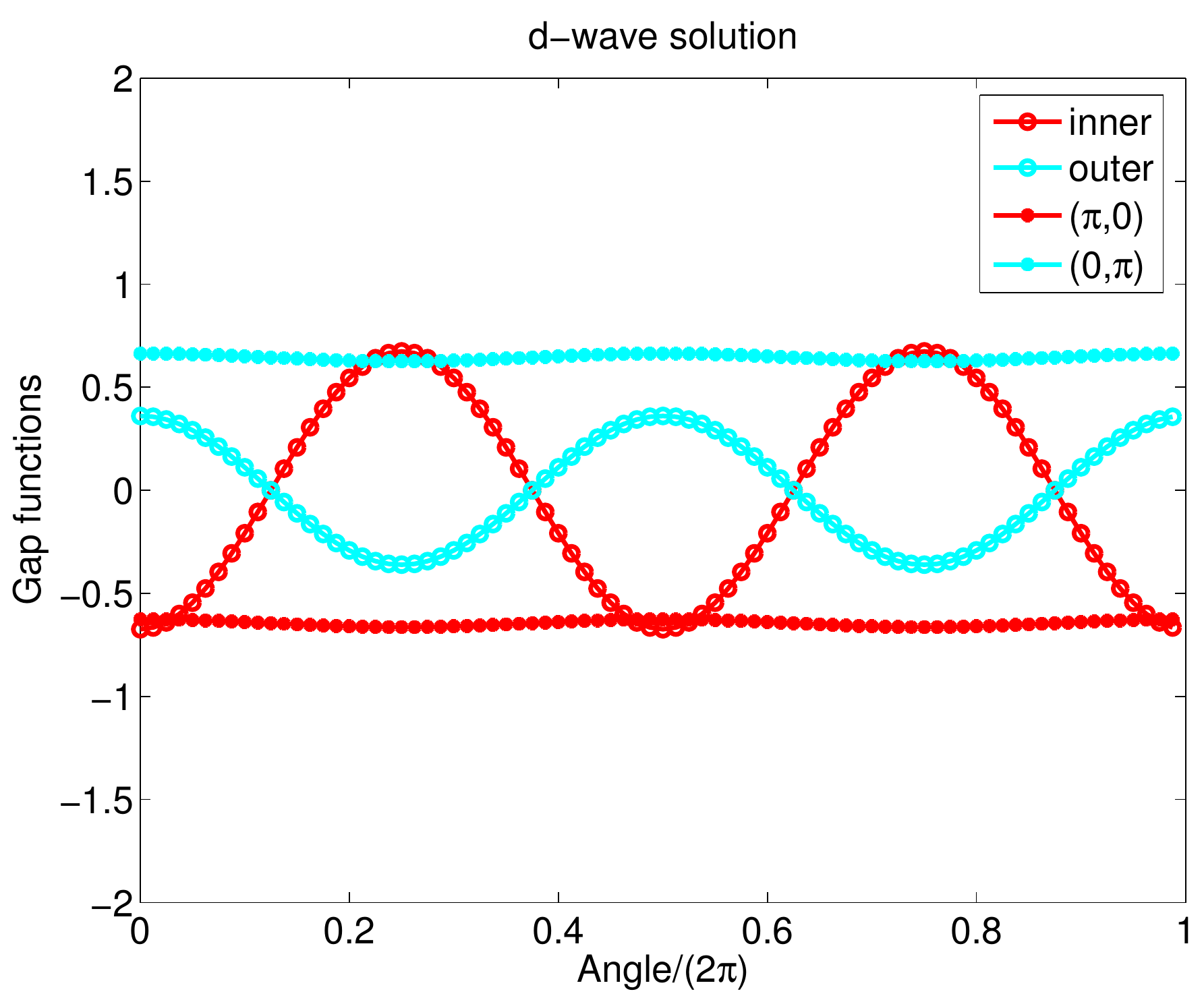}\\
\end{array}$
\caption{\label{fig:rev_2}
 Representative case of small/moderate electron doping, when both hole and electron pockets are present.
 Panel a -- the FS, panel b --  representative fits
 of the interactions by LAHA (the dots are RPA results, the lines are LAHA expressions,  Eqs (\ref{s_4})-(\ref{s_5_d})). Panels c and d --
  the eigenfunctions in $s-$wave and $d-$wave channels for the largest $\lambda^s$ and $\lambda^d$.  From Ref.~[\onlinecite{maiti_last}].}
\end{figure}

\begin{figure}[htp]
$\begin{array}{cc}
\includegraphics[width=0.45\columnwidth]{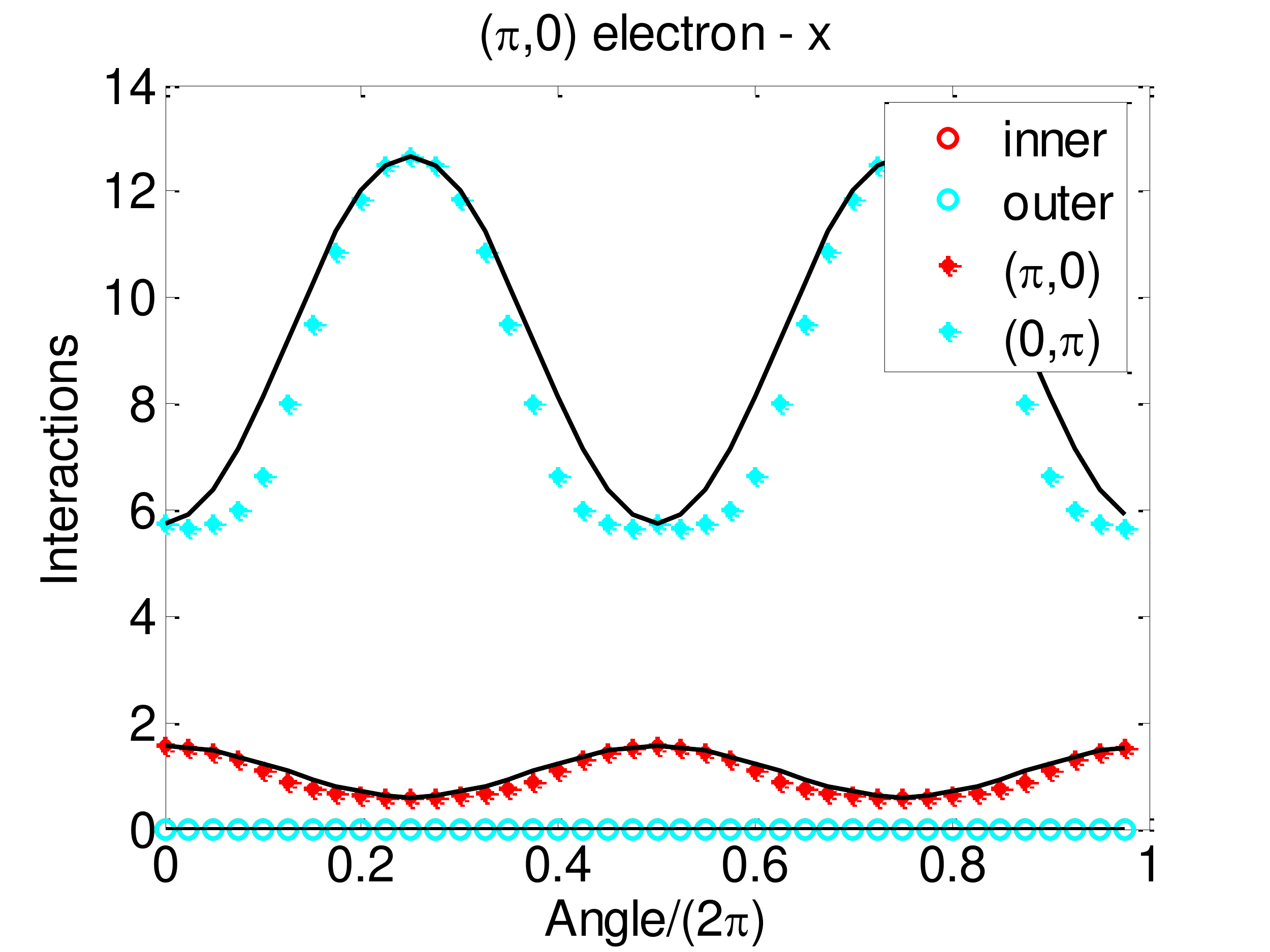}&
\includegraphics[width=0.45\columnwidth]{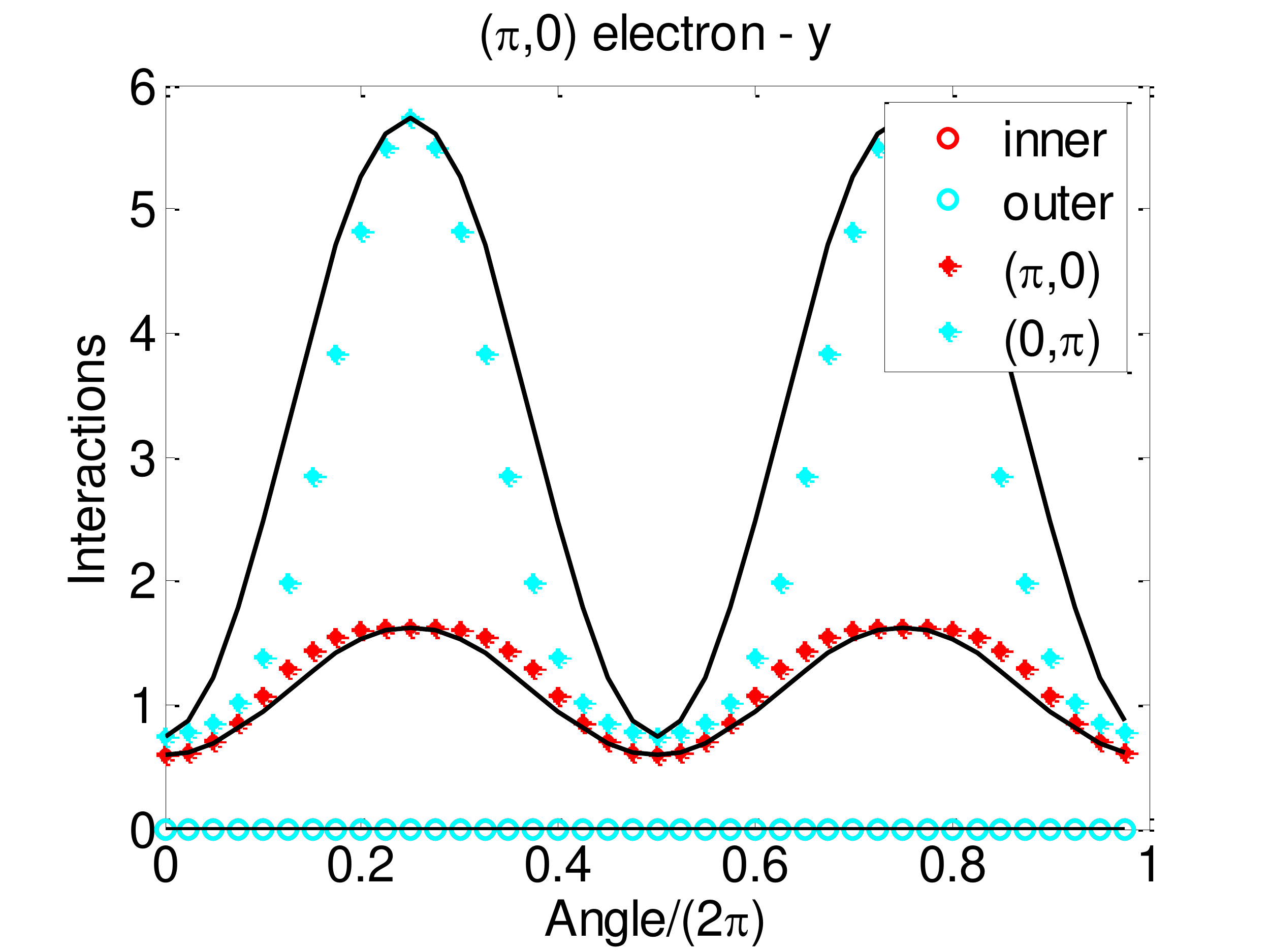}\\
\includegraphics[width=0.45\columnwidth]{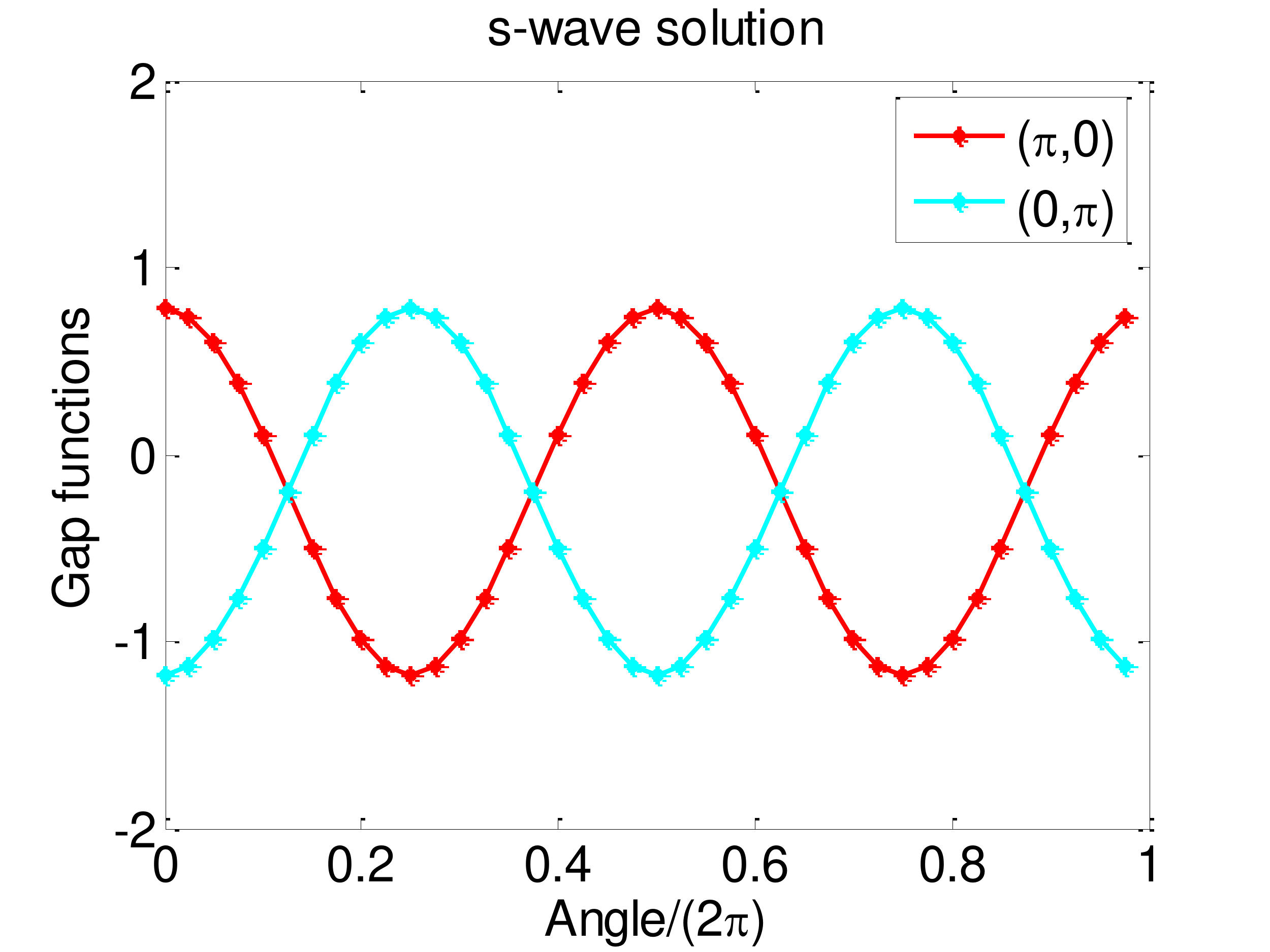}&
\includegraphics[width=0.45\columnwidth]{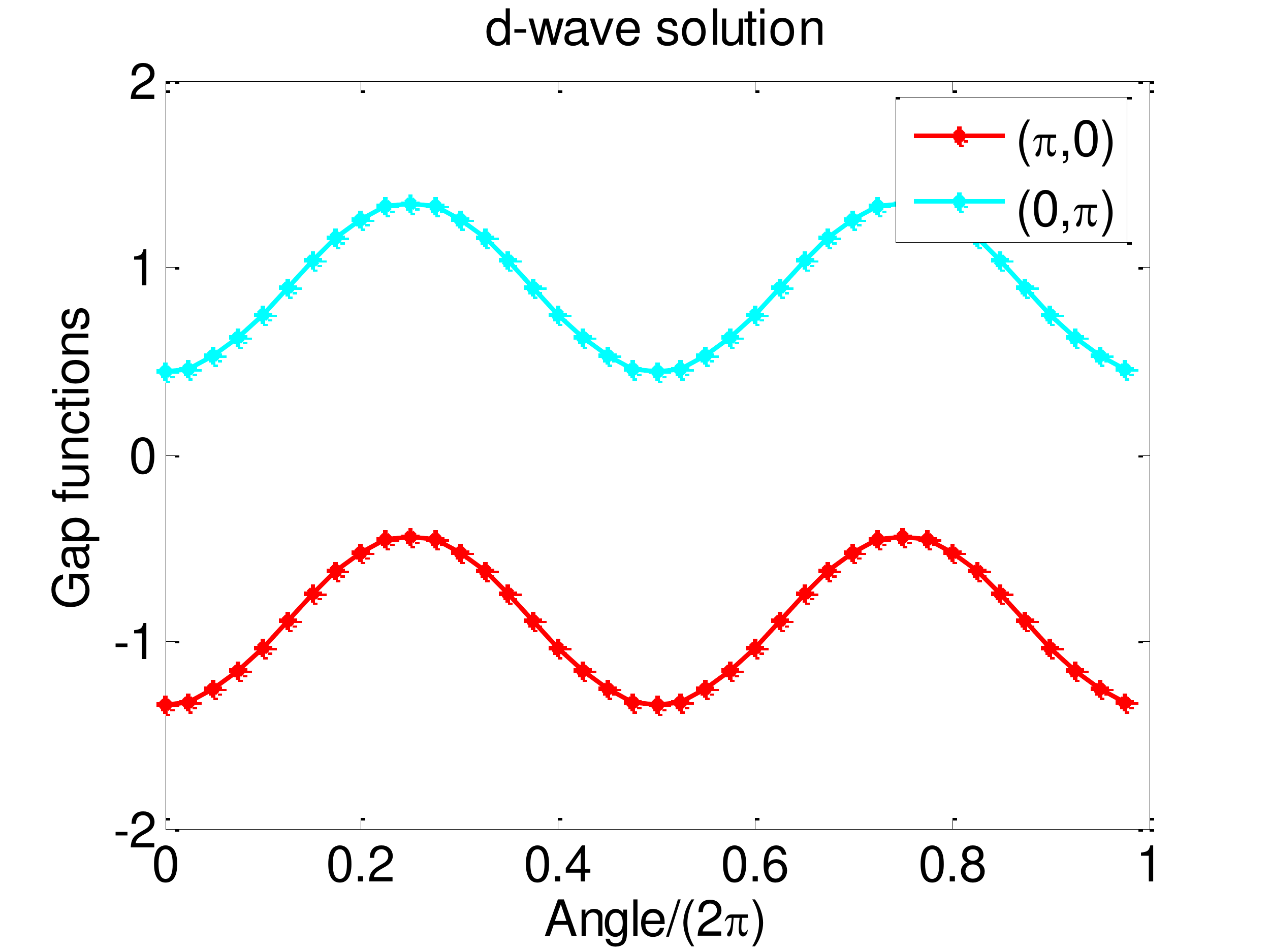}
\end{array}$
\caption{\label{fig:rev_3} The fits of the RPA
interactions by LAHA and the structure of $s-$wave and $d-$wave gaps
 for the case of heavy electron doping, when only electron
FSs are present.   From Ref.~[\onlinecite{maiti_last}].}
\end{figure}

There is another issue with RG related to angle-dependencies of the
interactions (see Sec. \ref{orb}).  A conventional logarithmical RG is applicable only when $\alpha$ and $\beta$ terms in (\ref{s_4})-(\ref{s_5_d}) are small.
 To  order $O(\alpha)$, $O(\beta)$, angle dependencies remains intact, only the overall factors $u_{ij}$ flow~~[\onlinecite{maiti_10}].  Beyond this order,
 logarithmical and non-logarithmical terms mix up and the selection of diagrams
 can no longer be rigorously justified.  One way to proceed in this situation is to keep the angle dependence intact by hand and use the same equations for $u_{ij}$ as if the interactions were angle-independent~~[\onlinecite{maiti_10}].  Another way is to partition each FS into patches and write a larger set of RG equations for the interactions between different patches.  This last approach is called functional RG (fRG)~~[\onlinecite{fRG,Thomale_10,dhl_AFESE,KFeAs_fRG}]. fRG is not an exact approach as it  mixes  logarithmical and non-logarithmical terms,
 but it is nevertheless a very powerful numerical technique to analyze the interplay between density-wave and superconducting instabilities in systems with angle-dependent interactions.

Until now, the results of RPA, fRG and logarithmical RG  fully agree with each other on (i) the interplay between SDW and SC  instabilities, (ii) the symmetry of the SC state at different dopings, and (iii)  the structure of the SC gap.
 This is a positive news
 because it likely implies that the underlying physics of FeSCs is quite robust.
 In particular, both fRG~~[\onlinecite{Thomale_10}] and conventional RG~~[\onlinecite{maiti_10}] show that in 4-band systems  SC may be the leading instability even without doping, while in 2-band and 5-band systems SDW is the leading instability at zero doping, and SC emerges only  upon doping.  RPA, fRG, and conventional RG all show
that $s^{\pm}$ gap has nodes along electron FS in a much wider parameter range in 4-pocket systems than in 5-pocket systems because
 the additional hole FS at $(\pi,\pi)$ tends to stabilize no-nodal gap.  An yet
 another example is the equivalence between RPA~~[\onlinecite{graser_11}]
 and fRG~~[\onlinecite{dhl_AFESE}] results for heavily electron doped systems, in which only electron pockets remain -- both approaches yield a d-wave gap with no nodes.

\section{Doping dependence of the couplings}
\label{sec:5}

In this section I assume that SC is the leading instability and
 briefly review how the gap symmetry and structure evolve with doping.
 RG, fRG, and RPA results on the doping evolution of the gap agree with each other, and for definiteness I will use RPA approach combined with the LAHA (see Sec. \ref{sec:3}).  The results differ for electron and hole dopings, and I  consider them separately.

\subsection{Electron doping}

For small and moderate electron dopings, the FS consists of 4 pockets -- two hole FS at $(0,0)$ and two electron FSs at $(0,\pi)$ and $(\pi,0)$.
 Typical fits by LAHA, the parameters extracted from the fits,
 and the solutions in s-wave and d-wave channels are shown in Fig. \ref{fig:rev_2}  and in Table \ref{tab:1}.
It turns out~~[\onlinecite{maiti_last}]
 that some system properties are sensitive to the choice of
the parameters, but some are quite universal.
 The parameter-sensitive properties are the presence or absence of accidental nodes in the $s$-wave gap (although for most of parameters
the gap does have nodes, as in Fig.~\ref{fig:rev_2}) and
 the gap symmetry itself, because for most of input parameters and dopings $\lambda^s$ and $\lambda^d$ remain comparable as long as both hole and electron FSs are present (see Table  \ref{tab:1}).
That $d-$wave state is a strong competitor in 4-pocket systems
 has been first emphasized in Refs.~~[\onlinecite{Graser,Kuroki}]. The authors of~~[\onlinecite{Graser}] hinted that different FeSCs may have different symmetry even
 for the same topology of the FS.

 The universal
observation is that the driving force for attraction in both
$s$-wave {\it and} $d$-wave channels is strong inter-pocket
electron-hole interaction ($u_{h_i e}$ and ${\tilde u}_{h_i e}$
terms)  {\it no matter how small the hole pockets are}.
 The gap structure actually changes only little with doping as long as both hole and electron pockets are present.

\begin{table*}[htp]
\caption{\label{tab:1} Some of the LAHA parameters  extracted from the LAHA fit
 in Figs. (\ref{fig:rev_2}) and (\ref{fig:rev_3})
 for electron doping. Blocks (i)  corresponds
to  Fig. (\ref{fig:rev_2}), block (ii) corresponds to  Fig. (\ref{fig:rev_3}) (no hole pockets). From Ref~~[\onlinecite{maiti_last}].}
\begin{ruledtabular}
\begin{tabular}{lcccccclccccccrccc}
& \multicolumn{6}{c}{(i)} & & \multicolumn{3}{c}{(ii)}\\
 \cline{2-7} \cline{9-11}
$s$-wave&$u_{h_1h_1}$&$u_{h_1e}$&$\alpha_{h_1 e}$&$u_{ee}$&$\alpha_{e e}$&$\lambda_s$& & $u_{e e}$&$\alpha_{e e}$&$\lambda_s$\\
&0.8&0.79&-0.19&0.91&0.05&0.25& &3.65&0.20&0.1\\
 \cline{2-7} \cline{9-11}
$d$-wave&$\tilde{u}_{h_1h_1}$&$\tilde{u}_{h_1e}$&$\tilde{\alpha}_{h_1 e}$&$\tilde{u}_{e e}$&$\tilde{\alpha}_{e e}$&$\lambda_d$& & $\tilde{u}_{e e}$&$\tilde{\alpha}_{e e}$&$\lambda_d$\\
&0.50&-0.39&-0.46&-0.04&1.5&0.37& &-2.57&0.29&5.9\\
\end{tabular}
\end{ruledtabular}
\end{table*}

The situation changes qualitatively once the hole pockets disappear
(Fig.~\ref{fig:rev_3}). It is clear from  Table~\ref{tab:1} that now the
$d$-wave channel becomes the dominant one. Comparing the LAHA parameters for the two
dopings, we see the reason: once the hole pockets disappear, a
direct $d$-wave electron-electron interaction ${\tilde u}_{ee}$
becomes strong and attractive. The argument why this happens
 is as follows:~~[\onlinecite{maiti_last}] ${\tilde u}_{ee}$ is an
antisymmetric combination of intra-pocket and inter-pocket
electron-electron interactions ${\tilde u}_{ee} = u_\mathrm{intra}^{ee} - u_\mathrm{inter}^{ee}$. Both $u_\mathrm{inter}^{ee}$ and
$u_\mathrm{intra}^{ee}$ are positive (repulsive), but the
sign of ${\tilde u}_{ee}$ depends on the interplay between
$u_\mathrm{inter}^{ee}$ and $u_\mathrm{intra}^{ee}$.
 As long as hole FSs are  present, SF are peaked near $\mathbf{q}=(0,\pi)$ and
$(\pi,0)$, which are an equal distance from the relevant momenta
$\mathbf{q}=0$ for $u_\mathrm{intra}^{ee}$ and $\mathbf{q}=(\pi,\pi)$ for $u_\mathrm{inter}^{ee}$.
In this situation, $u_\mathrm{intra}^{ee}$ and $u_\mathrm{inter}^{ee}$
 remain close in magnitude,
and ${\tilde u}_{ee}$ is small.
 Once the hole pockets disappear, the peak in the RPA spin
susceptibility shifts towards $(\pi,\pi)$ ~~[\onlinecite{graser_11}] and
$u_\mathrm{inter}^{ee}$ increases more due to the SF component than
$u_\mathrm{intra}^{ee}$. A negative $u_\mathrm{intra}^{ee} -
u_\mathrm{inter}^{ee}$ then gives rise to a ``plus-minus'' gap on
the two electron FSs. The gap changes sign under $k_x \to k_y$ and
therefore has $d_{x^2-y^2}$ symmetry. This pairing mechanism is
 essentially identical to spin-fluctuation scenario for d-wave pairing in the cuprates~~[\onlinecite{scalapino}].

A different proposal has been put forward in Refs.~~[\onlinecite{yu_11}] and ~[\onlinecite{bernevig}].   These authors argued that the gap symmetry may be nodeless $s-$wave (equal sign of the gap on the pockets at $(0,\pi)$ and $\pi,0)$.
 According to ~[\onlinecite{yu_11,bernevig}], s-wave pairing emerges,
in some range of parameters, if one uses for
 electron-electron interaction the orbital version of the
$J_1-J_2$ model.  Yet another proposal for strongly electron-doped FeSCs is $s^{++}$ pairing driven by orbital fluctuations~~[\onlinecite{kontani_se}].

\begin{figure}[htp]
$\begin{array}{cc}
\includegraphics[width=0.45\columnwidth]{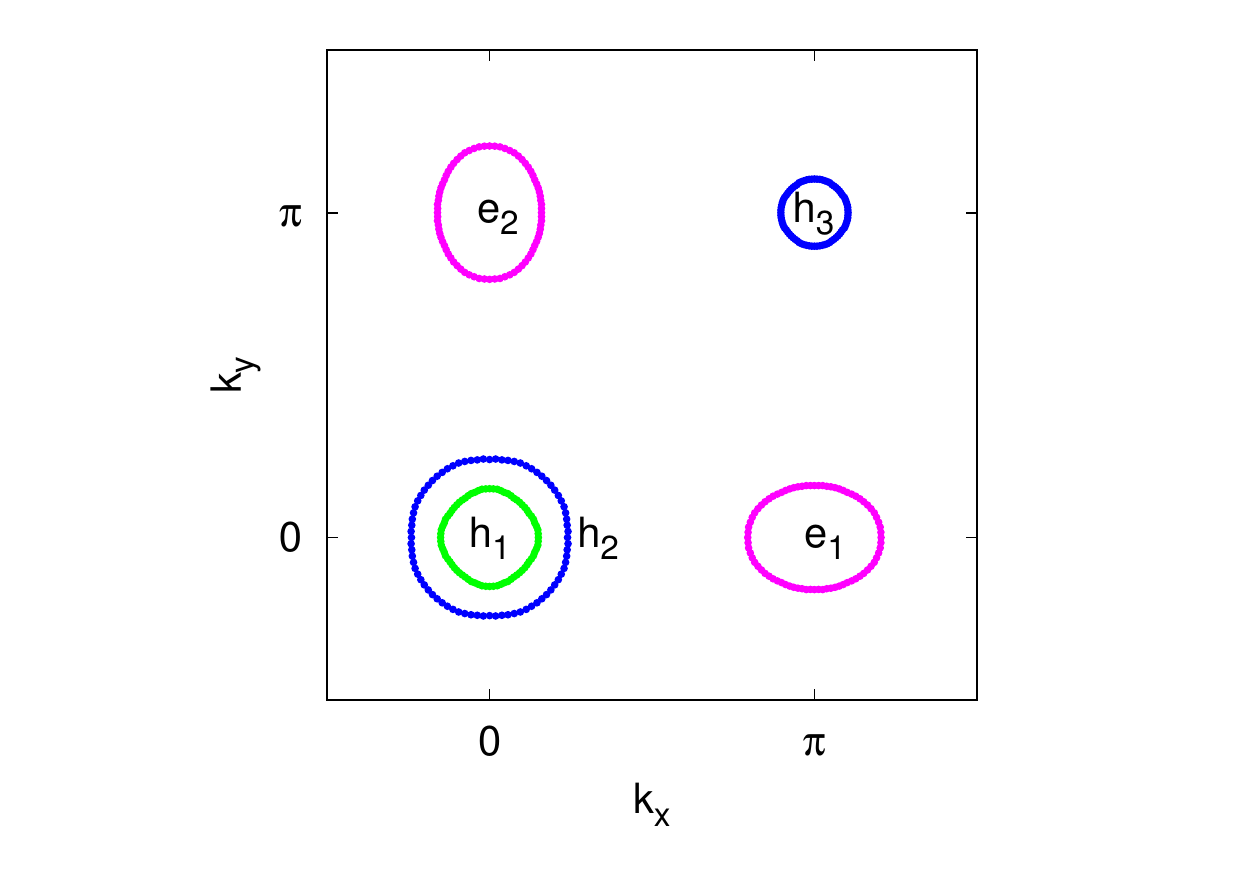}&
\includegraphics[width=0.45\columnwidth]{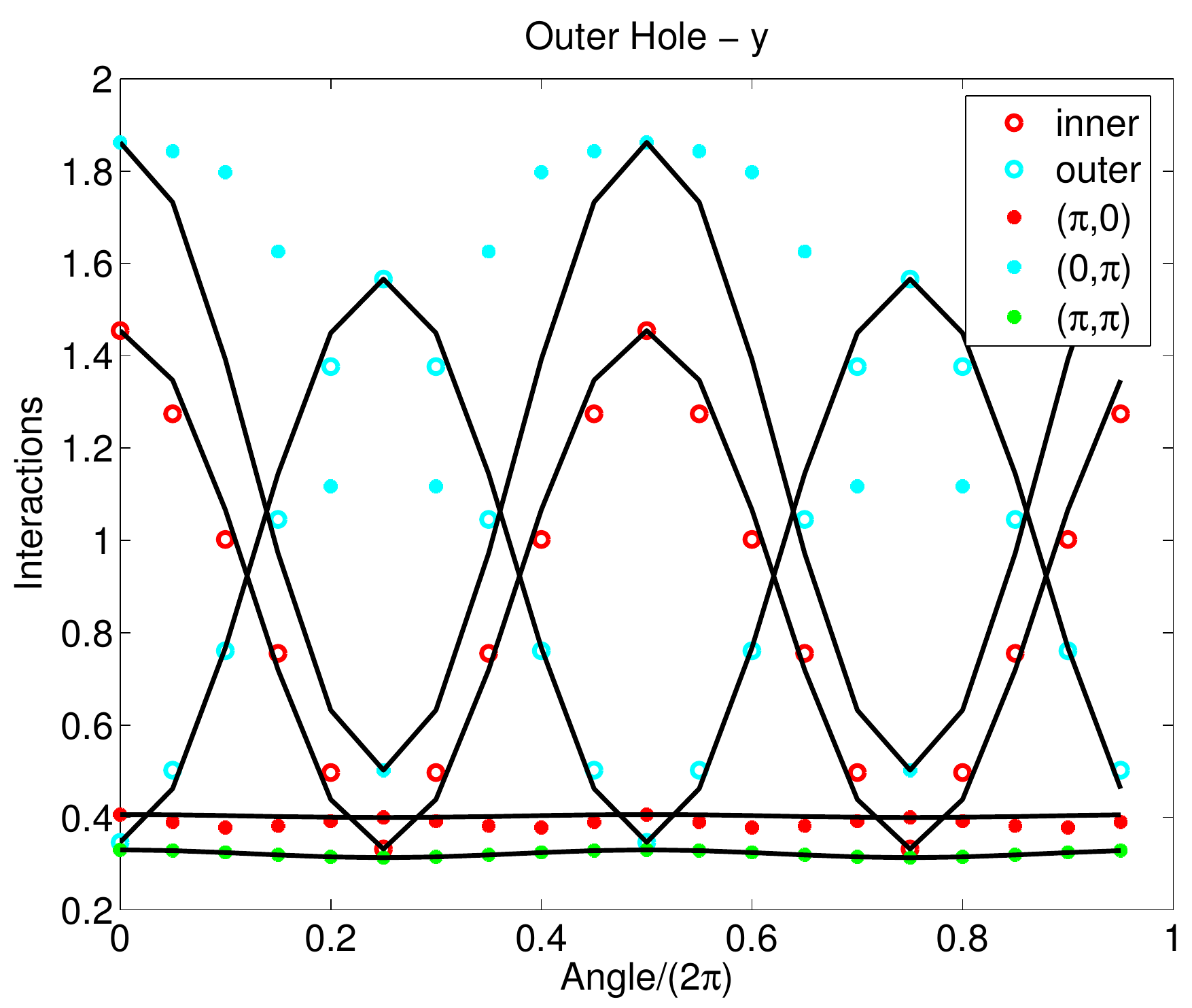}\\
\includegraphics[width=0.45\columnwidth]{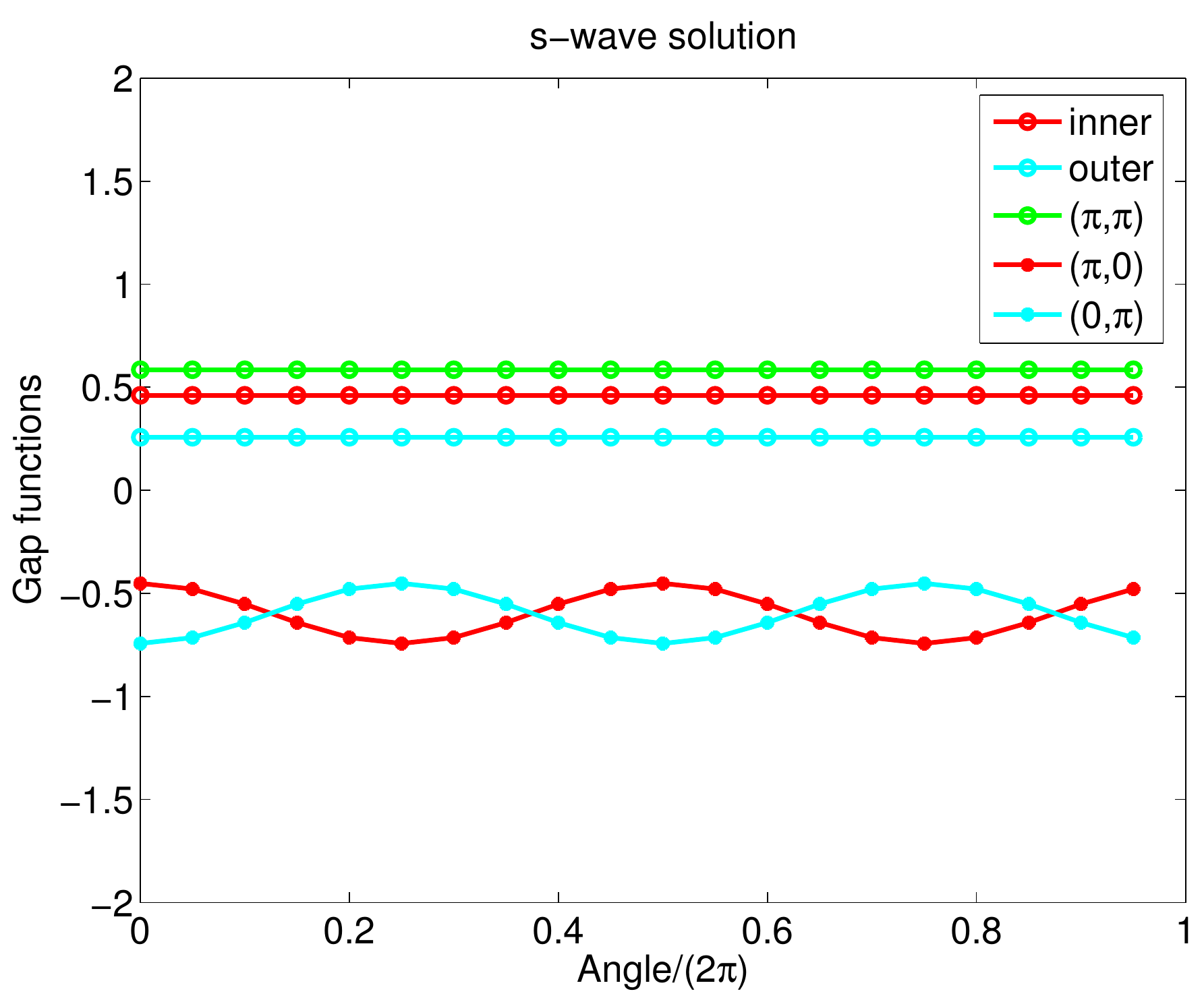}&
\includegraphics[width=0.45\columnwidth]{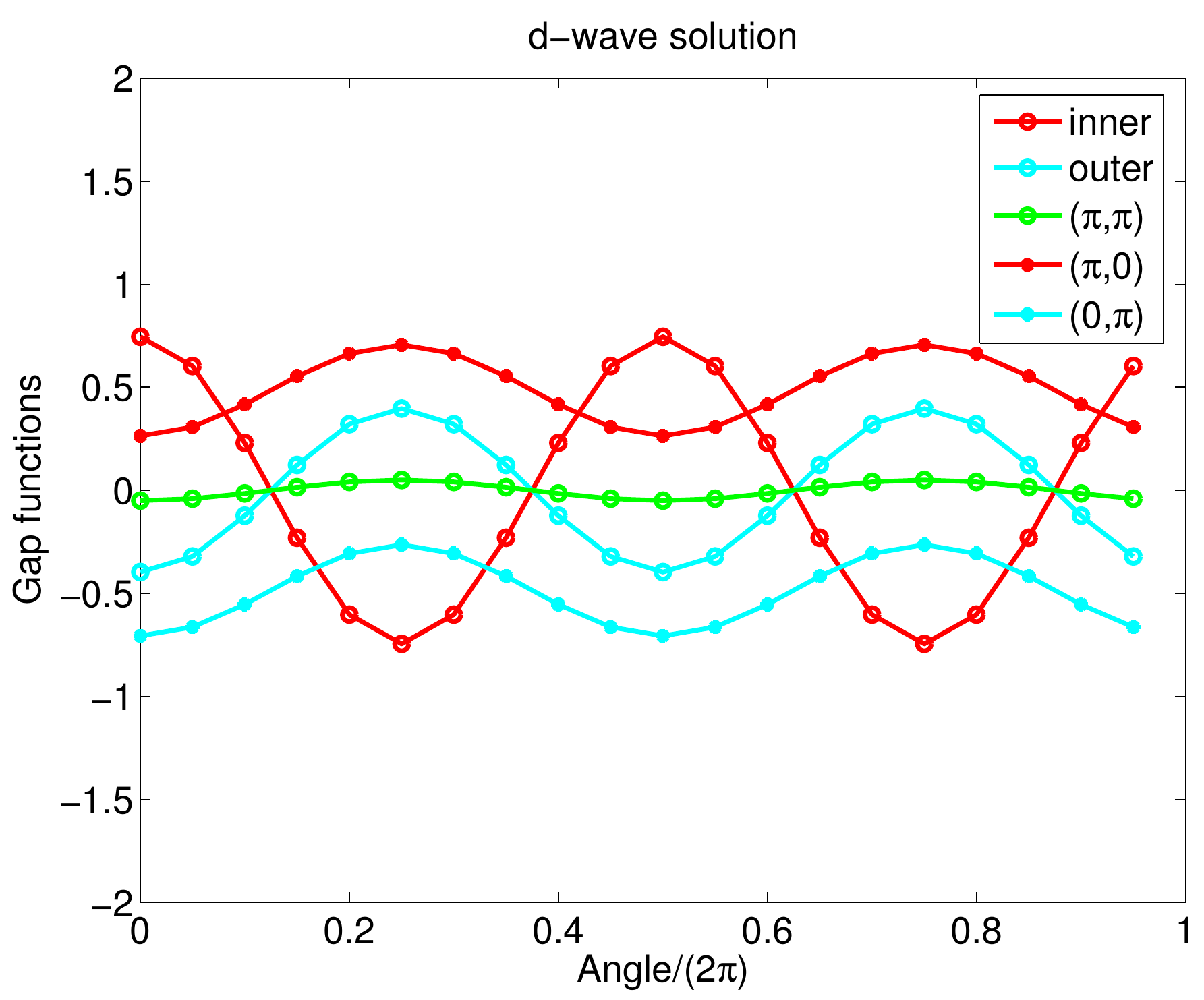}\\
\end{array}$
\caption{\label{fig:rev_4}
Representative case of small/moderate hole doping, when both hole and electron pockets are present.
 Panel a -- the FS, panel b --  representative fits
 of the interactions by LAHA (the dots are RPA results, the lines are LAHA expressions,  Eqs (\ref{s_4})-(\ref{s_5_d})). Panels c and d --
  the eigenfunctions in $s-$wave and $d-$wave channels for the largest $\lambda^s$ and $\lambda^d$.  From Ref.~[\onlinecite{maiti_last}].}
\end{figure}

\begin{figure}[htp]
$
\begin{array}{cc}
\includegraphics[width=0.45\columnwidth]{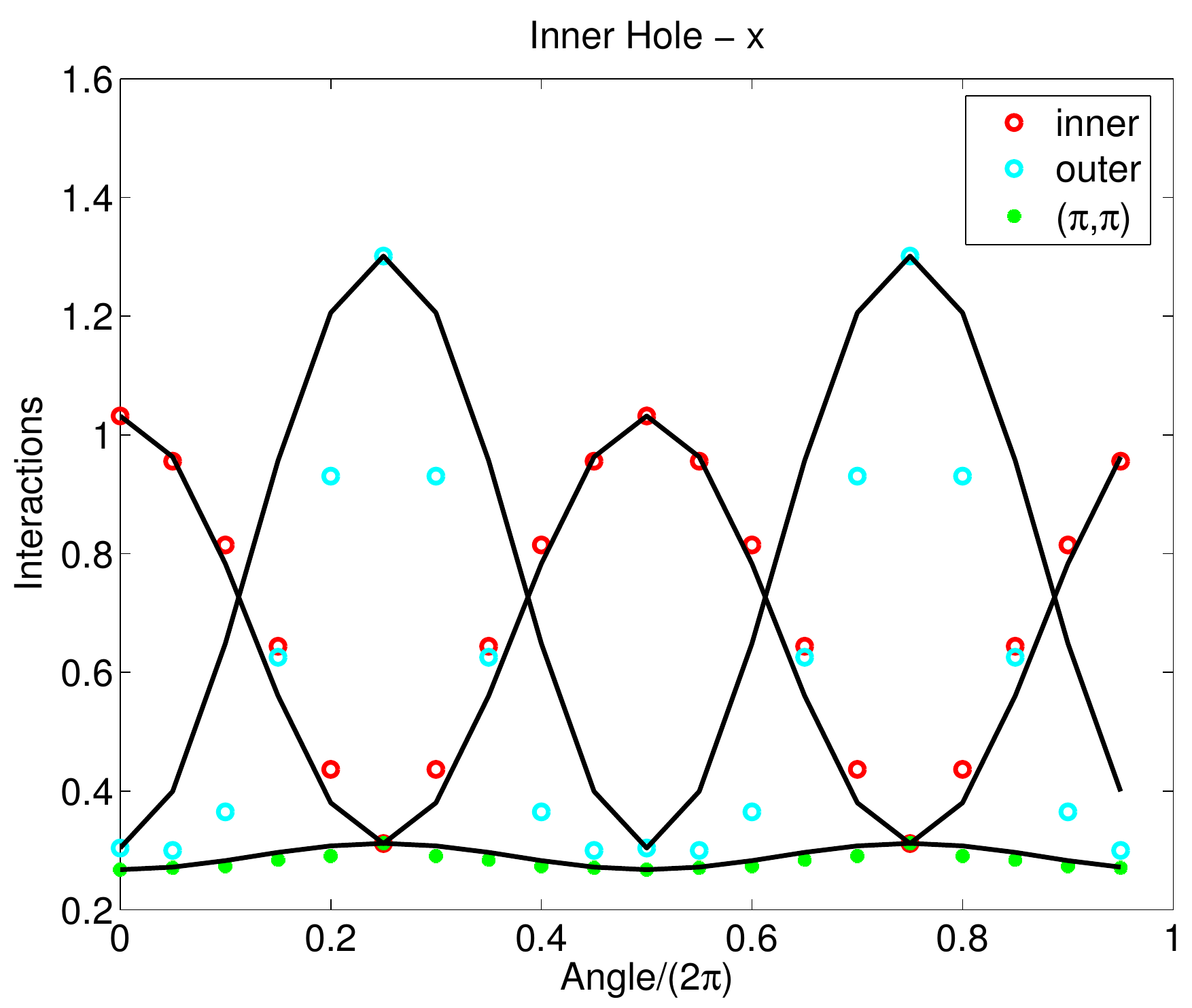}&
\includegraphics[width=0.45\columnwidth]{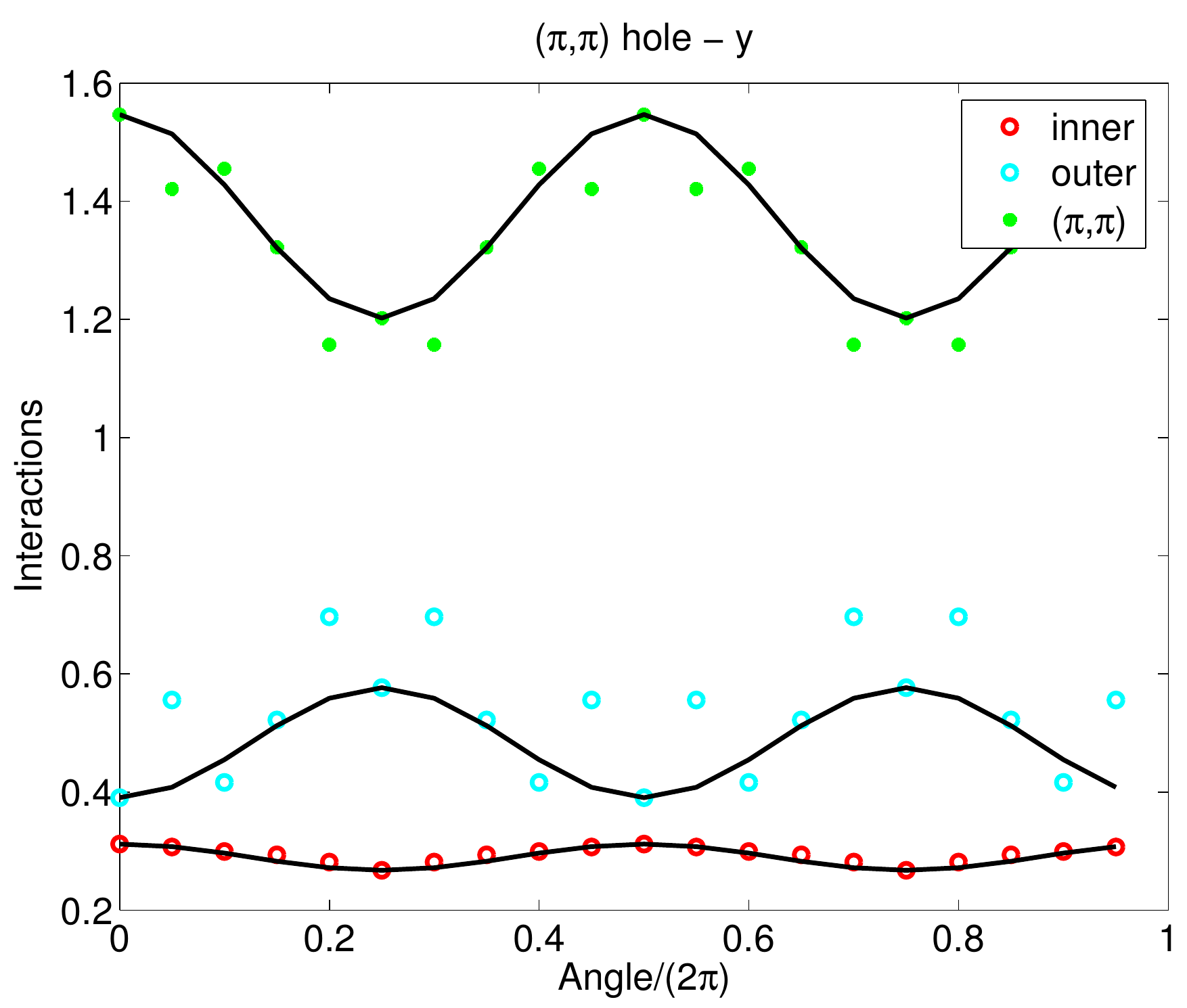}\\
\includegraphics[width=0.45\columnwidth]{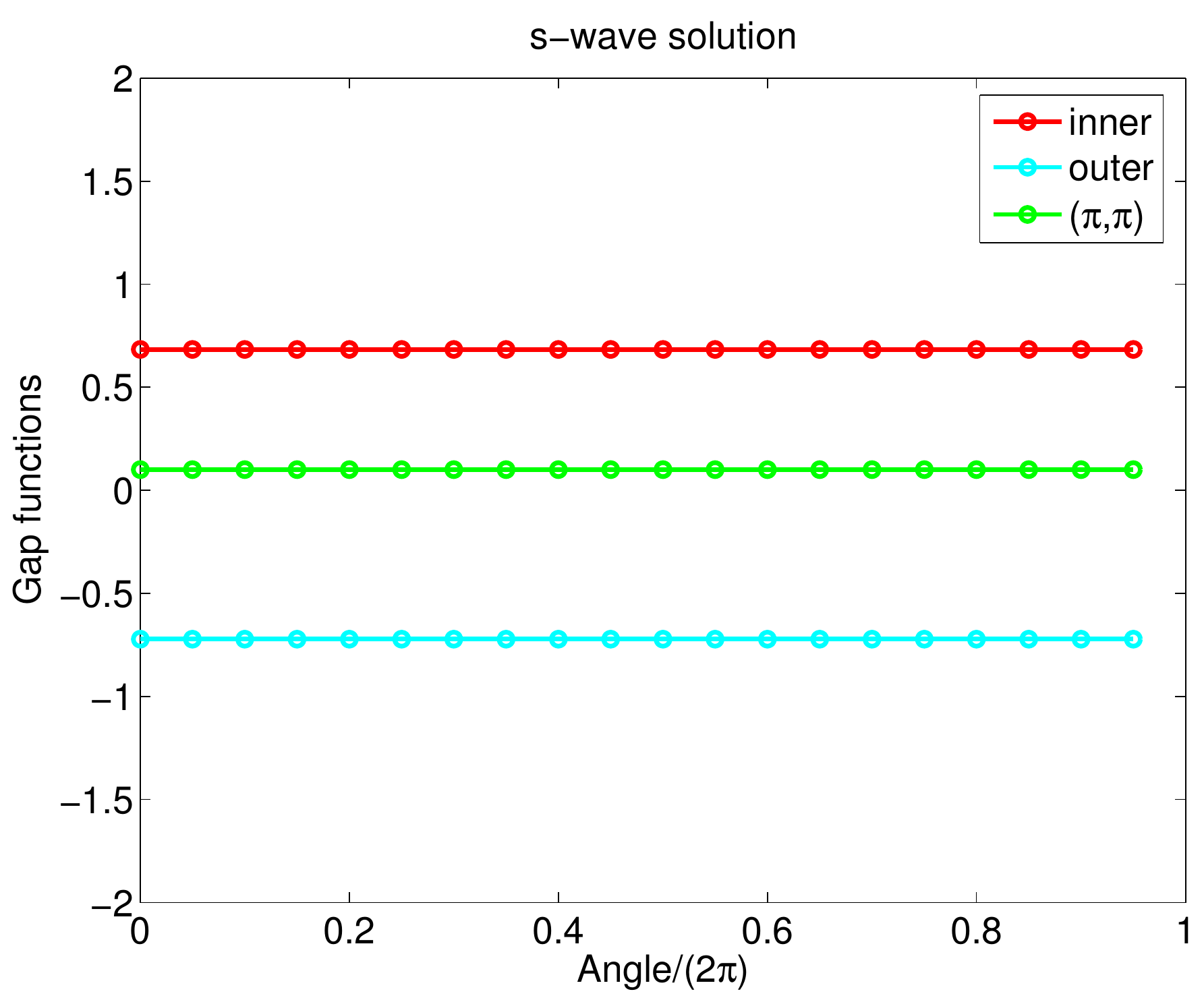}&
\includegraphics[width=0.45\columnwidth]{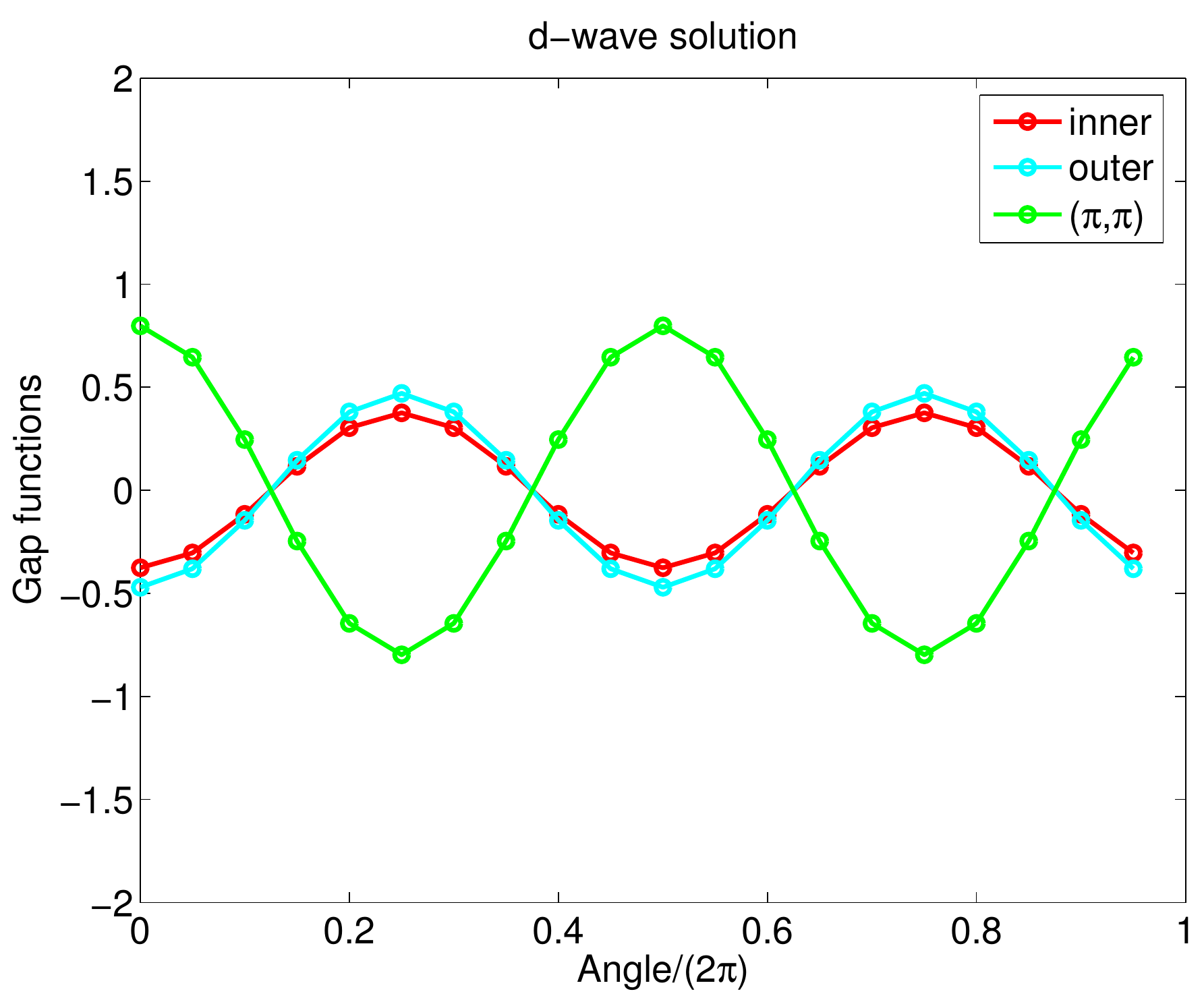}\\
\end{array}$
\caption{\label{fig:rev_5} The fits of the RPA
interactions by LAHA and the structure of $s-$wave and $d-$wave gaps in
 for strong hole doping ($\mu =-0.30eV$), when only hole
FSs are present.  From Ref.~[\onlinecite{maiti_last}].}
\end{figure}

\begin{table*}[htp]
\caption{\label{tab:2} Some of LAHA parameters  extracted from the fits in Figs.~\protect\ref{fig:rev_4} and \protect\ref{fig:rev_5} for hole doping. Block (i) corresponds to Fig. \protect\ref{fig:rev_4} (hole and electron pockets are present),  block (ii) corresponds to Fig. \protect\ref{fig:rev_5} ( no electron pockets).}
\begin{ruledtabular}
\begin{tabular}{lccccclcccccrccccc}
& \multicolumn{5}{c}{(i)}  & & \multicolumn{5}{c}{(iii)} \\
 \cline{2-6} \cline{8-12}
 $s-wave $&$u_{h_1h_1}$&$u_{h_1e}$&$\alpha_{h_1 e}$&$u_{ee}$&$\lambda_s$& & $u_{h_1h_1}$&$u_{h_1h_2}$&$u_{h_1
h_3}$&$u_{h_3 h_3}$&$\lambda_s$\\
&0.0.86&0.92&-0.18&1.00&0.58& &0.67&0.8&0.29&1.37&0.13\\
\cline{2-6} \cline{8-12}
$d-wave $&$\tilde{u}_{h_1h_1}$&$\tilde{u}_{h_1e}$&$\tilde{\alpha}_{h_1 e}$&$\tilde{u}_{e
e}$&$\lambda_d$& &$\tilde{u}_{h_1h_1}$&$\tilde{u}_{h_1 h_2}$&$\tilde{u}_{h_1 h_3}$&$\tilde{u}_{h_3 h_3}$&$\lambda_d$\\
&0.51&-0.45&-0.48&0.07&0.31& &0.36&-0.5&-0.02&-0.17&0.11\\
\end{tabular}
\end{ruledtabular}
\end{table*}

\subsection{Hole doping}

For small and moderate hole doping, the  FS
  contains 5 pockets --two hole pockets at $(0,0)$,
 two electron pockets at $(0,\pi)$ and $(\pi,0)$,
and one more hole pocket at $(\pi,\pi)$.
Representative FSs for hole doping,  typical fits by LAHA, the parameters extracted from the fit,
 and the solutions in s-wave and d-wave channels are shown in Fig. \ref{fig:rev_4} and in Table~\ref{tab:2}.
Just like for electron doping, there
 are universal and parameter-sensitive features.
The parameter-sensitive property is again the presence or absence of
accidental nodes in the $s$-wave gap along the electron FSs, although for
most of the parameters, the gap does not have nodes (see Fig.~\ref{fig:rev_4})
because the total $u_{he}$ increases once it acquires an
additional contribution $u_{h_3 e}$.

 There are two universal features. First,  the
$s$-wave eigenvalue is enhanced relative to a $d-$wave one and
 becomes  the leading instability  as long as both hole and electron pockets are present. Second,  the driving force for the
attraction in both $s$- and $d$- channels is again strong inter-pocket
electron-hole interaction ($u_{he}$ and ${\tilde u}_{he}$
terms), {\it no matter how small electron pockets are}.

The situation again  changes rapidly once electron pockets disappear, see Fig.~\ref{fig:rev_5}.  Now electron-hole interaction becomes irrelevant, and the attractive pairing interaction may only be due to intra and inter-pocket interactions involving  hole pockets.
LAHA analysis shows [\onlinecite{maiti_last}] that, at least for in some range of parameters,  there is an attraction in both
 $s-$wave and $d-$wave channels, and furthermore $\lambda_d \approx \lambda_s$, see Fig.~\ref{fig:rev_5}
The  near-equivalence of $s-$wave and $d-$wave eigenvalues was also found in recent unrestrictive RPA study~[\onlinecite{kuroki_11a}].
Within LAHA, the attractive
$\lambda_s$ is due to strong intra-pocket interaction between the two hole pockets centered at $(0,0)$.  The $s-$wave gap then changes sign
 between these two hole pockets.    The gap along $(\pi,\pi)$ pocket is induced  by a weaker inter-pocket interaction and is much smaller.
 LAHA neglects $\cos 4n \phi$  gap variations along the hole FSs (i.e., $s-$wave gaps are treated as angle-independent), but the theory can indeed be extended to include these terms.
The attractive $\lambda_d$  emerges by two reasons.
 First, the $d$-wave intra-pocket interaction ${\tilde
u}_{h_3 h_3}$ becomes negative, second, the
inter-pocket interaction ${\tilde u}_{h_1 h_2}$ between the two pockets at $(0,0)$ becomes larger in
magnitude than repulsive ${\tilde u}_{h_1 h_1}$ and ${\tilde
u}_{h_2 h_2}$ (see Table~\ref{tab:2}). The solutions with $\lambda_d>0$ then exist
separately for FSs $h_{1,2}$ and for $h_3$, the residual
inter-pocket interaction just sets the relative magnitudes and
phases between the gaps at $h_3$ and $h_{1,2}$.
The  $d$-wave gap with the same structure  has been obtain in  the fRG analysis at large hole doping~[\onlinecite{KFeAs_fRG}]. That study found less competition with $s-$wave than in PRA-based studies.

\section{Experimental situation}
\label{sec:6}

Experimental study of FeSCs is one of the major research topics in condensed-matter physics since 2008 and several detailed reviews have already appeared in the literature~~[\onlinecite{review,review_2,review_4,review_we,korshunov}].
 Here I briefly discuss the experimental situation concerning
the symmetry and the structure of the SC gap.

 As of today, there is no ``smoking gun'' experiment
which would carry the same weight as  phase-sensitive measurements of
$d_{x^2-y^2}$ gap symmetry in the cuprates~~[\onlinecite{d-wave}]. Still, there is enough experimental data to minimize the number of possible gap structures.

The theoretical proposals for the gap symmetry and structure are summarized
 in Sec. \ref{sec:8}. The proposed symmetry is different for weakly/moderately doped systems with hole and electron FSs and for strongly doped  systems where FSs of only one type are present. It  is then instructive to consider weak/moderate and strong doping separately.

\subsection{Moderate doping, gap symmetry}

The candidates are s-wave (either $s^\pm$ or $s^{++}$) or $d_{x^2-y^2}$ gap.
 The two behave very differently along the hole FSs centered at $(0,0)$ -- s-wave gap is nodeless with  $\cos 4 \phi$ variations, while d-wave gap has nodes along $k_x = \pm k_y$. ARPES measurements, both from synchrotron~~[\onlinecite{Evtush,122hole_ARPES,ARPES_1111,122electron_ARPES}] and using laser light~~[\onlinecite{laser_arpes}], show quite convincingly that the gap along hole FSs is nodeless in both hole and electron-doped FeSCs.  This unambiguously selects an s-wave.  Additional evidence in support of s-wave pairing comes from
 very flat low-T behavior of the penetration depth  in
 the highest $T_c$ 1111 FeSCs systems~~[\onlinecite{martin_09}].

\subsection{Moderate doping, $s^\pm$ vs $s^{++}$}

The distinction between  $s^\pm$ and $s^{++}$
 gaps is a more subtle issue, particularly given that both belong to the same $A_{1g}$ representation and also because in general $A_{1g}$
 gap on electron pockets may have strong oscillating component.
In general, the gaps on electron and hole FSs have non-equal magnitudes, and
 the issue whether the gap is $s^\pm$ or $s^{++}$ reduces to whether
 the gap averaged over an electron FS has the same sign
or opposite sign than the gap averaged over a hole FS.  This is not
 a fundamental  symmetry issue and, moreover, when $\cos 2 \phi$ oscillations are strong, one may switch from equal to opposite signs of the averaged gaps
 by a small change of parameters~~[\onlinecite{maiti_10}] or by adding impurities.~~[\onlinecite{efremov}] Still, when oscillations are not very strong, whether the eigenfunction has $s^\pm$ or $s^{++}$ character
  is essential because it determines, to a large extent, whether  the pairing is driven by spin  or by orbital fluctuations (see Sec.\ref{sec:4}).

The experimental data most frequently cited in support of $s^\pm$ gap is
 the observation of a magnetic  resonance in neutron scattering~~[\onlinecite{Cruz,resonance}].
If, as many researchers believe, the resonance is a spin exciton, it exists
 at a momentum $Q$ if the gaps at FS momenta $k_F$ and $k_F +Q$ are of opposite sign. Experimentally, the resonance is observed~~[\onlinecite{Cruz,resonance}] near $Q = (\pi,\pi)$ in the
folded BZ, which in this zone is precisely the distance between electron and hole FSs. The excitonic resonance then exists if the gap changes sign between hole and electron pockets and does not exist if the gap doesn't change sign.  A similar reasoning has been used in identifying the
  the resonance seen in the cuprates with a fingerprint of
 $d_{x^2-y^2}$ gap symmetry~~[\onlinecite{eschrig}]

From experimental perspective, the neutron peak is the resonance
 if it is narrow and is located below twice the gap value. The argument made by the supporters of $s^{++}$ scenario~~[\onlinecite{jap}] is that the observed neutron peak is more broad than the resonance seen in the cuprates, and that there is no firm evidence that the peak energy is below $2\Delta$ for the minimum gap.  For $s^{++}$ gap structure, there is no resonance, but there is a redistribution of the neutron spectral weight immediately above 2$\Delta$ what gives rise to a local maximum in the magnetic structure factor~~[\onlinecite{jap,kuroki_11,eremin_11}].
  Still, the majority of researchers do believe that the
 observed neutron peak is a resonance, and the fact that it is quite broad is at least partly due to $\cos 2 \phi$ gap variations along the electron FSs [\onlinecite{eremin_11}].

Another rather strong evidence in support of  $s^\pm$ gap
is the observed variation of the quasiparticle interference pattern in a magnetic field~~[\onlinecite{hanaguri}] although the interpretation of the data has been subject of debates~~[\onlinecite{debates}]. It  was also argued~~[\onlinecite{co-existence_Fe}] that the very presence of the co-existence region
between SC and stripe magnetism in FeSCs is a fingerprint of an $s^\pm$ gap, because for $s^{++}$ gap a first order transition between a pure magnetic and a pure SC state is a much more likely scenario.

\subsection{Moderate doping, nodal vs no-nodal $s^\pm$ gap}

Assume for definiteness that the pairing is driven by spin fluctuations and the
gap has $s^\pm$ structure. In 2D scenario, such
 gap has $\cos 2 \phi$ variations along electron FSs, which, according to theory, can be rather strong, particularly in electron-doped FeSCs.  Experimental data show that, whether or not the gap is nodeless or has nodes, depends on the material, on the doping, and on whether SC co-exists with SDW order.

\subsubsection {Hole doping}

 For hole-doped FeSCs (e.g. for Ba$_{1-x}$K$_x$Fe$_2$As$_2$) the data indicate
 that the gap is nodeless, away from the co-existence region.
  This is consistent with the theory (see Sec. \ref{sec:5}).
ARPES experiments do not show any angular variation of the gap along both hole and electron FSs~~[\onlinecite{Evtush,122hole_ARPES}], but it is not entirely clear whether ARPES can at present distinguish between the gaps  on the two electron FSs which in folded zone are both centered at $(\pi,\pi)$. Thermal conductivity data show that $\kappa/T$ tends to zero in the limit of $T=0$, in line with what one should expect for a nodeless SC~~[\onlinecite{reid_11}]. Specific heat data also show non-nodal behavior~~[\onlinecite{heat_hole}]. The interpretation of the penetration depth data requires more care as the data do show a power-law behavior $\lambda (T) - \lambda (0) \propto T^a$ with $a \sim 2$ (Refs. ~[\onlinecite{122hole_pen_depth}]).
 Such a behavior is expected for a SC with point nodes, but it is also
 expected in a wide range of $T$
 for a nodeless $s^\pm$ SC in the presence of modest inter-band scattering by non-magnetic impurities~~[\onlinecite{impurities}]. Penetration depth measurements on artificially irradiated samples~~[\onlinecite{kim_10}]  support the idea that the gap is nodeless and power-law
 $T^a$ behavior of   $\lambda (T) - \lambda (0)$ is due to impurities.

\subsubsection{Electron doping}

For electron-doped FeSCs, e.g., 122 materials like Ba(Fe$_{1-x}$Co$_x$)$_2$As$_2$
or 1111 materials like  NdFeAsO$_{1-x}$F$_x$,   ARPES shows no-nodal gap along hole FS~~[\onlinecite{ARPES_1111,122electron_ARPES}],
 but there are no data on the gap along each of
 the two electron FSs. At optimal doping, the data on both thermal conductivity~~[\onlinecite{thermal_opt,hashimoto}] and penetration depth~~[\onlinecite{hashimoto,pen_opt}] are consistent with no-nodal gap  However, the data for overdoped  Ba(Fe$_{1-x}$Co$_x$)$_2$As$_2$ indicate that gap nodes may develop: the behavior of $\lambda (T)$ becomes more steep, and $\kappa/T$ now tends to a finite value~~[\onlinecite{thermal_opt}], expected  for a SC with line nodes. The data also show  $\sqrt{H}$ behavior of $\kappa$ in a magnetic field~~[\onlinecite{thermal_opt}] also
 expected for a SC with line nodes~~[\onlinecite{volovik}], but it was argued that the behavior resembling $\sqrt{H}$ can be obtained even if $s^\pm$ gap has no nodes~~[\onlinecite{bang}].
  There is also clear anisotropy between in-plane
conductivity and conductivity along $z$ direction, what was interpreted~~[\onlinecite{thermal_opt}] as an indication that the nodes may be located near particular $k_z$.
Specific heat data in  overdoped   Ba(Fe$_{0.9}$Co$_{0.1}$)$_{2}$As$_{2}$ were also interpreted as evidence for the nodes.~~[\onlinecite{heat_el_over}]

The development of the nodes in $s^\pm$ gap upon electron doping is in line with the theory. The farther the system moves away from the
SDW phase, the weaker is the increase of  intra-band electron-hole interaction and hence the stronger is the competition from intra-band repulsion. As I discussed in Sec.\ref{sec:3}), the gap adjusts to this change by increasing its $\cos 2 \phi$ component in order to effectively
reduce the effect of the intra-band repulsion in the gap equation.

There is also experimental evidence for $\cos 2 \phi$ gap oscillations from
 the observed oscillations~~[\onlinecite{wen_10}] of the field-induced component of the specific heat $C(H,T)$ in superconducting FeTe$_{1-x}$Se$_x$ ($x \sim 0.5$).
  The measured $C(H,T)$ oscillates with the direction of the applied field as $\cos 4 \phi$. In theory, such an oscillation is related to the behavior of $\Delta^2 (\phi)$ (Ref. ~[\onlinecite{vekhter_10}]), hence $\cos 2 \phi$ gap oscillations in $\Delta$
lead to $\cos 4 \phi$ oscillations in $C(H,T)$.
The observed field and temperature dependence of the prefactor for  $\cos 4 \phi$ term are consistent with the idea that the oscillations are caused by
$\cos 2 \phi$ term in $\Delta$.  These data were also interpreted as evidence for no-nodal gap because if $\cos 2 \phi$ gap oscillations were strong and the gap had nodes at accidental points, the behavior  of $\Delta^2$ would be
 more complex than the observed $a+b \cos 4 \phi$.

For  LiFeAs, which is undoped but  has FS structure similar to electron-doped FeSCs, no-nodal behavior has  been observed in ARPES~~[\onlinecite{LiFeAs_ARPES2}],
 specific heat~~[\onlinecite{LiFeAs_ARPES1_nutron_scat}], penetration depth~~[\onlinecite{thermal_pen_LiFeAs}] and NMR~~[\onlinecite{buechner,ma_10}] measurements.
 The NMR data, however, were interpreted~~[\onlinecite{buechner}] as evidence for a possible B-type  $p$-wave  gap ($p_x + i p_y$).  Whether this is the case remains an open issue. In any event, the evidence for a no-nodal gap in this material is overwhelming.

\subsubsection{Co-existence region}

Taken at a face value, thermal conductivity and penetration depth data
 indicate that the gap becomes nodal in the co-existence regime in both
 hole-doped and in electron-doped FeSCs. The most striking evidence comes from thermal conductivity~~[\onlinecite{thermal_opt,reid_11}] -- in the co-existence regime $\kappa/T$ tends to a finite value at $T \to 0$ and shows $\sqrt{H}$ behavior, both typical for a SC with line nodes. There is no theoretical understanding at present why the SC gap develops nodes in the co-existence regime (for a toy two-pocket model theory predicts that the gap should have no-nodes~~[\onlinecite{parker}]), so this is another open issue.

\subsubsection{Isovalent doping}

Electron or hole doping is not the only way to change the properties of FeSCs. Another route is to replace one pnictide with the other. The most common replacement is As $\to$ P.  P-containing materials include the very first
FeSC -- LaFeOP, with $T_c \leq 5K$ (Ref. ~[\onlinecite{lafop}]), the family BaFe$_2$ As$_{1-x}$P$_{x}$ with the highest $T_c$ around 30K (Ref. ~[\onlinecite{matsuda}]), and  LiFeP~~[\onlinecite{private}].  Penetration depth, thermal conductivity, specific heat, and NMR
 data~~[\onlinecite{thermal_p}] in these materials all show the behavior consistent with line nodes. In particular, $\kappa$ scales linearly with $T$ at low $T$ and displays $\sqrt{H}$ behavior in a magnetic field, and $\lambda (T) - \lambda (0)$ is also linear in $T$ down to very low $T$.  Laser ARPES data
show~~[\onlinecite{laser_arpes}] that the gap along FS is nodeless, so the
 nodes likely are located on electron FSs.

On general grounds,  the existence of the nodes on electron FSs is in line with theory predictions particularly as BaFe$_2$ As$_{1-x}$P$_{x}$ has the same structure of 4 cylindrical FSs as electron-doped FeSCs for which nodes are most likely.
 It has been argued~~[\onlinecite{Kuroki}] that a replacement of As by P changes the hight of a pnictide with respect to Fe plane, what effectively reduces inter-pocket
electron-hole interaction, in which case the gap develops nodes to reduce
 the effect of intra-pocket repulsion.  However, this argument is only suggestive, and it is not entirely clear at the moment why all P-based FeSCs have
nodes. One way to analyze this semi-quantitatively is to study the correlation between $2\Delta/T_c$ on the hole FS and the presence of the nodes on electron FSs. This study shows~~[\onlinecite{maiti_delta}] that from this perspective P-based FeSCs are indeed the ``best case'' for the gap nodes.

Another open issue is the location of the nodes along z- direction.
 Oscillations of thermal conductivity
 with the direction of a magnetic field have been measured recently~~[\onlinecite{vekhter_11}], and
  $\cos 4 \phi$ component of these oscillations has
 been interpreted using the modified 2D form $\Delta_e (k_z) = \Delta_0 (1 + \alpha (k_z) \cos 2 \phi)$. The best fit to the data yields $\alpha(k_z) >1$ for some $k_z$ and $\alpha (k_z) <1$ for others,  in which case the nodes form
 patches along $k_z$. However, whether this is the only explanation of the data
 is debatable.

I caution  that although the nodes on electron FSs are most likely, they have not been directly observed yet, so it is still possible that the nodes are
 located on a hole FS, near particular $k_z$, as some of 3D theories suggest~~[\onlinecite{3D}]. Another possibility, which is also not entirely ruled out,
is that the system behavior near the surface, probed  by ARPES, is not the same as in the bulk. The
 probability that this is the case is not high, though, because ARPES data are obtained using a laser light which probes states located farther from the surface than in conventional synchrotron-based ARPES.

\subsection{Strongly doped FeSCs}

\subsubsection{Electron doping}

Strongly electron doped  materials is recently discovered family of
 A$_x$Fe$_{2-y}$Se$_2$
($A = K, Rb, Cs$)~~[\onlinecite{exp:AFESE,exp:AFESE_ARPES}] of which  K$_0.8$Fe$_{1.7}$Se$_2$ is the most studied material. $T_c$ in  A$_x$Fe$_{2-y}$Se$_2$ is rather high, almost 40K.  ARPES shows~~[\onlinecite{exp:AFESE_ARPES}] that
 only electron FSs are present in  A$_x$Fe$_{2-y}$Se$_2$, while
hole pockets are at least $60 meV$ from the FS, although hole dispersion above $60 meV$ is still clearly visible in ARPES.
 Two electron FSs are at $(0,\pi)$ and $(\pi,0)$, like in other FeSCs,
 and there is, possibly, another electron FS at $(0,0)$.
RPA, LAHA and fRG calculations for these systems predict that the gap should
 have a d-wave symmetry, at least for the case when the FSs are only at $(0,\pi)$ and $(\pi,0)$. A d-wave symmetry in this situation means that the gaps on the two electron FSs behave as $\Delta_0 (\pm 1 + \alpha \cos 2 \phi)$, and all calculations yield $\alpha <1$, i.e., no nodes (neglecting 3D effects). The theoretical  alternative is  $s^{++}$ symmetry by one reason~~[\onlinecite{kontani_se}] or the other~~[\onlinecite{yu_11,bernevig}]. At present, both ARPES and specific heat data point that the gap is nodeless, at least for most of $k_z$ values, but whether the gap is an $s-$wave or a nodeless $d-$wave remains to be seen.

\subsubsection{Hole doping}

Strongly hole doped system is  KFe$_2$As$_2$ ($T_c =3K$) which is at the opposite end from parent  BaFe$_2$As$_2$ in the family of  K$_x$Ba$_{1-x}$Fe$_2$As$_2$.
  According to ARPES~~[\onlinecite{KFeAs_ARPES_QO}], this system has  no electron pockets and has hole pockets at $(0,0)$.  There may also be additional hole pockets,
 but this is not entirely clear.
   Both thermal conductivity and penetration depth measurements clearly point to nodal behavior [\onlinecite{KFeAs_exp_nodal}].
  There is, however, no ``smoking gun''  symmetry-sensitive measurement, so whether the gap is a d-wave or an $s$-wave with nodes due to strong $\cos 4 \phi$ gap component on one of the FSs remains an open issue

\subsection{Summary}

Overall, the agreement between theory and experiment with respect to gap symmetry and structure is reasonably good.
Theory predicts that the gap symmetry in weakly and moderately doped FeSCs is
 an s-wave. This  is consistent with ARPES data.  Many theorist
  argue that an s-wave gap changes sign between hole and electron FSs.
Quasiparticle interference  and neutron scattering
data are consistent with this picture, if, indeed, the neutron peak is the resonance. Further, theory predicts that  $s^\pm$ gap has oscillations along electron FS, and these oscillations give rise to accidental nodes, which are more likely for systems with 2 hole and 2 electron cylindrical FSs than in systems with an additional hole cylindrical FS.  This is also generally consistent with the experiments, although at this moment the agreement is more qualitative than quantitative. For strongly electron-doped FeSCs, the theory based on RPA, LAHA, and fRG predicts a nodeless d-wave gap, except, perhaps, the region near $k_z = \pi/2$. This is neither confirmed not disproved by the experiments.
For strongly hole-doped FeSCs, with only hole pockets, experiments point to the presence of the nodes,
but whether the gap is a  $d-$wave or a nodal $s-$wave has not been firmly established.
Finally, in the co-existence region of SC + SDW, theory prediction for a 2-band toy model is a  nodeless $s^\pm$ gap.  This is consistent with the data for weak SDW order, but apparently inconsistent with the data for FeSCs with strong SDW order, which show behavior consistent with the nodes in the SC gap. What is the gap structure deep in the co-existence region remains to be seen.

\section{Conclusion}
\label{sec:7}

The analysis of the gap symmetry and structure in FeSCs is a fascinating subject
 because of multi-orbital/multi-band nature of these materials.  It is now well
 understood that in multi-band systems a  conventional notion that s-wave gap is nodeless, d-wave has 4 nodes, etc, does not work, and s-wave gap may have nodes, while d-wave gap may remain nodeless. Furthermore, nodes may appear or disappear, depending on doping and other external conditions.
 Another peculiarity of FeSCs is a
 close proximity between pairing eigenvalues $\lambda_i$ from different symmetry representations, e.g., $A_{1g}$ and $B_{1g}$. Because of  proximity,  even
the gap symmetry may change upon doping or other external perturbation.
Also,  below $T_c$, the system may lower the energy by moving into
 a mixed state with, e.g., $s+id$ gap function. To analyze this issue, one needs to go beyond the analysis presented in this review and solve the full non-linear gap equation.

The existence of superconductivity at the ``end points'' of the phase diagram,
 when only hole pockets or only electron pockets are present, is another
fascinating issue. There is reasonably good understanding of a d-wave pairing at strong hole doping, but there are competing proposals for the gap symmetry at strong electron doping, and so far experiments cannot distinguish between these proposals.  Finally, the gap structure in the co-existence regime
with SDW is yet another fascinating unexplored issue, and  comprehensive analysis of SC in the co-existence phase is clearly called for.

\section*{Acknowledgements}
I acknowledge helpful discussions with L. Benfatto, A. Bernevig, S. Budko, P. Canfield,  A. Carrington, A. Coldea,
 L. Digiorgi, R. Fernandes, S. Graser, H. Ding, W. Hanke, P. Hirschfeld, K. Honerkamp, D. Efremov, I. Eremin, J. Knolle, H. Kontani, K. Kuroki, D-H. Lee, T. Maier, S. Maiti, Y. Matsuda, I. Mazin, K. Moller, R. Prozorov, D. Scalapino, T. Shibauchi, Q. Si, J. Schmalian, H. Takagi, Z.~Tesanovic, R. Thomale, M. Vavilov,  A. Vorontsov, and H.H. Wen. I am thankful to S. Sachdev for careful reading of the manuscript and useful comments. This work was supported by NSF-DMR-0906953.

\section{Summary points}
\label{sec:8}

 \begin{itemize}
\item
For weakly and moderately electron-doped FeSCs, $s^\pm$ and $d_{x^2-y^2}$ pairing
  channels are nearly degenerate. The driving force for the
 pairing in both channels  is
 inter-pocket electron-hole interaction, enhanced by spin fluctuations.
 d-wave gap has nodes at $k_x = \pm k_y$ on the hole FSs,
 $s^\pm$ gap has $\cos 2 \phi$ variations on electron FSs and may have nodes.
 The probability that $s^\pm$ gap has nodes increases with electron doping.
\item
For strongly electron doped FeSCs, when only electron pockets remain, some results show that the gap has a $d-$wave symmetry, which in this situation implies that angle-independent component of the gap changes sign between the two electron pockets.
 The driving mechanism for the pairing is a direct $d-$wave attraction between
 electron pockets, again enhanced by spin fluctuations.
 The gap is nodeless, but  may acquire nodes near $k_z =\pi/2$ when 3D effects and/or hybridization between electron pockets are included. A competing theory
 proposal, based on orbital $J-1-J-2$ model,  is that the gap is sign-preserving s-wave.
\item
For weakly and moderately hole-doped FeSCs, $s^\pm$ pairing is  the dominantinstability. The driving force is again inter-pocket electron-hole interaction,
 enhanced by spin fluctuations. The gap has $\cos 2 \phi$ variation along
 electron FSs but likely does not have nodes.
\item
For strongly hole-doped FeSCs, when only hole FSs remain, the gap symmetry may be $d-$wave, with nodes at $k_x = \pm k_y$ on all hole FSs.  The d-wave pairing
 mechanism is the combination of a d-wave attraction within the hole pocket at $(\pi,\pi)$ and a strong intra-pocket d-wave interaction between the two hole pockets centered at $(0,0)$.   A competing theory
 proposal   is that the gap is s-wave, with potential nodes due to strong $\cos 4 \phi$ variation along hole FSs centered at $(0,0)$.
\item
For real 3D systems, one scenario is that the 2D picture survives, and nodes, if present, are ``vertical'' nodes, located on electronic FSs at given $k_x$ and $k_y$ but arbitrary $k_z$. Another scenario is that nodes on electron FSs exists
 only in some ranges of $k_z$. And the third, more radical scenario, is that
 the nodes appear as ``horizontal'' nodes near particular $k_z$, either
 on electron or on hole FSs.
\item
An alternative scenario for all FeSCs is that the gap is a conventional, sign-preserving $s^{++}$ gap, but still may have $\cos 2 \phi$ oscillations on electron FSs.  Such a gap appears if  inter-pocket electron-hole interaction  $u_{he}$ is negative (attractive) and is enhanced by charge (orbital) fluctuations and by phonons
\end{itemize}

\section{A list of future issues}
\label{sec:9}
\begin{itemize}
\item
What would be the ``smoking gun'' experiment to distinguish between $s^\pm$ and $s^{++}$ gap?
\item
Why all P-based FeSCs appear to have nodes, and where these nodes are located?
\item
What is the structure of the SC gap in the co-existence region with SDW order?
\item
What is the pairing mechanism and the gap symmetry and structure in heavily
 electron-doped  K$_x$Fe$_{2-y}$Se$_2$ and in heavily hole-doped KFe$_2$As$_2$?
\item
Is there a mixed $s+id$ SC state in FeSCs at $T=0$?
\item
Is there a FeSC with a p-wave symmetry?
\end{itemize}


\begin{thebibliography}{5}

\bibitem{bib:Hosono} Y. Kamihara, T. Watanabe, M. Hirano, H.
Hosono, J. Am. Chem. Soc. \textbf{130}, 3296(2008).
\bibitem{bib:X Chen} X. H. Chen, T. Wu, G. Wu,
R. H. Liu, H. Chen, D. F. Fang, Nature \textbf{453}, 761(2008).
\bibitem{bib:G Chen} G. F. Chen, Z. Li, D. Wu, G. Li, W. Z. Hu {\it et al}
 Phys. Rev. Lett. \textbf{100}, 247002 (2008).
\bibitem{bib:ZA Ren}Z.-A. Ren, G.-C. Che, X.-L. Dong, J. Yang, W. Lu
{\it et al}
 Europhys. Lett. \textbf{83}, 17002(2008)
\bibitem{bib:Rotter} M. Rotter, M. Tegel, D. Johrendt,  Phys. Rev. Lett. \textbf{101}, 107006 (2008)
\bibitem{bib:Sasmal}K. Sasmal, B. Lv, B. Lorenz, A. M. Guloy, F. Chen, Y.-Y. Xue, and
C.-W. Chu, Phys. Rev. Lett. \textbf{101}, 107007 (2008),
\bibitem{bib:Ni}N. Ni, A. Thaler, J. Q. Yan, A. Kracher, E. Colombier, S.
L. Bud'ko, P. C. Canfield, Phys. Rev. B \textbf{82}, 024519
(2010).
\bibitem{nakai}
Y. Nakai, T. Iye, S. Kitagawa, K. Ishida, H. Ikeda, S. Kasahara, H. Shishido, T. Shibauchi, Y. Matsuda, T. Terashima,
Phys. Rev. Lett. 105, 107003 (2010).
\bibitem{bib:Wang}X.C.Wang, Q.Q. Liu, Y.X. Lv, W.B. Gao, L.X.Yang, R.C.Yu, F.Y.Li, and C.Q. Jin, arXiv:0806.4688v3;
 S. V. Borisenko, V. B. Zabolotnyy, D. V. Evtushinsky, T. K. Kim, I. V. Morozov {\it et al}
  Phys. Rev. Lett. 105, 067002 (2010).

\bibitem{bib:Mizuguchi}Y. Mizuguchi, F. Tomioka, S. Tsuda,
T. Yamaguchi, and Y. Takano, Appl. Phys. Lett. \textbf{93}, 152505
(2008), F. C. Hsu et al., Proc. Natl. Acad. Sci. U.S.A.
\textbf{105}, 14 262 (2008), M. H. Fang, H. M. Pham, B. Qian, T. J. Liu, E. K. Vehstedt, Y. Liu, L. Spinu, and Z. Q. Mao,  Phys. Rev. B
\textbf{78}, 224503 (2008), G. F. Chen, Z. G. Chen, J. Dong, W. Z. Hu, G. Li, X. D. Zhang, P. Zheng, J. L. Luo, and N. L. Wang,
 Phys. Rev. B \textbf{79}, 140509(R) (2009).

\bibitem{exp:AFESE} J. Guo, S. Jin, G. Wang, S. Wang, K. Zhu, T. Zhou, M. He, and X. Chen, Phys. Rev. B \textbf{82}, 180520(R) (2010).

\bibitem{exp:AFESE_ARPES} T. Qian, X.-P. Wang, W.-C. Jin, P. Zhang, P. Richard, G. Xu, X. Dai, Z. Fang, J.-G. Guo, X.-L. Chen, H. Ding,
Phys. Rev. Lett. 106, 187001 (2011).
\bibitem{Cruz} For the latest results on magnetic measaurements, see D. S. Inosov, J. T. Park, P. Bourges, D. L. Sun, Y. Sidis, A. Schneidewind, K. Hradil, D. Haug, C. T. Lin, B. Keimer,  and V. Hinkov,
 Nature Physics \textbf{6}, 178-181 (2010)
and references therein.
\bibitem{bib:Kamihara}Y. Kamihara, T. Watanabe, M. Hirano, and H.
Hosono,  J. Am. Chem. Soc. \textbf{128}, 10012 (2006).
\bibitem{review}
 D.C. Johnston, Adv. Phys., {\bf 59}, 803 (2010)
\bibitem{review_2}
 J-P Paglione and R.L. Greene, Nature Phys. {\bf 6}, 645 (2010).
\bibitem{review_3} I.I. Mazin, Nature {\bf 464}, 183 (2010).
\bibitem{review_4}  H.H. Wen and S. Li, Annu. Rev. Condens. Matter Phys.,  {\bf 2}, 121 (2011).
\bibitem{Graser}S. Graser, T. A. Maier, P. J. Hirshfeld, D. J. Scalapino, New
J. Phys. \textbf{11}, 025016 (2009).
\bibitem{peter} A. F. Kemper, T. A. Maier, S. Graser, H.-P. Cheng, P. J. Hirschfeld, and D. J. Scalapino, New J. Phys. {\bf 12},  073030 (2010).
\bibitem{Kuroki_2}K. Kuroki, H. Usui, S. Onari, R. Arita, and H. Aoki, Phys. Rev. B
79, 224511 (2009).
\bibitem{rev_physica}  A. V. Chubukov
Physica C \textbf{469}, 640(2009).
\bibitem{mazin_schmalian} I.I. Mazin and J. Schmalian,  Physica C,
\textbf{469}, 614 (2009).
\bibitem{review_we} D.N. Basov and A.V. Chubukov, Nature Physics {\bf 7}, 241 (2011).
\bibitem{korshunov} P.J. Hirschfeld, M.M. Korshunov, and I.I. Mazin, arXiv:1106.3712.
\bibitem{Onnes}
H.~{Kamerlingh}~Onnes. Comm. Phys. Lab. Univ. Leiden (1911).
\bibitem{BCS}
J.~Bardeen, L.~N. Cooper, and J.~R. Schrieffer.
 Phys. Rev., \textbf{108}, 1175 (1957).
\bibitem{mgb2}
J.~Kortus, I.~I. Mazin, K.~D. Belashchenko, V.~P. Antropov, and L.~L. Boyer.
Phys. Rev. Lett., \textbf{86}, 4656 (2001).
\bibitem{3he}
P.~W. Anderson and P.~Morel.
Phys. Rev., \textbf{123}, 1911--1934 (1961).
\bibitem{bednortz}
J.~G. Bednorz and K.~A. M\"{u}ller, Zeitschrift f\"{u}r Physik B Condensed Matter, \textbf{64}, 189
  (1986).
\bibitem{stat_phys} E. M. Lifshitz
and L. P. Pitaevski, \emph{%
Statistical Physics}, (Pergamon Press, 1980).
\bibitem{KL} W. Kohn and J. M. Luttinger,  Phys. Rev. Letters, {\bf 15}, 524 (1965).
\bibitem{d-wave}  see e.g., B. Goss Levi,  Physics Today, January 1996,
 Page 19 and references therein.
\bibitem{Mazin} I. I. Mazin, D. J. Singh, M. D. Johannes, and M.
H. Du, Phys. Rev. Lett. \textbf{101}, 057003 (2008).
\bibitem{Kuroki}K. Kuroki, S. Onari, R. Arita, H. Usui, Y. Tanaka, H. Kontani, and H. Aoki Phys. Rev. Lett. \textbf{101}, 087004 (2008).
\bibitem{jap} Seiichiro Onari and  Hiroshi Kontani,  Phys. Rev. Lett. 103, 177001 (2009).
\bibitem{ku}   W.G. Yin, C. C. Lee, and Wei Ku, Phys. Rev. Lett. 105, 107004 (2010).
\bibitem{rafael_we} R. Fernandes, A.V. Chubukov, I. Eremin, J. Knolle, and J. Schmalian, in preparation.
\bibitem{p_lee}  P. A. Lee and X.-G. Wen, Phys. Rev. B
\textbf{78}, 144517 (2008).
\bibitem{KFeAs_ARPES_QO} T. Sato, K. Nakayama, Y. Sekiba, P. Richard, Y.-M. Xu {\it et al}
  Phys. Rev. Lett. \textbf{103}, 047002 (2009);
  T. Terashima \textit{et al.}, J. Phys. Soc. Japan \textbf{79}, 053702 (2010).
\bibitem{KFeAs_exp_nodal} J. K. Dong, S. Y. Zhou, T. Y. Guan, H. Zhang, Y. F. Dai
 {\it et al}
  Phys. Rev. Lett. \textbf{104}, 087005 (2010);
  K. Hashimoto, A. Serafin, S. Tonegawa, R. Katsumata, R. Okazaki \textit{et al.},
  Phys. Rev. B 82, 014526 (2010).
  \bibitem{laser_arpes}  T. Shimojima, F. Sakaguchi, K. Ishizaka, Y. Ishida, T. Kiss \emph{et. al.}, (unpublished)
\bibitem{buechner} P. M. R. Brydon, M. Daghofer, C. Timm, J. van den Brink, Phys. Rev. B 83, 060501 (2011).
\bibitem{ding}
H.~Ding, P.~Richard, K.~Nakayama, K.~Sugawara, T.~Arakane {\it et al}
  Europhys. Lett., \textbf{83}, 47001  (2008).
\bibitem{chen}
T.~Y. Chen, Z.~Tesanovic, R.~H. Liu, X.~H. Chen, and C.~L. Chien, Nature, \textbf{453}, 1224 (2008).
\bibitem{osborn1}
A.~D. Christianson, E.~A. Goremychkin, R.~Osborn, S.~Rosenkranz, M.~D. Lumsden,
  C.~D. Malliakas, I.~S. Todorov, H.~Claus, D.~Y. Chung, M.~G. Kanatzidis,
  R.~I. Bewley, and T.~Guidi, Nature, \textbf{456}, 930 (2008).
\bibitem{carrington1}
L.~Malone, J.~D. Fletcher, A.~Serafin, A.~Carrington, N.~D. Zhigadlo,
  Z.~Bukowski, S.~Katrych, and J.~Karpinski, Phys. Rev. B 79, 140501(R) (2009).
\bibitem{nodes} S. Kasahara {\it et al},
Phys. Rev. B 81, 184519, (2010).
\bibitem{carrington}
J.~D. Fletcher, A.~Serafin, L.~Malone, J.~Analytis, J.-H. Chu, A.~S. Erickson,
  I.~R. Fisher, and A.~Carrington, Phys. Rev. Lett. 102, 147001 (2009).
\bibitem{proz}
R.~T. Gordon, N.~Ni, C.~Martin, M.~A. Tanatar, M.~D. Vannette, H.~Kim {\it et al}
Phys. Rev. Lett., \textbf{102}, 127004 (2009).
\bibitem{proz_1}
C.~Martin, R.~T. Gordon, M.~A. Tanatar, H.~Kim, N.~Ni, S.~L. Bud'ko {\it et al}
 Phys. Rev. B 80, 020501(R) (2009).
\bibitem{moler} L. Luan, O. M. Auslaender, T. M. Lippman, C. W. Hicks, B. Kalisky {\it et al}
Phys. Rev. B 81, 100501(R) (2010).
\bibitem{Li} M. Yi, D. H. Lu, J. G. Analytis, J.-H. Chu, S.-K. Mo {\it et al}
  Phys. Rev. B \textbf{80}, 024515
(2009)
\bibitem{bib:Suchitra} S. E. Sebastian, J. Gillett, N. Harrison, P. H. C. Lau, C. H. Mielke, and G. G. Lonzarich,  J. Phys.: Condens. Matter \textbf{20} 422203(2008).
\bibitem{Evtush} D. V. Evtushinsky, D. S. Inosov, V. B. Zabolotnyy, M. S. Viazovska, R. Khasanov {\it et al}
 New J. Phys. \textbf{11}, 055069 (2009);
\bibitem{bib:first}
D.J. Singh and M.-H. Du, Phys. Rev. Lett. \textbf{100}, 237003
(2008); M.J. Calderon, B. Valenzuela, and  E. Bascones, Phys. Rev.
B 80, 094531 (2009).
\bibitem{Boeri} L. Boeri, O. V. Dolgov, and A. A. Golubov, Phys. Rev. Lett. \textbf{101}, 026403 (2008)
\bibitem{mazin} I.I. Mazin, arXiv:1102.3655  (unpublished).
\bibitem{3D} S. Graser, A. F. Kemper, T. A. Maier, H.-P. Cheng, P. J. Hirschfeld, and D. J. Scalapino, Phys. Rev. B 81, 214503 (2010).
\bibitem{norman} M. Norman, Physics {\bf 1}, 21 (2008).
\bibitem{ising} see e.g. R. M. Fernandes, L. H. VanBebber, S. Bhattacharya, P. Chandra, V. Keppens {\it et al}
 Phys. Rev. Lett. 105, 157003 (2010);
 C. Xu, M. Mueller, and S. Sachdev, Phys. Rev. B 78, 020501(R) (2008); C. Fang, H. Yao, W.-F. Tsai, J.P. Hu, and S. A. Kivelson,
  Phys. Rev. B 77 224509 (2008).
\bibitem{vekhter_10} A.B.Vorontsov and I.Vekhter, Phys. Rev. Lett. 105, 187004
 (2010);  A.V. Chubukov and I. Eremin,
 Phys. Rev. B 82, 060504(R) (2010).
\bibitem{mike} Ar. Abanov, A. V. Chubukov, and M. R. Norman, Phys. Rev. B 78, 220507(R) (2008).
\bibitem{maiti_10} S. Maiti and A.V. Chubukov, Phys. Rev. B \textbf{82}, 214515 (2010).
\bibitem{fRG}
Fa Wang, H. Zhai, Y. Ran, A. Vishwanath, and D-H Lee, Phys. Rev. Lett.
\textbf{102}, 047005 (2009); C. Platt, C.  Honerkamp, and Werner Hanke,
New J. Phys. \textbf{11}, 055058 (2009); R. Thomale, C. Platt, J-P. Hu, C. Honerkamp, and B. A. Bernevig, Phys. Rev. B \textbf{80}, 180505 (2009). For general discussion on fRG see, e.g.,
M. Salmhofer {\it et al}, Prog. Theor. .Phys. {\bf 112}, 943 (2004).
\bibitem{maiti_last} S. Maiti, M.M. Korshunov, T.A. Maier, P.J. Hirschfeld, and A.V. Chubukov, arXiv:1104.1814.
\bibitem{cvv}  A. V. Chubukov, M. G. Vavilov, A. B. Vorontsov,
Phys. Rev. B \textbf{80}, 140515(R)(2009).
\bibitem{orbital_J} Q. Si and E. Abrahams, Phys. Rev. Lett. {\bf 101}, 076401 (2008).
\bibitem{ref:Cao} C. Cao, P.J. Hirschfeld, H.-P. Cheng, Phys. Rev. B \textbf{77}, 220506(R) (2008).
\bibitem{cee} A. V. Chubukov, D. Efremov, and I. Eremin,
Phys. Rev. B \textbf{78}, 134512 (2008).
\bibitem{mcmillan} W. L. McMillan, Phys. Rev. 167, 331 (1968); N.N. Bogolubov,
V.V. Tolmachev, and D.V. Shirkov, Consultants Bureau, 1959.
\bibitem{podolsky} Daniel Podolsky, Hae-Young Kee, and Yong Baek Kim,
Europhysics Letters 88, 17004 (2009).
\bibitem{rice} K. Le Hur and T. M. Rice, Annals of Physics 324 (2009) 1452
\bibitem{berk_schrieffer} N.F. Berk and J. R. Schrieffer, Phys. REv. Lett {\bf 17}, 433 (1966).
\bibitem{scalapino}   See, e.g.,   P. Monthoux and D. Pines, Phys. Rev. B 47, 6069 (1993); D.J. Scalapino, Phys. Rep. \textbf{250},
329 (1995);  D. Manske, I. Eremin, and K. H. Bennemann, Phys. Rev. B 62, 13922 (2000); A. Abanov, A. V. Chubukov, and J. Schmalian, Adv. Phys. 52, 119 (2003).
\bibitem{chub_maslov}  D.L. Maslov and A.V. Chubukov, Phys. Rev. B 81, 045110 (2010).
\bibitem{lara} L. Benfatto, M. Capone, S. Caprara, C. Castellani, and C. Di Castro,  Phys. Rev. B 78, 140502(R) (2008).
\bibitem{nematic}  J-H. Chu, J G. Analytis,  K. De Greve, P. L. McMahon, Z. Islam, Yoshihisa Yamamoto, and Ian R. Fisher
 Science {\bf 329}, 824 (2010); M.A. Tartar,
  E. C. Blomberg, A. Kreyssig, M. G. Kim, N. Ni {\it et al}
 Phys. Rev. B {\bf 81}, 184508 (2010);
  T.M. Chuang     * T.-M. Chuang, M. P. Allan, J. Lee, Yang Xie {\it et al}
, Science {\bf 327}, 181 (2010);
 S. Kasahara {\it et al}, unpublished.

\bibitem{zlatko} V. Stanev, J. Kang, Z. Tesanovic, Phys. Rev. B \textbf{78}, 184509 (2008); V. Stanev, B. S. Alexandrov, P. Nikoli\'{c} and Z. Te\v{s}anovi\'{c}, arXiv:1006.0447;  V. Cvetkovic and Z. Tesanovic,  Phys. Rev. B
\textbf{80}, 024512(2009).

\bibitem{Thomale_10} R. Thomale, C. Platt, W. Hanke, and B. A. Bernevig, Phys. Rev. Lett. 106, 187003 (2011).
\bibitem{tom_09}  T.A. Maier, S. Graser,  D.J. Scalapino, and P.J. Hirschfeld,
 \prb \textbf{79} 224510 (2009).

\bibitem{yu_11} R. Yu, P. Goswami, Q. Si, P. Nikolic, and J-X Zhu, arXiv:1103.3259.

\bibitem{bernevig} A. Bernevig,  private coimmunication.

\bibitem{kontani_se} T. Saito, S. Onari, and H. Kontani,
 Phys. Rev. B 83, 140512(R) (2011).

\bibitem{graser_11}  T.A. Maier, S. Graser, P. J. Hirschfeld, and D. J. Scalapino,
Phys. Rev. B \textbf{83}, 100515(R) (2011)

\bibitem{dhl_AFESE} Fa. Wang, F. Yang, M. Gao, Z-Y. Lu, T. Xiang, and D-H. Lee. Europhys. Lett. \textbf{93}, 57003 (2011).

\bibitem{KFeAs_fRG} R. Thomale,  C. Platt, W. Hanke, J-P. Hu, and B. A.  Bernevig,   arXiv:1101.3593.

\bibitem{kuroki_11a}  K. Suzuki, H. Usui, and K. Kuroki  arXiv:1108.0657.

\bibitem{122hole_ARPES}  K. Nakayama, T. Sato, P. Richard, Y.-M. Xu, T. Kawahara, K. Umezawa, T. Qian, M. Neupane, G. F. Chen, H. Ding, and T. Takahashi,
 Phys. Rev. B \textbf{83}, 020501 (2011);  Y-M. Xu, Y-B. Huang, X-Y. Cui, E. Razzoli, M. Radovic {\it et al}
  Nature Physics {\bf 7}, 198-202 (2011).
\bibitem{ARPES_1111} T. Kondo, A. F. Santander-Syro, O. Copie, C. Liu, M. E. Tillman, {\it et al}
  Phys. Rev. Lett. {\bf 101}, 147003 (2008).
\bibitem{122electron_ARPES} Y. Sekiba, T Sato, K Nakayama, K Terashima, P Richard {\it et al}
 New J. Phys. \textbf{11}, 025020 (2009); K. Terashima \emph{et. al.}, Proceedings of the National Academy of Sciences of the USA (PNAS) \textbf{106}, 7330 (2009).
\bibitem{martin_09}
C. Martin, M. E. Tillman, H. Kim, M. A. Tanatar, S. K. Kim {\it et al}
 Phys. Rev. Lett. 102, 247002 (2009).

\bibitem{efremov}  D.V. Efremov, M.M. Korshunov, O.V. Dolgov, A.A. Golubov, and P.J. Hirschfeld,  arXiv:1104.3840.

\bibitem{resonance}  A. D. Christianson, E. A. Goremychkin, R. Osborn, S. Rosenkranz, M. D. Lumsden,  \emph{et. al.},
Nature \textbf{456}, 930 (2008);  R. Osborn, J.-P. Castellan, S. Rosenkranz, E. A. Goremychkin, D. Y. Chung,
 {\it et al}, arXiv:1106:0771.

\bibitem{eschrig} see e.g. Ar. Abanov, A. V. Chubukov, and J. Schmalian
 Journal of Electron Spectroscopy and Related Phenomena, {\bf 117}, 129 (2001);
  M. Eschrig, Adv. Phys. {\bf 55}, 47 (2006), and references therein.

\bibitem{kuroki_11} Y. Nagai and K. Kuroki,  arXiv:1103.0586;  T.A. Maier, S. Graser, P.J. Hirschfeld, D.J. Scalapino, Phys. Rev. B 83, 220505(R) (2011).

\bibitem{eremin_11} S. Maiti, J. Knolle, I. Eremin, and A.V. Chubukov, arXiv:1108:0266

\bibitem{hanaguri} T. Hanaguri,  S. Niitaka,  K. Kuroki, and H. Takagi
 Science {\bf 328}, 474 (2010).

\bibitem{debates} I.I. Mazin amd D.J. Singh, arXiv 1007.0047; T. Hanaguri, S. Niitaka, K. Kuroki, and H. Takagi, arXiv:1007.0307.

\bibitem{co-existence_Fe}  R. M. Fernandes, D. K. Pratt, W. Tian, J. Zarestky, A. Kreyssig \emph{et. al.},
    Phys. Rev. B {\bf 81}, 140501(R) (2010);  R.M. Fernandes and J. Schmalian, Phys. Rev. B 82, 014521 (2010); A.B.Vorontsov, M.G.Vavilov, and A.V.Chubukov, Phys. Rev. B 81, 174538 (2010);  M. G. Vavilov, A. V. Chubukov, and A. B. Vorontsov, Supercond. Sci. Technol. 23, 054011 (2010).

\bibitem{reid_11}  J.-Ph. Reid, M. A. Tanatar, X. G. Luo, H. Shakeripour, S. René de Cotret {\it et al} . arXiv:1105.2232.

\bibitem{heat_hole}
P. Popovich, A. V. Boris, O. V. Dolgov, A. A. Golubov, D. L. Sun, C. T. Lin, R. K. Kremer, and B. Keimer,
  Phys. Rev. Lett. 105, 027003 (2010).

\bibitem{122hole_pen_depth} C. Martin, R. T. Gordon, M. A. Tanatar, H. Kim, N. Ni, {\it et al},
 Phys. Rev. B 80, 020501(R) (2009);
 R. Khasanov,  D. V. Evtushinsky, A. Amato, H.-H. Klauss, H. Luetkens, {\it et al}
   Phys. Rev. Lett. \textbf{102}, 187005(2009).

\bibitem{impurities}  A.B. Vorontsov, M.G. Vavilov, and A.V. Chubukov,  Phys. Rev. B {\bf 79}, 140507(R) (2009); O.V. Dolgov, A.A. Golubov, D. Parker, New Journal of Physics, {\bf 11}, 075012 (2009); Y. Bang, Europhys. Letters,  {\bf 86}, 47001 (2009).

\bibitem{kim_10}  H. Kim,  R. T. Gordon, M. A. Tanatar, J. Hua, U. Welp {\it et al}
Phys. Rev. B 82, 060518 (2010).

\bibitem{thermal_opt} J.-Ph. Reid
M. A. Tanatar, X. G. Luo, H. Shakeripour, N. Doiron-Leyraud {\it et al}
 Phys. Rev. B 82, 064501 (2010);
M.A. Tanatar, J.-Ph. Reid, H. Shakeripour, X. G. Luo, N. Doiron-Leyraud {\it et al},
  Phys. Rev. Lett. 104, 067002 (2010).

\bibitem{volovik} G.E. Volovik, JEPT Lett., {\bf 58}, 469 (1993).

\bibitem{bang} Y. Bang, Phys. Rev. Lett., 104, 217001 (2010).

\bibitem{hashimoto} K. Hashimoto,  T. Shibauchi, T. Kato, K. Ikada, R. Okazaki  {\it et al},
  Phys. Rev. Lett. 102, 017002 (2009).

\bibitem{pen_opt} R.R. Gordon,  H. Kim, N. Salovich, R. W. Giannetta, R. M. Fernandes  {\it et. al.}
 L. Luan, T. M. Lippman, C. W. Hicks, J. A. Bert, O. M. Auslaender {\it et. al.},
   Phys. Rev. Lett. 106, 067001 (2011).

\bibitem{heat_el_over}  D-J. Jang, A. B. Vorontsov, I. Vekhter, K. Gofryk, Z. Yang {\it et al},
New Journal of Physics 13, 023036 (2011).

\bibitem{wen_10}  B. Zeng, G. Mu, H. Q. Luo, T. Xiang, H. Yang {\it et al},
 arXiv:1007.3597.

\bibitem{LiFeAs_ARPES2} S.V. Borisenko, V. B. Zabolotnyy, D. V. Evtushinsky, T. K. Kim, I. V. Morozov {\it et al}
 Phys. Rev. Lett. \textbf{105}, 067002 (2010).

\bibitem{LiFeAs_ARPES1_nutron_scat} D. S. Inosov, J. S. White, D. V. Evtushinsky, I. V. Morozov, A. Cameron  \emph{et. al.}, Phys. Rev. Lett. \textbf{104}, 187001 (2010).

\bibitem{thermal_pen_LiFeAs} H. Kim, M. A. Tanatar, Y. J. Song, Y. S. Kwon, and R. Prozorov,
 Phys. Rev. B 83, 100502 (2011).

\bibitem{ma_10} L. Ma, J. Zhang, G. F. Chen, and W. Yu
 Phys. Rev. B 82, 180501(R) (2010)

\bibitem{parker} D. Parker, M. G. Vavilov, A. V. Chubukov, and I. I. Mazin,  Phys. Rev. B {\bf 80}, 100508 (2009).

\bibitem{lafop} A.I. Coldea, J. D. Fletcher, A. Carrington, J. G. Analytis, A. F. Bangura,
 {\it et al}, Phys. Rev. Lett. {\bf 101}, 216402 (2008).

\bibitem{matsuda} S. Kasahara, T. Shibauchi, K. Hashimoto, K. Ikada, S. Tonegawa
  {\it et al}, Phys. Rev. B 81, 184519, (2010).

\bibitem{private} T. Shibauchi, private communication.

\bibitem{thermal_p}
J.D. Fletcher,  A. Serafin, L. Malone, J. G. Analytis, J.-H. Chu,
 {\it et al}
  Phys. Rev. Lett. 102, 147001 (2009);
 K. Hashimoto, A. Serafin, S. Tonegawa, R. Katsumata, R. Okazaki
   {\it et al}, Phys. Rev. B 82, 014526 (2010);
  M. Yamashita, N. Nakata, Y. Senshu, S. Tonegawa, K. Ikada,
    {\it et al}, Phys. Rev. B 80, 220509(R) (2009);
    J. S. Kim, P. J. Hirschfeld, G. R. Stewart, S. Kasahara, T. Shibauchi, T. Terashima, and Y. Matsuda,
     arXiv:1002.3355;
     Y. Nakai,  T. Iye, S. Kitagawa, K. Ishida, H. Ikeda,
       {\it et al}, Phys. Rev. Lett. 105, 107003 (2010).

\bibitem{maiti_delta} S. Maiti and A.V. Chubukov, arXiv:1104.2923.

\bibitem{vekhter_11}  M. Yamashita, Y. Senshu, T. Shibauchi, S. Kasahara, K. Hashimoto,
  {\it et al},  arXiv:1103.0885.

\end{thebibliography}
\end{document}